\documentclass[3p,review]{elsarticle}

\usepackage{tabularray}
\usepackage{multirow}
\usepackage{subcaption}
\usepackage{amssymb}
\usepackage{lineno}
\usepackage{mhchem}
\usepackage[acronym]{glossaries}
\usepackage{booktabs}
\usepackage{longtable}
\usepackage{hyperref}
\usepackage{color}
\usepackage[section]{placeins}
\usepackage{xcolor}

\newif\ifmarked
\markedfalse   

\ifmarked\else
  \AtBeginDocument{%
    \renewcommand{\textcolor}[2]{#2}%
    \renewcommand{\color}[1]{}%
  }
\fi

\makeatletter
\newcommand{\leftcaptions}{%
  \setlength{\LTcapwidth}{\textwidth}%
  \def\LT@makecaption##1##2##3{%
    \LT@mcol\LT@cols l{\hbox to\z@{%
      \parbox[t]\LTcapwidth{%
        \reset@font\footnotesize
        \vskip\abovecaptionskip
        \noindent ##1{##2: }##3\endgraf}%
      \hss}}%
  }%
}
\makeatother

\journal{Acta Astronautica}

\begin{document}
\begin{frontmatter}

\title{Radiation Hardness Assurance of Optical Coatings for the COMPASSO Iodine Reference}
%%%%%ONLY FOR ARXIV!!!
\tnotetext[license]{\copyright\ 2026. This manuscript version is made available
under the CC-BY-NC-ND 4.0 license
\url{https://creativecommons.org/licenses/by-nc-nd/4.0/}.
The published version is available at \url{https://doi.org/10.1016/j.actaastro.2026.10.051}.}
%%%%%%%%%%%%%%%%%%%%%%%%%%%%%%%%%%%%%%%%%%%%%%%%%%

\author[iauz,p3run]{Maciej Sznajder\corref{cor1}}
\ead{M.Sznajder@ia.uz.zgora.pl}

\author[dlr]{Tim Blomberg}
\ead{Tim.Blomberg@dlr.de}

\author[dlrry]{Thomas Renger}
\ead{Thomas.Renger@dlr.de}

\author[isi]{Jan Hrabina}
\ead{hrabina@isibrno.cz}

\author[isi]{Jindrich Oulehla}
\ead{oulehla@isibrno.cz}

\author[isi]{Pavel Pokorny}
\ead{pokorny@isibrno.cz}

\author[isi]{Ondrej Cip}
\ead{ocip@isibrno.cz}

\author[isi]{Josef Lazar}
\ead{joe@isibrno.cz}

\author[dlr]{Markus Oswald}
\ead{Markus.Oswald@dlr.de}

\author[dlr]{Jan Wüst}
\ead{Jan.Wuest@dlr.de}

\author[dlr]{Andre Bußmeier}
\ead{Andre.Bussmeier@dlr.de}

\author[dlr]{Frederik Kuschewski}
\ead{Frederik.Kuschewski@dlr.de}

\author[dlr]{Niklas Röder}
\ead{Niklas.Roeder@dlr.de}

\author[dlr]{Johanna Popp}
\ead{Johanna.Popp@dlr.de}

\author[dlrry]{Jonas Pollex}
\ead{Jonas.Pollex@dlr.de}

\author[dlr]{Thilo Schuldt}
\ead{Thilo.Schuldt@dlr.de}

\author[dlr,uulm,iqst]{Claus Braxmaier}
\ead{Claus.Braxmaier@dlr.de}

\cortext[cor1]{Corresponding author}

\affiliation[iauz]{
  organization={Janusz Gil Institute of Astronomy, University of Zielona Góra},
  addressline={Licealna 9},
  city={Zielona Góra},
  postcode={65-417},
  country={Poland}
}

\affiliation[p3run]{
  organization={PW Sznajder, P3RUN$^{TM}$ - Radiation Hardness Assurance \& Data Science Expertise},
  addressline={Dolina Zielona 19A},
  city={Zielona Góra},
  postcode={65-154},
  country={Poland}
}

\affiliation[dlr]{
  organization={DLR Institute of Quantum Technologies},
  addressline={Wilhelm-Runge-Str. 10},
  city={Ulm},
  postcode={89081},
  country={Germany},
}

\affiliation[isi]{
  organization={Institute of Scientific Instruments, Czech Academy of Sciences},
  addressline={Kralovopolska 147},
  city={Brno},
  postcode={61200},
  country={Czech Republic},
}

\affiliation[dlrry]{
  organization={DLR Institute of Space Systems},
  addressline={Robert-Hooke-Str. 7},
  city={Bremen},
  postcode={28359},
  country={Germany},
}

\affiliation[uulm]{
  organization={Ulm University, Institute of Microelectronics},
  addressline={Albert-Einstein-Allee 43},
  city={Ulm},
  postcode={89081},
  country={Germany},
}
%New additional affiliation for Claus Braxmaier
\affiliation[iqst]{
  organization={\textcolor{red}{Ulm University, Center for Integrated Quantum Science and Technology (IQST)}},
  addressline={\textcolor{red}{Albert-Einstein-Allee 11}},
  city={\textcolor{red}{Ulm}},
  postcode={\textcolor{red}{89081}},
  country={\textcolor{red}{Germany}},
}

\begin{abstract}
%Context
The COMPASSO payload demonstrates compact iodine‑based optical frequency references on the International Space Station (ISS), requiring optical windows and antireflective coatings that remain stable in the Low Earth Orbit (LEO) radiation environment. Radiation analysis was performed to quantify how trapped electrons, protons, and ions affect the fused‑silica iodine cell and its coatings under mission‑relevant shielding and exposure scenarios.
  
%Aims
The aim was to determine whether selected coatings meet the COMPASSO optical requirements after realistic LEO irradiation scenarios, identify degradation mechanisms linked to deposition technology and materials, and provide selection and qualification guidance for coatings used in spaceborne iodine spectroscopy and precision optical‑clock payloads.
  
%Methods
The nominal and worst‑case corpuscular environments were derived for the LEO ISS orbit using AE8/AP8, CREME96, ESP, and ISO15390 models to define electron, proton, and heavy‑ion flux and fluence spectra. {\color{red}{The ion component of the environment was assessed by simulation only. The irradiation test used electrons and protons.}} Eight coating stacks produced by different deposition methods were exposed to proton and electron irradiation corresponding to the nominal and worst-case mission loads. Pre‑ and post‑irradiation optical characterization included reflectance and transmittance measurements, together with visual and microscopic inspection to assess coating integrity.

%Results
Laboratory irradiation tests led to the selection of one coating, a four-layer $\mathrm{Ta_2O_5/SiO_2}$ stack, which was then used to produce the iodine cell optical window. The deposited and transferred particle-fluence magnitudes were calculated for the selected coating and all particle species. Also, using GRAS software, the total ionizing dose magnitudes for all layers of the selected {\textcolor{red}{iodine cell}} coating were derived.

%Conclusions
{\color{red}{The selected iodine cell coating will not be affected by corpuscular radiation for years in the LEO ISS radiation environment. GRAS analysis revealed that the TID levels expected at the ISS Bartolomeo platform will not affect its optical properties.}}
\end{abstract}

%\glsresetall

%\begin{graphicalabstract}
%\centering
%graphics here
%\end{graphicalabstract}

%\begin{highlights}
%  \item Comprehensive Radiation Hardness Assurance of COMPASSO optical components for LEO environment, combining AE8/AP8, CREME96, ESP and ISO15390 models to define nominal and worst case test spectra.
%  \item Experimental irradiation of eight coating stacks with representative LEO proton and electron fluences, followed by pre- and post-irradiation optical characterization.
%  \item Practical recommendations for coating selection and qualification protocols to support spaceborne iodine spectroscopy and precision optical-clock payloads. 
%\end{highlights}


\end{frontmatter}

%\linenumbers

%---------------------------------------------------------------------------
\section{Introduction}

Optical systems used in space rely on resilient components in harsh environments. One key aspect is optical coatings, which have to withstand the radiation environment in space. Therefore, optics and its coatings must undergo analysis and testing regarding irradiation and the potential influence of low energy ($<$1MeV) electrons and protons on optical parameters such as reflectance and transmittance \cite{WER18,KUH18}. These aspects are also considered in future space-borne technologies, like satellite-operated optical clocks. These technologies are interesting for future generations of Global Navigation Satellite System (GNSS), such as Galileo, and for new GNSS architectures, such as the Kepler concept developed at DLR \cite{GIO19}.  

As part of a general technology development roadmap for future Galileo, the COMPASSO mission will demonstrate the function of optical {\color{red}{clock}} technologies in space. It will be the first in-orbit verification of compact and highly stable laser-optical clocks \cite{SCH21}. The payload consists of two subsystems, the so-called \textit{iodine-based optical frequency references}, or shortly Iodine Reference (IR). The payload is attached to the Bartolomeo platform outside the Columbus module of the International Space Station (ISS) on slot 2A, see Figure 2a in Steimle C. et al. \cite{STE19}. The iodine references utilise the Doppler-free spectroscopy technique to resolve the iodine hyperfine structure near a wavelength of 532 nm \cite{SCH21}. 

\begin{figure}[!htbp] 
  \begin{center}
    \includegraphics[width=0.9\textwidth]{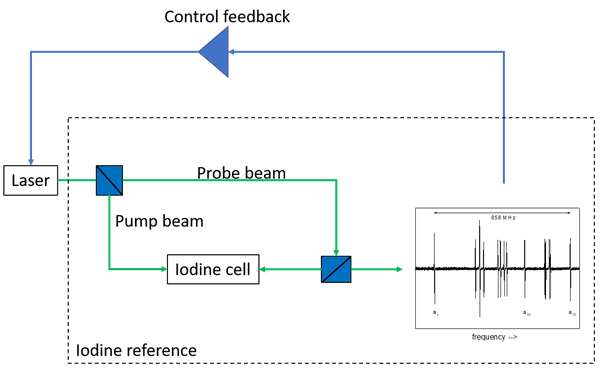}
  \end{center}
  \caption{Schematic representation of the iodine reference. Two counter-propagating laser beams (pump and probe) are guided through an iodine cell and are used to perform Doppler-free detection of hyperfine spectra. The selected hyperfine transition is used for frequency control of the laser.}\label{fig:iodine_principle}
\end{figure}

Based on a technique, two counter-propagating laser beams pass a cell filled with molecular iodine. The Pump-beam (higher intensity) excites the iodine molecules, which the weaker Probe-beam probes. This leads to the hyperfine structure signal necessary for laser frequency stabilisation \cite{SCH21,KUS24}. Figure \ref{fig:iodine_principle} shows a schematic representation of the IR.

\begin{figure}[!htbp] 
  \begin{center}
    \includegraphics[width=1.0\textwidth]{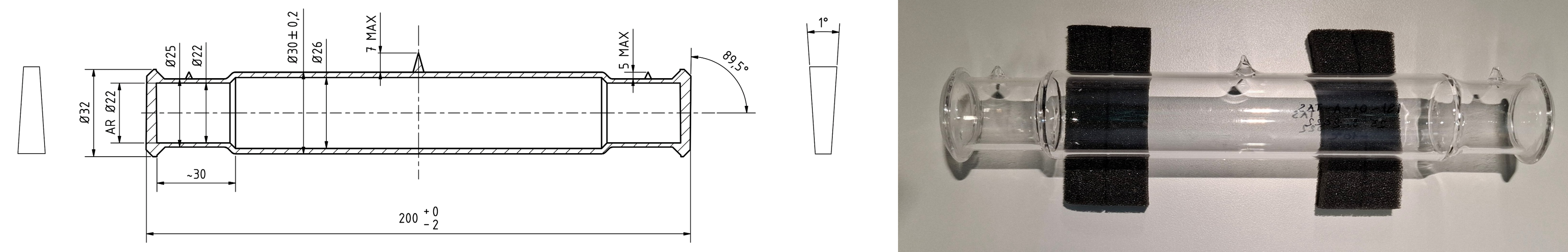}
  \end{center}
  \caption{The design of the {\color{red}{COMPASSO}} setup iodine cell. {\color{red}{Left the technical drawing and right a photograph of the cell.}} The cell is completely made of fused silica glass and its windows are equipped with AR coatings on both inner and outer surfaces (active AR diameter 22 mm).}
  \label{fig:compasso_setup_iodine_cell}
\end{figure}

One key part of the spectroscopy setup is the iodine cell. Its design is shown in Fig. \ref{fig:compasso_setup_iodine_cell}. The cell is made of fused silica glass to minimise the risk of slow iodine contamination caused by the long-term release of impurities from the cell wall \cite{Hrabina2017}. The wedged optical windows are equipped with antireflective coatings on both the inner and outer sides. This ensures minimal optical losses and unwanted back reflections in the setup. The cell is filled with molecular iodine \ce{^{127}I_{2}}, with a certain amount of iodine corresponding to 1.4 Pa (-10$^{\circ}$C saturation point temperature). Using this approach allows us to simplify the setup and omit the cell cold finger and the necessary temperature controller \cite{Hrabina2014}.

Available coatings technologies offered us eight different combinations of deposition methods and coating materials (details in Section \ref{sec:coatings}). These possible coatings were tested regarding proton and electron irradiation to ensure that the cell and its coated windows can withstand the radiation environment in space. 

%---------------------------------------------------------------------------
\section{Radiation analysis}

Radiation analysis (RA) is a process of the Radiation Hardness Assurance (RHA) activity which {\color{red}{leads to ensuring}} that spacecraft's onboard Electrical, Electronic and Electromechanical (EEE) components as well as materials hold acceptable properties change within operational space environment with predefined space mission parameters. The environment determines the magnitudes of electromagnetic and corpuscular radiation. First, degrades mostly outer surfaces of a spacecraft, while the second, ages outer and inner satellite components. The amount of radiation is determined by solar activity and its approx. 11 years cycle. 

The radiation environment of the COMPASSO payload is calculated for the LEO orbit of the ISS, where it is exposed to both corpuscular and electromagnetic radiation. Particles are trapped by the Earth's magnetic field, constituting the so-called inner and outer Van-Allen radiation belts \cite{FSC69,LH19,RCU20}. The inner belt spans to approx. 2.4 Earth radii $R_{\mathrm{E}}$, while the outer one from 2.8 $R_{\mathrm{E}}$ up to 12 $R_{\mathrm{E}}$. Both are separated from each other by the so-called \textit{slot} \cite{SA12}. Occasionally, a third belt is formed, which is mainly filled with MeV-range electrons \cite{BKH13}. The distances stated above are given in the equatorial plane. The ISS is influenced by the inner belt, which is populated with electrons and ions, which consist mainly of protons \cite{FSC69}.

Two environments were considered, the nominal and the worst corpuscular radiation environment, denoted as NCRE and WCRE, respectively. Both are related to space weather conditions. The more severe the space weather, the stronger the Earth's magnetosphere disturbance, and hence, the larger the particle flux spectra magnitude recorded for a selected orbit. Therefore, the WCRE environment represents more harsh radiation conditions for the payload components as opposed to the NCRE environment. 

\subsection{Nominal Corpuscular Radiation Environment}

For the NCRE environment, the ISS orbit of 400 km altitude and an inclination of $\mathrm{51.6^{\circ}}$ was chosen \cite{iss_2010, iss_2015}. 

\subsubsection{Flux Spectra} \label{sec:flux_spectra_ncre}

\begin{figure}[!htbp] 
  \begin{center}
    \includegraphics[width=1.0\textwidth]{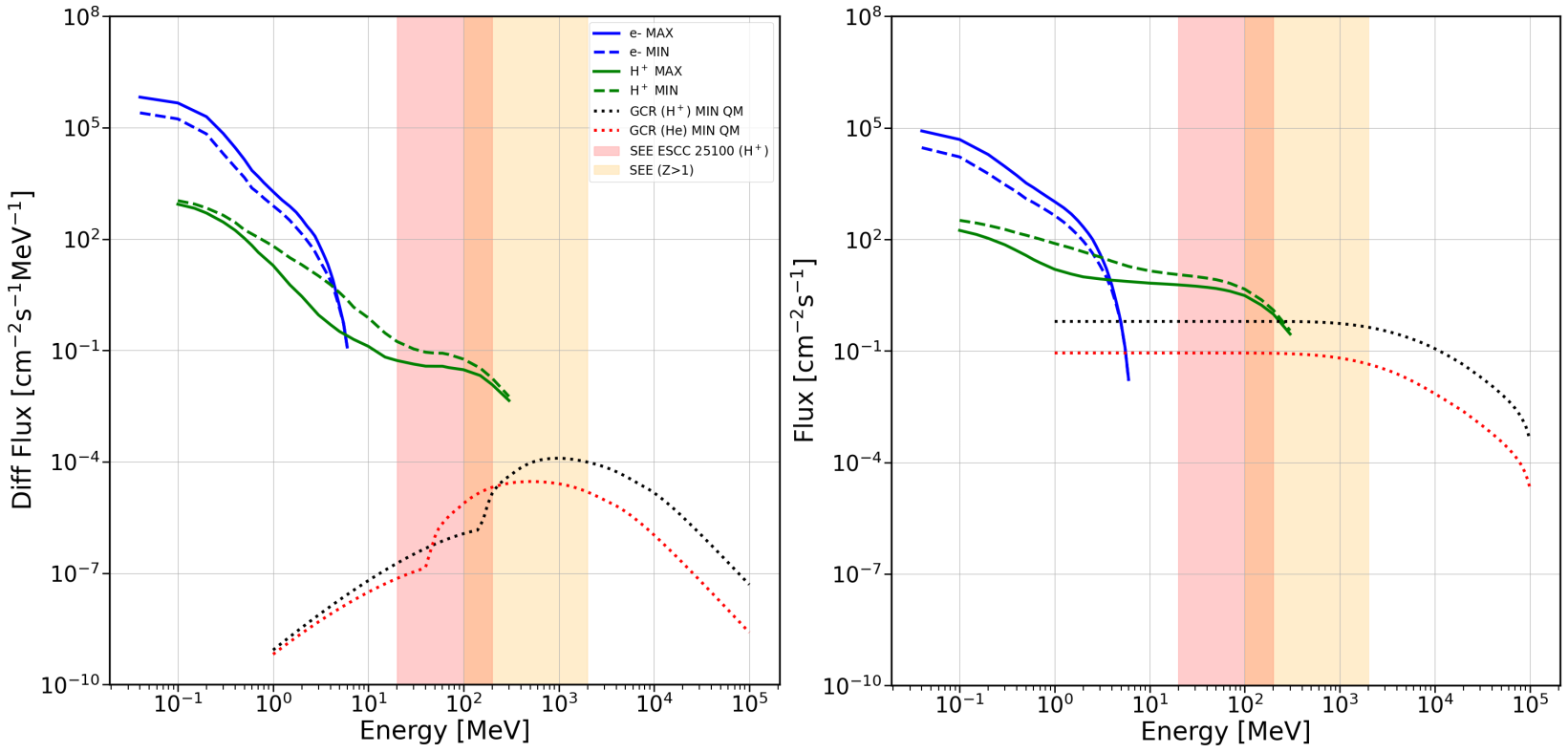}
  \end{center}
  \caption{Differential (left) and integral (right) corpuscular radiation flux for LEO ISS orbit of the NCRE.} \label{fig:ncre_flux_1}
\end{figure}

Radiation belt trapped electrons and protons fluxes, both differential and integral, were calculated by use of the AE8 and AP8 models, respectively. Both are recommended by the ECSS standard \cite{ecss-04C}. Fluxes are shown in Figure \ref{fig:ncre_flux_1}. The left panel depicts the differential, while the right one shows the integral flux, respectively. Electron flux for a case of the solar maximum activity (MAX) is drawn by use of a solid blue line, while for a case of the solar minimum activity (MIN) is drawn by use of the dashed blue line. Electron flux magnitude is larger for the solar maximum than for the solar minimum activity state. Similarly, proton flux was considered for MIN and MAX solar activities and drawn using dashed and solid lines. It can be spotted that the magnitude of the proton flux for the MAX solar activity is lower than for the MIN solar activity for the whole available energy range of the AP8 model. This can be explained by the fact that during solar MAX activity, increased intensity of the UV part of the solar spectrum causes the Earth's atmosphere to expand. This increases energy loss and scattering of trapped protons, resulting in a decrease of proton flux magnitude at selected altitude \cite{ap8min_ap8max}. Proton energy range from 20 MeV to 200 MeV, marked with pink color in Figure \ref{fig:ncre_flux_1}, encapsulates energies recommended by the ESCC standard \cite{escc-25100} for the Single Event Effect (SEE) radiation tests. SEE tests with heavy ions are made with a typical Linear Energy Transfer (LET) range from a few to almost a hundred MeV $\mathrm{cm^{2}}$ $\mathrm{mg^{-1}}$ in Si. This LET range corresponds to ion energy range from $\sim$100 MeV to $\sim$2000 MeV, marked with orange color. Both energy ranges are marked consistently in figures throughout the whole manuscript. For further RHA activities, the electron flux for the solar maximum state and the proton flux for the solar minimum state are used. Larger flux magnitudes result in a more conservative approach, i.e., larger radiation loads in COMPASSO components under study.  

\begin{figure}[!htbp] 
  \begin{center}
    \includegraphics[width=1.0\textwidth]{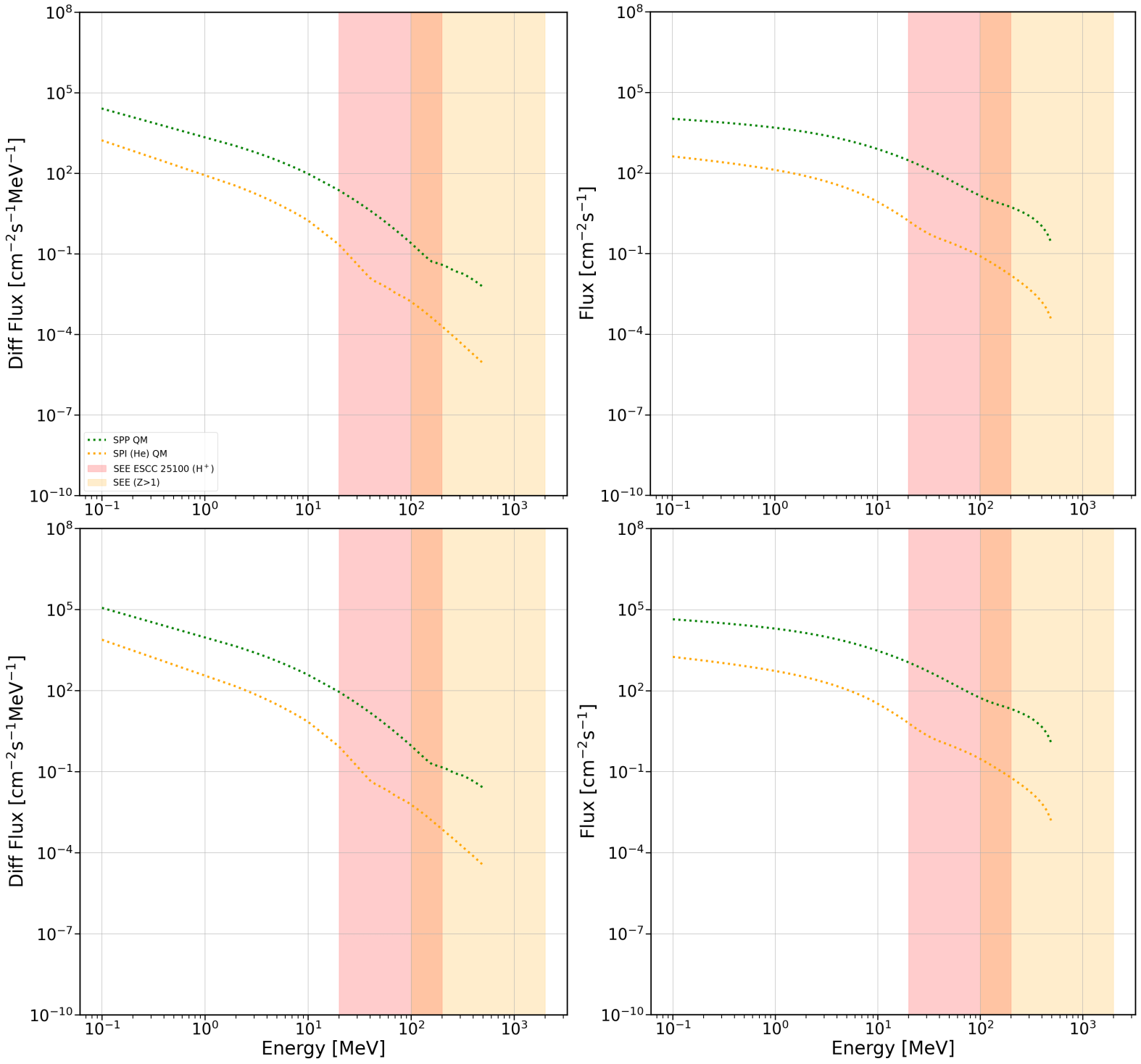}
  \end{center}
  \caption{Differential (upper and lower left) and integral (upper and lower right) corpuscular radiation flux spectra for LEO ISS orbit. CREME96 worst day (upper left and right) and peak 5min (lower left and right) SPP and SPI fluxes of the OCT 1989 event are presented.} \label{fig:ncre_wd_p5_cpp_spi}
\end{figure}

Solar peak proton (SPP) and ion (SPI) fluxes were calculated using the CREME96 model \cite{TAB97} recommended by the ECSS standard \cite{ecss-04C}. Fluxes were computed for the \textit{worst day} and \textit{peak 5-min} average flux of the OCT 1989 event \cite{SAN92}. Results for both cases are presented in Figure \ref{fig:ncre_wd_p5_cpp_spi}. The differential flux is shown on two left, and the integral flux is on two right plots of the figure. Here, only the fluxes of He ions are plotted for ions with $Z$$>$1, as this ion species exhibits the highest flux magnitude among all such ions in the interplanetary solar plasma. Fluxes were computed for the quiet state of the Earth's magnetosphere (QM). The so-called rigidity cutoffs of solar ions were calculated using the St{\o}rmer method \cite{STO30,AST81}. This method is recommended by the ECSS standard \cite{ecss-04C}. {\color{red}{The St{\o}rmer}} method uses a dipole approximation for the Earth's magnetosphere. It is a fair choice for considering calm solar activity states, therefore, {\color{red}{it is used for the}} NCRE. The \textit{peak 5-min} average SPP and SPI fluxes are taken for further MC calculations in the context of the TID, TNID, and particle fluence estimations since they have higher magnitude than those for the \textit{worst day} case. Hence, they represent more conservative approach for estimating radiation loads in COMPASSO components under study.   

Proton and helium ion fluxes for the Galactic Cosmic Rays (GCR) were calculated using the ISO15390:2004 model recommended by the ECSS standard \cite{ecss-04C}. Also here, quiet state of the magnetosphere (QM) was used. The GCR proton flux is drawn with black dotted line and the helium with red dotted line, respectively.   

A satellite orbiting the Earth in the LEO environment is exposed to a flux of Atomic Oxygen (ATOX). ATOX is produced by the photodissociation effect caused by the UV light of the Sun's electromagnetic radiation. When a satellite, e.g., the ISS, orbits the Earth with an average speed of approx. 7.6 km/s, all of its RAM direction surfaces are hit by atmospheric particles. RAM refers to the side of the spacecraft that faces the direction of its motion as it travels through space. These include ATOX. The energy of a single surface ATOX hit is approx. 5 eV. ATOX flux magnitude is dependent on a satellite orbit, solar activity state, and the angle between the flight direction and a normal to the material surface facing directly the radiation environment \cite{ADR10}. For ISS orbit, one can expect the ATOX flux magnitude of about $10^{14}$ $\mathrm{cm^{-2}s^{-1}}$ for an intermediate solar activity \cite{ADR10}.

\subsubsection{Fluence Spectra}

\begin{figure}[!htbp] 
  \begin{center}
    \includegraphics[width=1.0\textwidth]{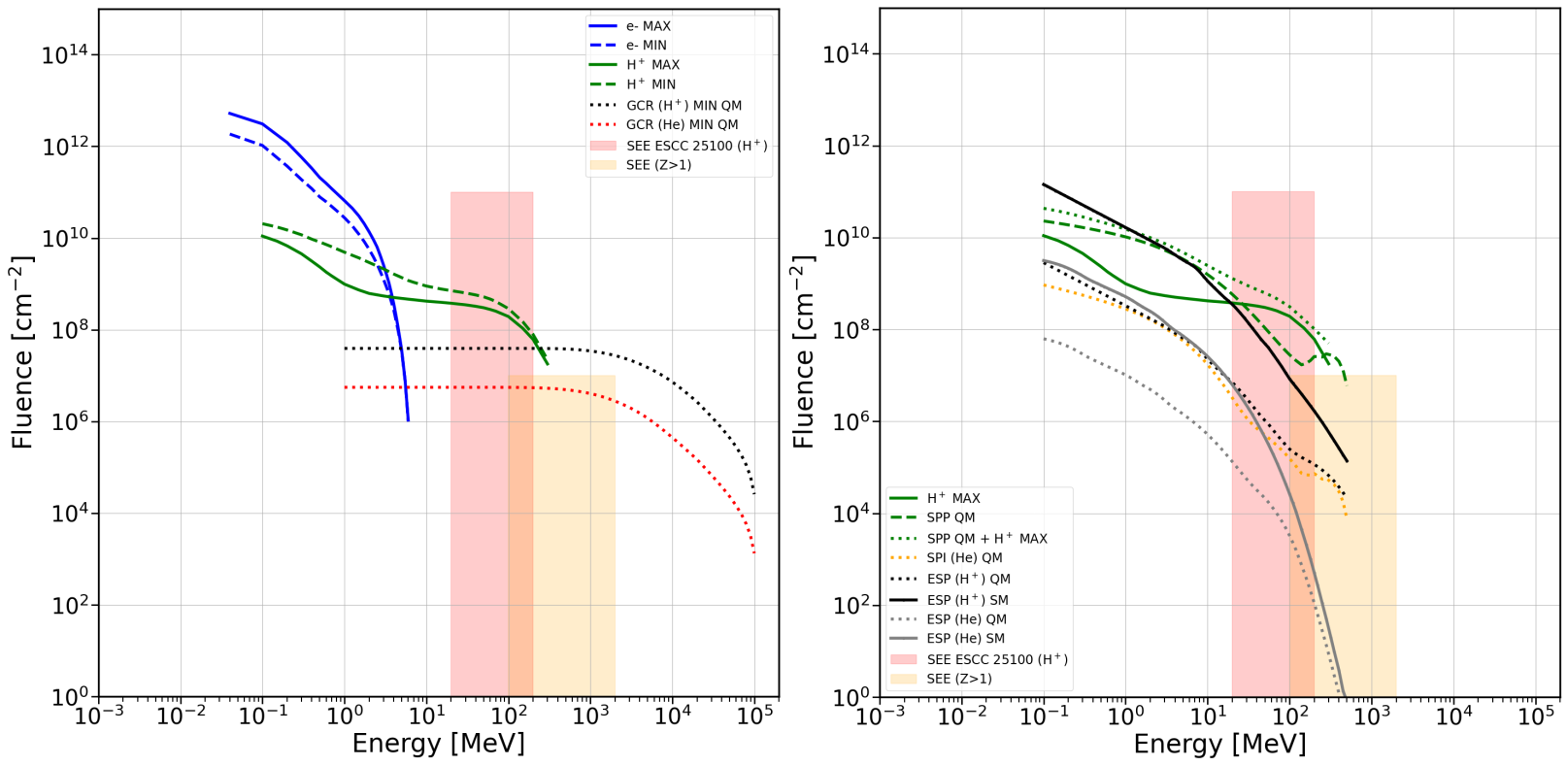}
  \end{center}
  \caption{Left plot, fluence spectra of the trapped electrons, protons and GCR for NCRE. Right plot, proton fluence spectra calculated from AP8, CREME96 and ESP models. Helium ions spectra are calculated from CREME96 and ESP models.} \label{fig:ncre_fluence}
\end{figure}

Fluence is defined as {\color{red}{the flux spectrum integrated over time}} \cite{ecss-04C}. Hence, here, the flux spectra are integrated over the mission time, i.e., 2 years. Fluence spectra are shown in Figure \ref{fig:ncre_fluence}. The left figure shows spectra of trapped electrons (blue lines), trapped protons (green lines), and GCR (black and red lines). Here, solar minimum and maximum states and quiet Earth's magnetosphere state were used to derive the fluence. {\color{red}{The light}} red fluence bar between 20 MeV and 200 MeV {\color{red}{corresponds}} to proton energy range recommended by the ESCC 25100 standard \cite{escc-25100} which gives guidelines for Single Event Effects (SEE) testing. The orange and the light red bar heights refer to magnitudes of heavy ions fluence ($10^7$ $\mathrm{cm^{-2}}$) and proton fluence ($10^{11}$ $\mathrm{cm^{-2}}$) recommended by the standard. Here, it must be emphasized that the recommendations are made only for EEE components and SEE testing. As seen from the figure, trapped electrons for the maximum activity state of the Sun have approx. 3-orders of magnitude higher fluence than the fluence of trapped protons for minimum activity state of the Sun. Next, GCR for protons and He ions have fluence magnitude of $\sim$$10^7$ $\mathrm{cm^{-2}}$ and $\sim$$10^6$ $\mathrm{cm^{-2}}$, respectively. Hence, it is approx. 3- and 4-orders of magnitude smaller than for trapped protons. 

Lastly, SPP and SPI (He ions) fluence, for QM state and based on \textit{peak 5-min} flux, has a magnitude of $\approx$$4$$\times$$10^{10}$ $\mathrm{cm^{-2}}$ and $\approx$$9$$\times$$10^{8}$ $\mathrm{cm^{-2}}$, respectively. Proton fluence spectrum results from SPP and trapped protons and is {\color{red}{drawn}} using dotted green line. It is denoted as \textit{SPP QM + $H^+$ MAX}. Here, the spectrum has a magnitude almost two times larger than its constituents. That's an expected result since both, the SPP (dashed green line) and the trapped $\mathrm{H^+}$ spectra (solid green line), have similar fluence magnitudes. Therefore, the combined spectrum is used for further calculations. The figure also shows fluence spectra computed using the Emission of Solar Protons (ESP) model in two cases with stormy and quiet magnetosphere states. The model is used to derive proton fluence spectra from solar particle events \cite{ecss-04C}. Proton fluence calculated from CREME96 model has a magnitude larger than {\color{red}{that}} calculated from {\color{red}{the}} ESP model in QM state but smaller in case of SM state. {\color{red}{An analogous}} situation {\color{red}{applies to}} the SPI, however, in the energy range from 20 MeV to 200 MeV for protons and from 100 MeV to 2000 MeV for SPI, the fluence calculated from CREME96 model has larger magnitude than fluence estimated from {\color{red}{the}} ESP model. Both energy ranges are relevant for radiation testing of EEE components as well as, in case of COMPASSO payload, optical components located behind an {\color{red}{Al-shield}}. Hence, CREME96 based spectra are considered for further calculations.

\subsubsection{NCRE Particle Spectra for Further RHA Activities}

Flux and fluence spectra used in the further RHA activities are summarized in {\color{red}{Table \ref{tab:ncre_spectra_recap}}}. The first column contains the particle species, the second column holds the model (or spectrum) used to calculate flux and fluence, the third column indicates the solar activity state, and the last column provides a reference to a spectrum figure.

\begin{table}[!htbp] 
	\centering
	\scriptsize
	\begin{tabular}{ l|c|c|l } 
		Species & Model/Spectrum & Sun state & Spectrum reference \\ \hline
		$\mathrm{e^-}$ & AE8 MAX     & Maximum & Figure \ref{fig:ncre_flux_1} and Figure \ref{fig:ncre_fluence} (solid blue line) \\
                & & & \\
		$\mathrm{H^+}$               & CREME96 (peak 5-min) \& AP8 MIN & Maximum & Figure \ref{fig:ncre_flux_1} (dashed green line) and \\
                &  & (CREME96 model)  & Figure \ref{fig:ncre_wd_p5_cpp_spi} and Figure \ref{fig:ncre_fluence} (dotted green line) \\
                & & & \\
                & & & \\
		He ions          & CREME96 (peak 5-min) & Maximum & Figure \ref{fig:ncre_wd_p5_cpp_spi} and Figure \ref{fig:ncre_fluence} (dotted orange line) \\
                & & & \\
		GCR ($\mathrm{H^+}$) & ISO15390:2004 & Minimum & Figure \ref{fig:ncre_flux_1} and Figure \ref{fig:ncre_fluence} (dotted black line)\\
                & & & \\
		GCR ($\mathrm{{He}}$ ions) & ISO15390:2004 & Minimum & Figure \ref{fig:ncre_flux_1} and Figure \ref{fig:ncre_fluence} (dotted red line)\\
		\hline
	\end{tabular}
	\caption{NCRE flux and fluence spectra used for further RHA activities. Here, quiet Earth's magnetosphere state is considered.}
  \label{tab:ncre_spectra_recap}
\end{table}

\subsection{Worst Case Corpuscular Radiation Environment}

The WCRE of the COMPASSO payload is calculated for the LEO ISS orbit with an altitude of 500 km and an inclination of $\mathrm{51.6^{\circ}}$. Such an altitude choice was dictated by the system requirements (SR) of the COMPASSO project.

\subsubsection{Flux Spectra}
\begin{figure}[!htbp] 
  \begin{center}
    \includegraphics[width=1.0\textwidth]{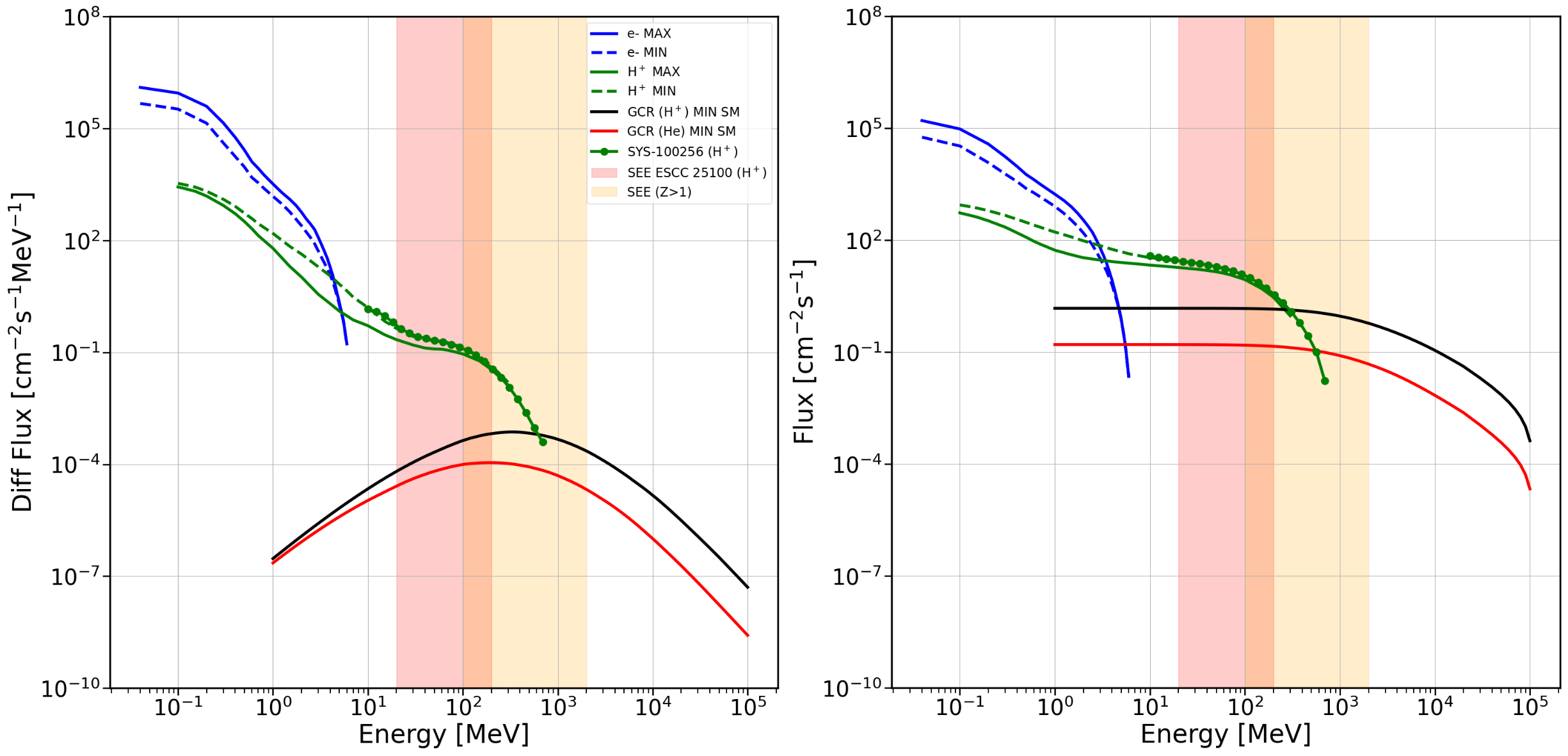}
  \end{center}
  \caption{Differential (left) and integral (right) corpuscular radiation flux for LEO ISS orbit of the WCRE.} \label{fig:wcre_iss_500km}
\end{figure}

The flux of trapped protons is specified by the COMPASSO system requirements, and spectra are calculated based on the AP8 MIN model (spectra are named as SYS-100256). However, SYS-100256 spectra start from an energy of 10 MeV, while spectra calculated by use of the SPENVIS system (also AP8 MIN model) start from 0.1 MeV. Both spectra are presented in Figure \ref{fig:wcre_iss_500km} and are plotted with dashed and solid-dotted green lines. The left panel depicts the differential, while the right one shows the integral flux, respectively. The SYS-100256 spectra are taken for further calculations. For the COMPASSO payload, protons with kinetic energy lower than 10 MeV are fully blocked by its radiation shield. COMPASSO system requirements don't specify the spectrum of trapped electrons. Hence, such a spectrum is calculated from the AE8 MAX model for the maximum solar activity state. Here, the maximum Sun activity has been chosen since the electron spectrum has a higher magnitude than during the solar minimum activity state. An opposite situation is with proton flux magnitudes, i.e., flux magnitudes are higher for minimum solar activity states rather than during maximum states. This effect is explained in Section \ref{sec:flux_spectra_ncre}. The spectrum is presented in Figure \ref{fig:wcre_iss_500km}. It is plotted with the solid blue line, and for comparison, the electron AE8 MIN spectrum is plotted with a dashed blue line.       

\begin{figure}[!htbp] 
  \begin{center}
    \includegraphics[width=1.0\textwidth]{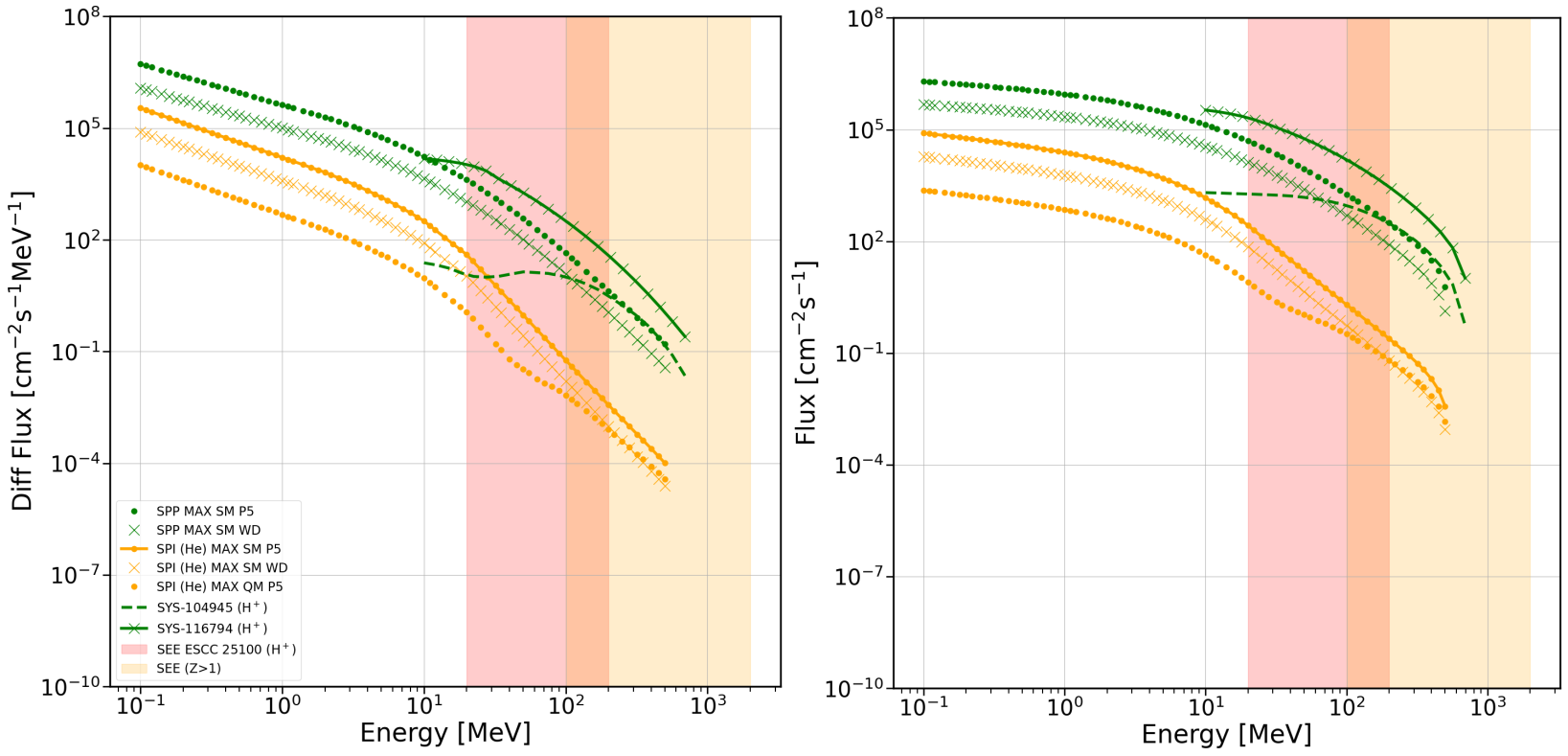}
  \end{center}
  \caption{Differential (left) and integral (right) corpuscular SPP and SPI radiation flux for LEO
 ISS orbit of the WCRE.} \label{fig:wcre_spp_spi_flux}
\end{figure}

The project's system requirements specify a peak trapped proton flux (here named as SYS-104945). It is a flux received by the ISS when it \textit{passes through the South Atlantic Anomaly (SAA) on 50\% of its orbits for a duration of 5-10 minutes, the Epoch 1964 geomagnetic field is used.} Also as it is emphasized in the SR document, the spectrum is based on the AP8 MIN model flux and the CREME model flux. The spectrum is shown in Figure \ref{fig:wcre_spp_spi_flux} by use of a dashed green line. 

System requirements also define the so-called solar-peak (extreme) proton flux for the WCRE case (named as SYS-116794). There, the flux is derived from the extreme solar proton environment of the October 1989 event and reported in the NASA \textit{Space Station Ionizing Radiation Environment} document \cite{NAS94}. The spectrum is plotted in Figure \ref{fig:wcre_spp_spi_flux} using a green solid-dotted line, where, for comparison, spectra from the CREME96 model and for \textit{peak 5-min} (P5), the \textit{worst day} (WD) average fluxes, and the stormy state of the magnetosphere (SM) are plotted using a green dotted and $x$-sign lines, respectively.  

SPI fluxes of He ions and ions with $Z$$>$2 for WCRE case are calculated using the CREME96 model recommended by the ECSS standard \cite{ecss-04C}. Fluxes were computed for the P5 and WD average flux of the OCT 1989 event. Spectra are shown in Figure \ref{fig:wcre_spp_spi_flux}. Differential and integral SPI (He) fluxes for the P5 and the WD spectra are plotted by use of orange solid-dotted and $x$-symbol lines, respectively. Additionally, the QM state and P5 flux are considered, and they are plotted using a dotted orange line. Here, further calculations consider a combination of SM and QM states. This choice is discussed in Section \ref{sec:fluence_spectra}. 

Similarly to NCRE, proton and helium ion fluxes for the GCRs were calculated using the ISO15390:2004 model. However, only the stormy state of the magnetosphere SM is considered. It brings larger flux magnitudes than in the case where the QM state is used. Also, here, only a minimum solar activity state is used to derive the spectra. The flux is shown in Figure \ref{fig:wcre_iss_500km}. Black lines represent the GCR proton flux, and the helium flux by red lines, respectively.

ATOX flux for an orbit with an altitude of 500 km is $10^{14}$ $\mathrm{cm^{-2}s^{-1}}$ \cite{ADR10} for times during maximum solar activity state. The influence of the ATOX on the COMPASSO payload is discussed in Section \ref{sec:impact_other_rad_sources}.

\subsubsection{Fluence Spectra} \label{sec:fluence_spectra}

\begin{figure}[!htbp] 
  \begin{center}
    \includegraphics[width=1.0\textwidth]{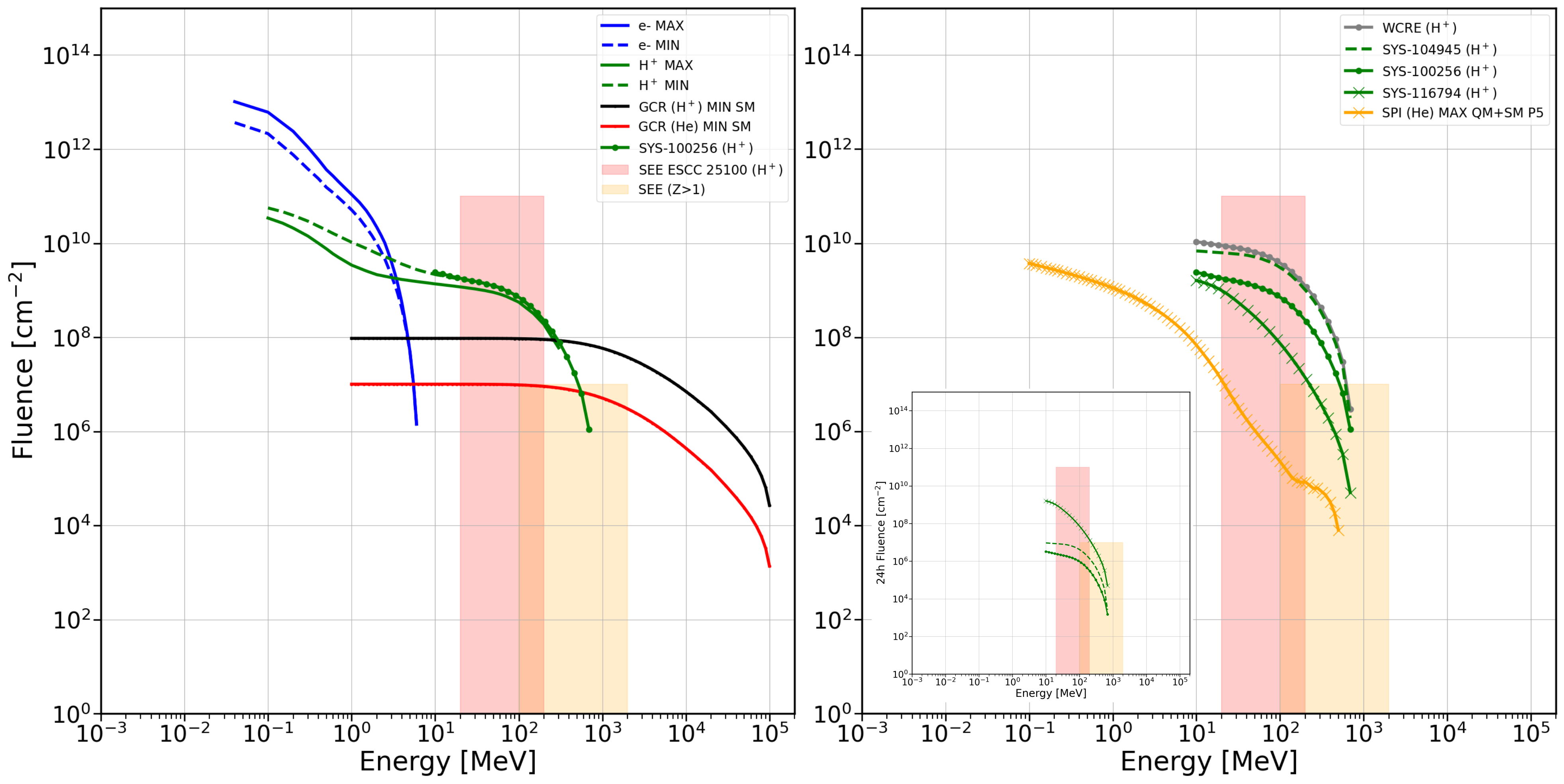}
  \end{center}
  \caption{Fluence spectra of the trapped electrons and protons together with GCR (left) and SPP accompanied with SPI (right) for the WCRE environment.} \label{fig:wcre_fluence}
\end{figure}

Fluence spectra are obtained using the integral flux spectra presented in Figure \ref{fig:wcre_iss_500km} and Figure \ref{fig:wcre_spp_spi_flux}. They are shown in Figure \ref{fig:wcre_fluence}. The left plot depicts the spectrum of trapped electrons (blue lines), trapped protons (green lines), and GCR (black and red lines). Here, solar minimum and maximum states as well as Earth's SM state were used to derive the fluence. Trapped electrons for the maximum activity state of the Sun have approx. 2-orders of magnitude higher fluence than the fluence of trapped protons. Trapped proton fluence derived from proton flux from the SYS-100256 (solid-dotted green line) is complied with the one from the AP8 MIN model implemented in the SPENVIS system. Next, GCR for protons and He ions and SM state and minimum activity state of the Sun have fluence of $\sim$$10^{8}$ $\mathrm{cm^{-2}}$ and $\sim$$10^{7}$ $\mathrm{cm^{-2}}$ for E$>$1 MeV, respectively. 

The proton fluence calculated from the flux SYS-104945 is determined according to the requirement, stating again that the payload is exposed to SSA radiation conditions for 50\% of its orbits for a duration of 5-10 minutes. Here, a 10-minute exposure time is assumed for higher redundancy of the resulting fluence. It is depicted using a dashed green line in Figure \ref{fig:wcre_fluence} (right).

Proton fluence spectrum derived from solar peak (extreme) proton flux SYS-116794 is drawn using a solid-x green line of the right plot in Figure \ref{fig:wcre_fluence}. Here, again, to calculate the fluence, the exposure time of the payload to the radiation must be determined. The project requirement states that the payload \textit{is experienced over each approximately 10-minute period per half orbit when the ISS is at the magnetic latitude extremes of its orbit for the worst 6-8 orbits over the 24-hour day}. Here, for further calculations, 8 orbits are considered. Also, the requirement states that \textit{the extreme environment occurs once over an 11-year solar cycle period, and lasts for approximately 24 hours.} Here, it is assumed that such extreme space weather occurs once during the mission time.   

During the complete mission time, the proton fluence spectrum is a result of a proportional combination of proton spectra defined by SYS-100256, SYS-104945, and SYS-116794 and denoted as WCRE ($\mathrm{H^+}$). It is presented in Figure \ref{fig:wcre_fluence} using a gray solid-dotted line. As seen from the figure, the spectrum lies just above the SYS-104945 proton spectrum.

It needs to be emphasized that particle fluence spectra are calculated for the whole mission period. Here, the SYS-116794 spectrum brings the smallest contribution to the overall proton fluence magnitude. However, taking 24 hours period, so the time during which the extreme conditions take place, the spectrum brings the highest contribution to the overall proton fluence, see plot placed within a small window in Figure \ref{fig:wcre_fluence}.

Fluence of SPI (He ions) is calculated from a combination of CREME96 \textit{peak 5-min} flux spectra derived for quiet and stormy Earth's magnetosphere, see Figure \ref{fig:wcre_spp_spi_flux} dotted and solid-dotted orange lines, respectively. Here, the strategy was to use the same approach as for the proton fluence spectra calculation procedure, i.e., the fluence is a result of time integration of the SPI flux for SM, which lasts 24 hours, and QM, which lasts for the rest of the mission time, see Figure \ref{fig:wcre_fluence}, solid-x orange line.

\subsubsection{WCRE Spectra for Further RHA Activities}

Flux and fluence spectra used in the further RHA activities are summarized in Table \ref{tab:wcre_spectra_recap}. The first column contains the particle species, the second column holds the model (or spectrum) used to calculate flux and fluence, the third column indicates the solar activity state, the fourth column holds the Earth's magnetosphere state used to generate the spectra, and the last column provides a reference to a spectrum figure.    

\begin{table}[!htbp] 
	\centering
	\scriptsize
	\begin{tabular}{ l|c|c|c|l } 
		Species & Model/Spectrum & Sun state & Magn. state & Flux \& Fluence spectra \\ \hline
		$\mathrm{e^-}$ & AE8     & Maximum & --- & Figure \ref{fig:wcre_iss_500km} and Figure \ref{fig:wcre_fluence} (solid blue line) \\ 
                & & & & \\
		$\mathrm{H^+}$           & WCRE ($\mathrm{H^+}$) & --- & --- & Figure \ref{fig:wcre_iss_500km} (solid-dotted green line) \\
                & & & & Figure \ref{fig:wcre_spp_spi_flux} (solid-x and dashed green lines) \\
                & & & & Figure \ref{fig:wcre_fluence} (solid-dotted gray line) \\
                & & & & \\
		He ions          & CREME96 (peak 5-min) & Maximum & QM+SM & Figure \ref{fig:wcre_spp_spi_flux} (dotted and solid-dotted orange lines) \\
                & & & & Figure \ref{fig:wcre_fluence} (solid-x orange line) \\
                & & & & \\
                & & & & \\
		GCR ($\mathrm{H^+}$) & ISO15390:2004 & Minimum & SM & Figure \ref{fig:wcre_iss_500km} (solid black line) \\
                & & & & Figure \ref{fig:wcre_fluence} (solid black line) \\
                & & & & \\
		GCR ($\mathrm{{He}}$ ions) & ISO15390:2004 & Minimum & SM & Figure \ref{fig:wcre_iss_500km} (solid red line) \\
                & & & & Figure \ref{fig:wcre_fluence} (solid red line) \\
		\hline
	\end{tabular}
	\caption{WCRE flux and fluence spectra used for further RHA activities. Spectra are defined by a model, Sun activity and the Earth's magnetosphere states.}
        \label{tab:wcre_spectra_recap}
\end{table}

\subsection{Spectrum of Electromagnetic Radiation}

\begin{figure}[!htbp] 
  \begin{center}
    \includegraphics[width=1.0\textwidth]{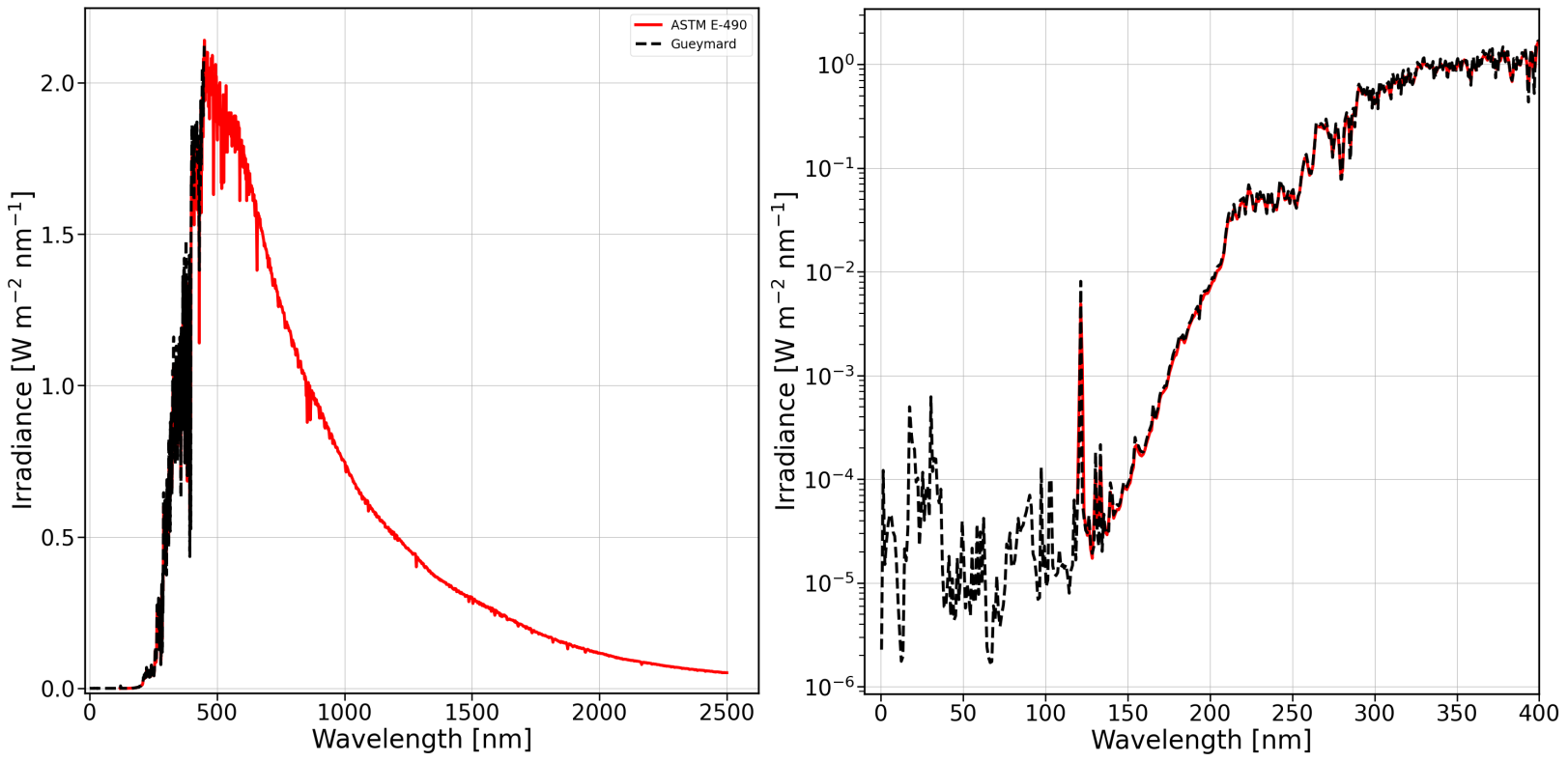}
  \end{center}
  \caption{Spectrum of the electromagnetic radiation derived from the ASTM 490 (red line) standard \cite{ASTM} and the Gueymard data set (black dashed line) \cite{GUE04}.} \label{fig:emr_spectra}
\end{figure}

The spectrum of electromagnetic radiation (EMR) recommended by the ECSS standard \cite{ecss-04C} is the ASTM\footnote{ASTM - American Society for Testing and Materials.} E-490 standard spectrum \cite{ASTM}. It is defined at 1 AU distance from the Sun and outside of the atmosphere, i.e., Air Mass 0 (AM0). The spectrum is defined from 119.5 nm to $10^6$ nm. Its Total Solar Irradiance (TSI) is 1362 W$\mathrm{m^{-2}}$, and it is referenced as 1 solar constant (SC). As already stated, the spectrum starts from 119.5 nm, hence, there is no information about TSI magnitude for soft X-rays (or XUV, from 0.1 nm to 10 nm) and for a complete Extreme UV range (EUV, from 10 nm to 121 nm). Therefore, the Gueymard data set \cite{GUE04} is used to obtain a complete solar EMR spectrum. It is defined from 0.5 nm to 450 nm. Both spectra, the E-490 and from the Gueymard data set, are shown in Figure \ref{fig:emr_spectra}. 

A quantitative comparison between the two spectra is provided in Table \ref{tab:tsi_emr_spectra}. Clearly, there is only a 4\% difference in their TSI magnitude. That is in a wavelength range where both have defined the irradiance, i.e., from 119.5 nm to 450 nm. The Gueymard spectrum, below 119.5 nm, has a TSI of 0.0053 W $\mathrm{m^{-2}}$. However, photons in that $\lambda$-range, and in particular in the UV range, can irreversibly degrade optical components \cite{NEU72, SEM95}. Please note that ECSS standard \cite{ecss-04C} defines UV range from 100 nm to 400 nm. This range covers approx. 8\% of 1 SC. That is 110.9 W $\mathrm{m^{-2}}$ when based on the Gueymard spectrum. The impact of EMR on the COMPASSO payload is discussed in Section \ref{sec:impact_other_rad_sources}.

\begin{table}[!htbp]
	\scriptsize
	\begin{center}
	\begin{tabular}{ l|c|c } 
		Spectrum & TSI $<$119.5 nm; 450 nm$>$ [$\mathrm{W\ m^{-2}}$] & TSI $<$0.5 nm; 119.5 nm$>$ [$\mathrm{W\ m^{-2}}$] \\ \hline
		ASTM E-490        & 192.43  & --- \\
		Gueymard data set & 201.15  & 0.0053  \\
		\hline
	\end{tabular}
	\caption{TSI values for different wavelength ranges of ASTM E-490 and Gueymard spectra.}
	\label{tab:tsi_emr_spectra}
	\end{center}
\end{table} 

%---------------------------------------------------------------------------
\section{Impact of Other Radiation Sources on COMPASSO Payload} \label{sec:impact_other_rad_sources}
\subsection{Atomic Oxygen}\label{sec:degradation_atomic_oxygen}
Materials {\textcolor{red}{exposed to ATOX erode through two distinct mechanisms. The first is the physical phenomenon of}} sputtering \cite{SIGP69,MACD70}, i.e., target material atoms are ripped off their bulk due to a hit event with an incident ion. {\textcolor{red}{Sputtering is a threshold process, i.e., no material is removed until the energy of the incident particles exceeds a value characteristic of the given gas/target combination. Above this threshold, the amount of sputtered material increases with the energy of the incoming particles \cite{ADR89}.}} {\textcolor{red}{The second mechanism is chemical, in the LEO environment ATOX is a highly reactive oxidant that impacts the surface with a kinetic energy of approx. 5 eV and undergoes an oxidation reaction with the exposed material \cite{ADR10}. Whether this chemical attack results in net material loss depends on the nature of the oxide that is formed. For polymers, carbon, and for metals such as silver or osmium, volatile or poorly adherent oxides are produced, which continuously expose fresh material and hence lead to erosion. In contrast, for metals such as aluminium, copper, and steel, as well as for silicon, the reaction product is a dense, adherent, and non-porous oxide that remains on the surface and protects the underlying bulk from further oxidation \cite{ADR10}.}} 

Estimated ATOX flux for both the NCRE and the WCRE equals $10^{14}$ ATOX $\mathrm{cm^{-2}s^{-1}}$. Hence, ATOX fluence for a period of 2 years mission time equals $6.3\times 10^{21}$ ATOX $\mathrm{cm^{-2}}$. It is assumed{\textcolor{red}{, for a more conservative approach,}} that during the whole mission time, an angle between the normal of one of the COMPASSO payload outer shell surfaces and the ISS RAM direction equals $0^{\circ}$. Hence, {\textcolor{red}{resulting fluence represents its maximum}} possible magnitude. One of the project requirements states that each COMPASSO space sub-systems shall be designed to withstand an average atomic oxygen yearly fluence of $5\times10^{21}$ ATOX $\mathrm{cm^{-2}}$. Hence, the {\color{red}{fluence magnitude estimated here}} is approx. 37\% smaller than that from the requirement. For {\textcolor{red}{aluminium}} used as a target material the so-called Erosion Yield (EY), which informs how much material volume is being {\textcolor{red}{removed}} by a single ATOX, equals 0 $\mathrm{cm^{3}}$ ATOX atom$^{-1}$ \cite{ldef1}. \textcolor{red}{It should be noted that the erosion yield is not a fixed material constant. Flight experiments performed during, e.g., the MISSE missions have shown that the ATOX erosion yield of a given material can vary with the accumulated ATOX fluence, with the co-incident solar electromagnetic exposure, and with the temperature \cite{DGB19}.} \textcolor{red}{\textit{Vanishing} macroscopic erosion yield does not imply that aluminium is chemically inert towards ATOX. On the contrary, the exposed pure aluminium reacts readily with the impinging oxygen atoms, but in doing so it instantaneously forms an extremely thin, dense, and tightly adherent passivation layer of aluminium oxide (\ce{Al2O3}). Since this oxide is non-porous and protective, it blocks the further inward diffusion of oxygen atoms towards the underlying metal and is therefore self-limiting, so that no continued oxidation or net thickness loss is observed on a macroscopic scale \cite{ADR10}. It is this self-passivation mechanism, rather than a genuine absence of chemical reactivity, that is responsible for the zero measured erosion yield of aluminium.} For the COMPASSO payload, the outer shell is made of {\textcolor{red}{7 mm aluminium}}, and there are no {\color{red}{gaps or holes}} that could allow ATOX to enter the payload and harm its components. Also, for the LEO ISS orbit, ATOX particles {\textcolor{red}{with their kinetic energy of 5 eV have}} a negligible penetration depth in {\textcolor{red}{aluminium}}, i.e., approx. 0.2 nm. This value was derived by use of the SRIM \cite{ZZB10} software. Therefore, ATOX brings no degradation risk to the COMPASSO components. 

\subsection{Electromagnetic Radiation}\label{sec:degradation_emr}
Electromagnetic radiation brings no harm to COMPASSO components. It is effectively blocked by the system's aluminium shell. This can be verified using the Beer-Lambert law. There, a light intensity $I=I(\lambda,x)$ is calculated as a function of photon wavelength $\lambda$, and its penetration depth $x$ within a target material:

\begin{equation}
    I = I_{\mathrm{0}}e^{-\alpha x}, \qquad \alpha = \frac{4\pi k}{\lambda}.
\end{equation}

\begin{figure}[!htbp] 
  \begin{center}
    \includegraphics[width=1.0\textwidth]{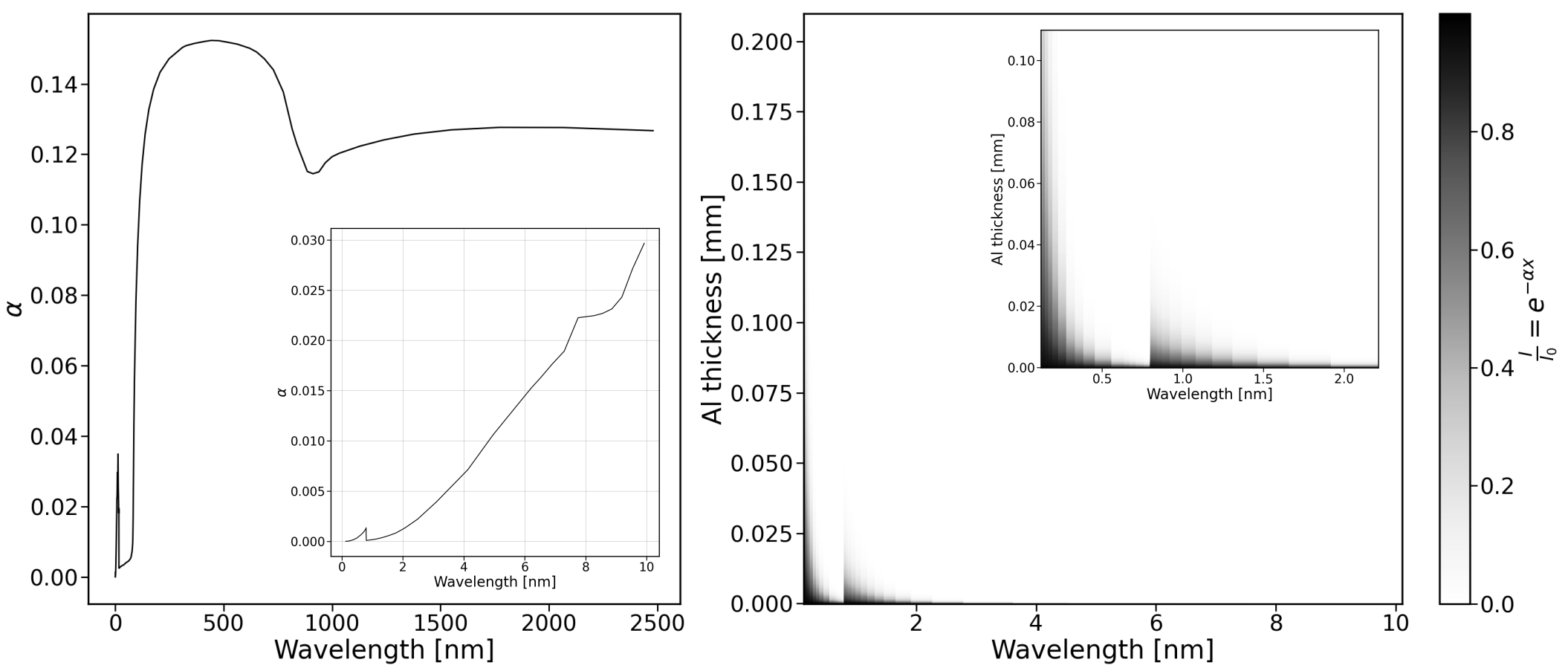}
  \end{center}
  \caption{Attenuation coefficient $\alpha$ as a function of wavelength with a magnified range from 0.1 nm to 10 nm (left). Relative change of a light intensity $\frac{I}{I_{\mathrm{0}}}$ as a function of aluminium shell thickness and wavelength range. Here, the function is only shown in the range from 0.1 nm to 10 nm and in a magnified range from 0.1 nm to 2 nm.} 
  \label{fig:emr_beer_lambert}
\end{figure}

\noindent Here, $\alpha$ is the absorption coefficient of a material, which is a function of the wavelength and the so-called extinction coefficient $k$. Figure \ref{fig:emr_beer_lambert}, left plot, shows the absorption coefficient for a wavelength range from 0.1 nm to 2500 nm and, in a zoom-in window, from 0.1 nm to 10 nm. Only in that range, the light intensity $I$ has non-zero values, see the right plot of the figure, where the $\frac{I}{I_{\mathrm{0}}}$ is shown. Clearly, a material made of aluminium with a thickness of only 0.2 mm would effectively stop all incident electromagnetic radiation in the soft X-ray and XUV range (0.1 nm to 10 nm). 

%---------------------------------------------------------------------------
\section{Spectroscopy Subunit and the Iodine Cell}

\begin{figure}[!htbp] 
  \begin{center}
    \includegraphics[width=1.0\textwidth]{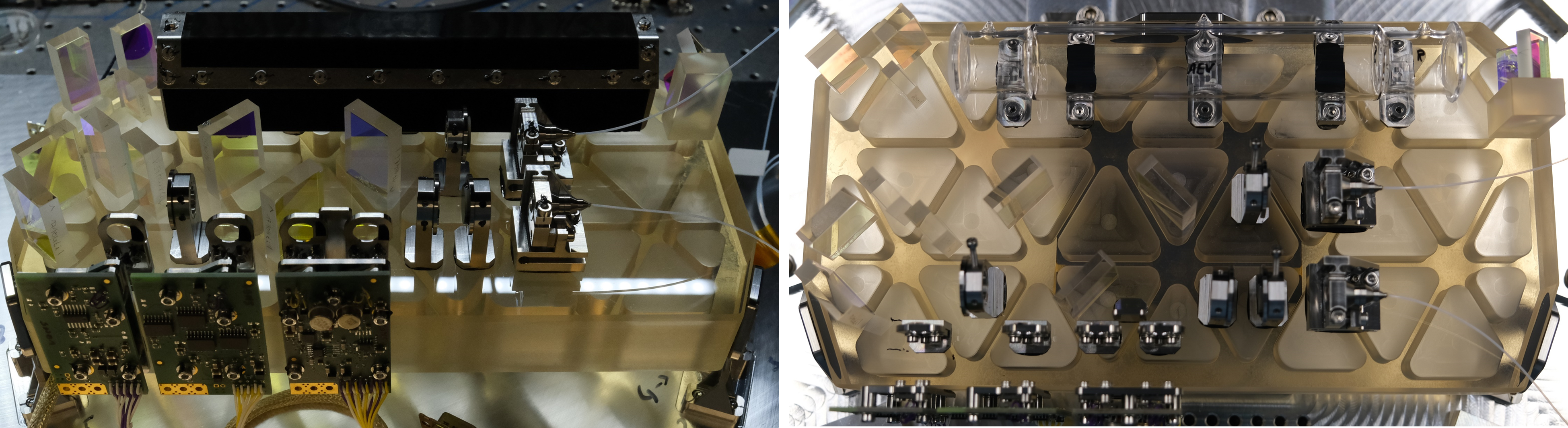}
  \end{center}
  \caption{Spectroscopy subunit with its optical components fixed to the zerodur bench. On the left the iodine cell is covered by a lid. On the right the iodine cell is exposed.} 
  \label{fig:spec_board}
\end{figure}

\begin{figure}[!htbp] 
  \begin{center}
    \includegraphics[width=1.0\textwidth]{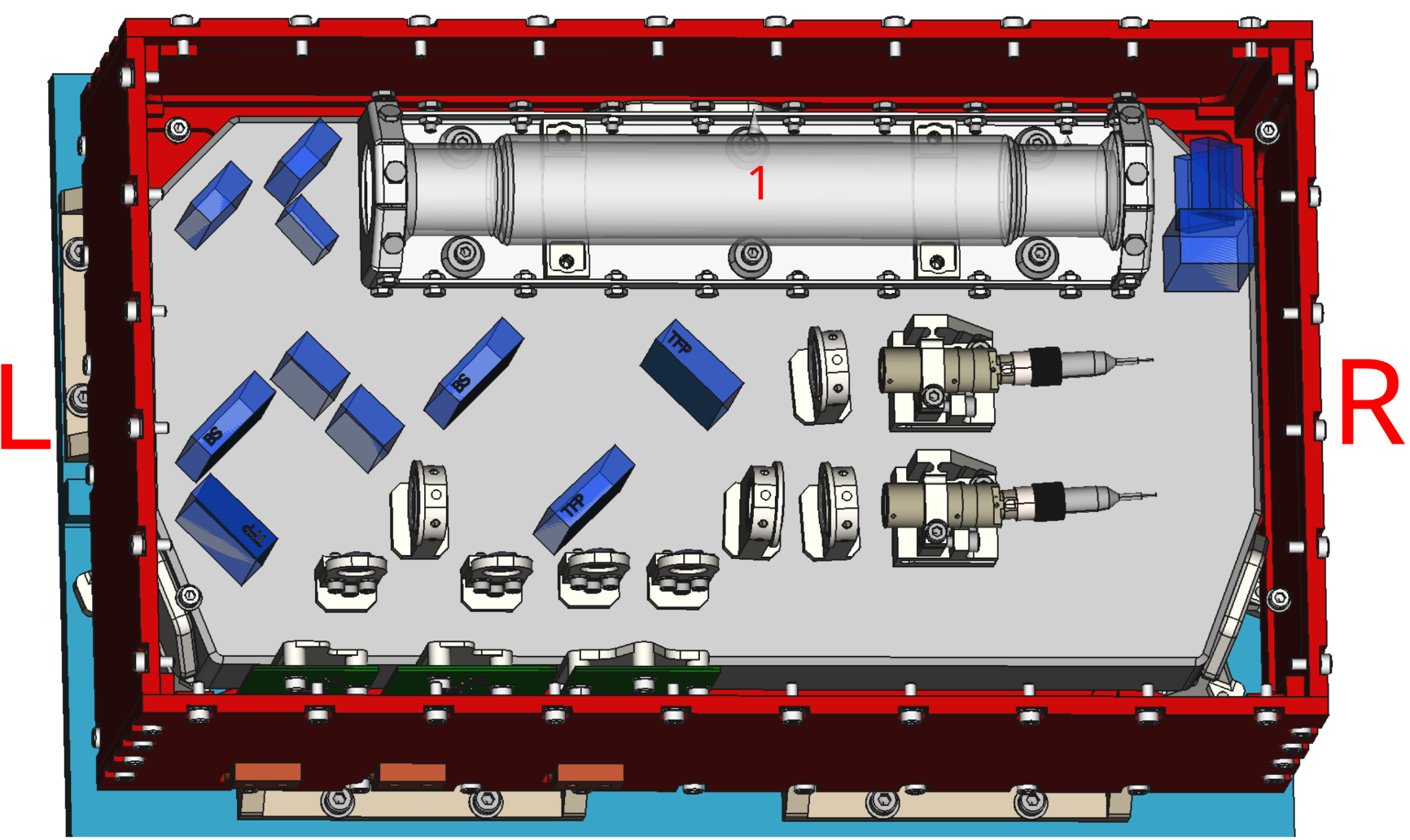}
  \end{center}
  \caption{Position {\color{red}{of the iodine cell}} of COMPASSO spectroscopy subunit {\color{red}{is}} marked with red label {\color{red}{"1"}}.} 
  \label{fig:spec_board_CAD}
\end{figure}

The iodine cell is part of a spectroscopy subunit and is shown in Figure \ref{fig:spec_board}. The cell is covered with a lid, with openings for two optical windows located on the left- and right-hand side of the component, respectively. The left side of Figure \ref{fig:spec_board} shows the subunit with the covered and the right with the exposed iodine cell. Additionally, {\color{red}{a CAD}} model of the subunit is shown in Figure \ref{fig:spec_board_CAD}. The cell is marked with the label "\textit{1}", and furthermore, this label is used as a reference when {\color{red}{the}} TID radiation loads within the cell and its window coatings are reported, see Section \ref{sec:optical_window_corp_radiation}.

%FIGURE GOES HERE.

\subsection{Corpuscular Radiation Environment}
Electrons, protons, helium ions, and other element ions can degrade optical coatings and their substrate, which results in an irreversible change of their thermo-optical properties. Hence, RA requires the estimation of the fluence magnitude of particles deposited and transferred through each layer of an optical material under study. To achieve that goal, a differential particle flux spectrum needs to be determined in front of the very first layer of the material. Additionally, a set of particle energies is required, corresponding to penetration depths that equal the coating thicknesses. These energies are needed to calculate deposited and transferred particle fluence magnitudes in each coating and the substrate of the material. The following RA strategy was made. It is assumed that radiation, when going through the outer COMPASSO shell and a thermal shield, will fill isotropically the interior of the spectroscopy subunit, i.e., the volume seen in Figure \ref{fig:spec_board} and Figure \ref{fig:spec_board_CAD}. However, this assumption is applied only to optical components and is used solely to estimate the fluence of deposited and transferred particles for optical coatings. The calculation of TID and TNID does not rely on this assumption. It is essential to {\color{red}{emphasize}} that the isotropic radiation field within the spectroscopy subunit affects the calculated particle fluence magnitudes. Using this assumption, components located away from the thermal shield receive a larger particle fluence than they would without it. Hence, the isotropic radiation field represents a more conservative approach for RA of optical components within the considered subunit.

\begin{figure}[!htbp] 
  \begin{center}
    \includegraphics[width=1.0\textwidth]{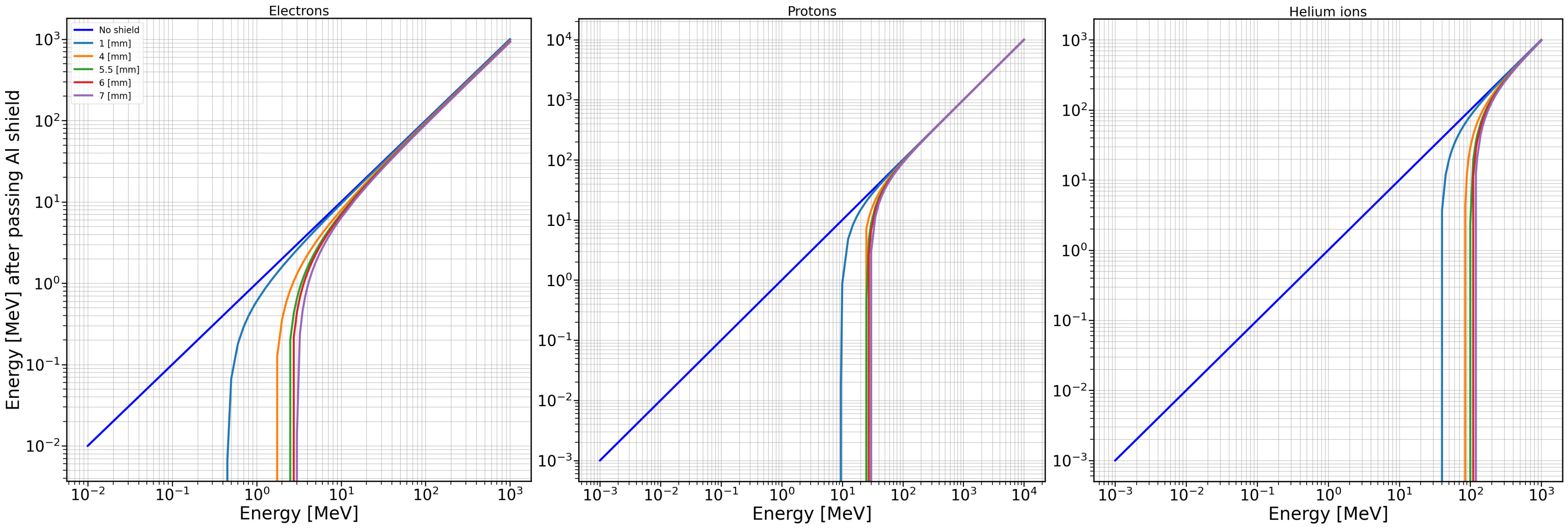}
  \end{center}
  \caption{Electrons, protons, and helium ion energies before and after passing the Al-shield with a given thickness.} \label{fig:kinetic_energy}
\end{figure}

All COMPASSO payload parts, being placed between an optical component under study and the space radiation environment, act as a shield. Hence, such parts stop or reduce the magnitude of the corpuscular radiation flux. Also, the incident particle kinetic energy decreases for those particles that pass the shield. This effect is presented in Figure \ref{fig:kinetic_energy}. It displays three plots showing the kinetic energy of particles before and after passing through various Al shield thicknesses. Figures correspond, from left to right, to electrons, protons, and helium ions, respectively. The X-axis denotes particle kinetic energy before the radiation hits a shield, while the Y-axis corresponds to particle kinetic energy after passing the shield. The most substantial energy reduction is observed for helium ions. This is related to the largest stopping power of He ions in Al from all three considered particle species. For the COMPASSO spectroscopy subunit, the shield has a combined thickness of 7~mm of Al. It consists of the payload outer shell and a thermal shield. Plots indicate important {\color{red}{consequences}} for the ageing of optical components. Particles, after passing the shield, have a spectrum with energies from just a few {\color{red}{eV}}. Hence, all optical coatings (and their interfaces) with a thickness of even a few nm will be populated with incident particles.        

\begin{figure}[!htbp] 
  \begin{center}
    \includegraphics[width=0.7\textwidth]{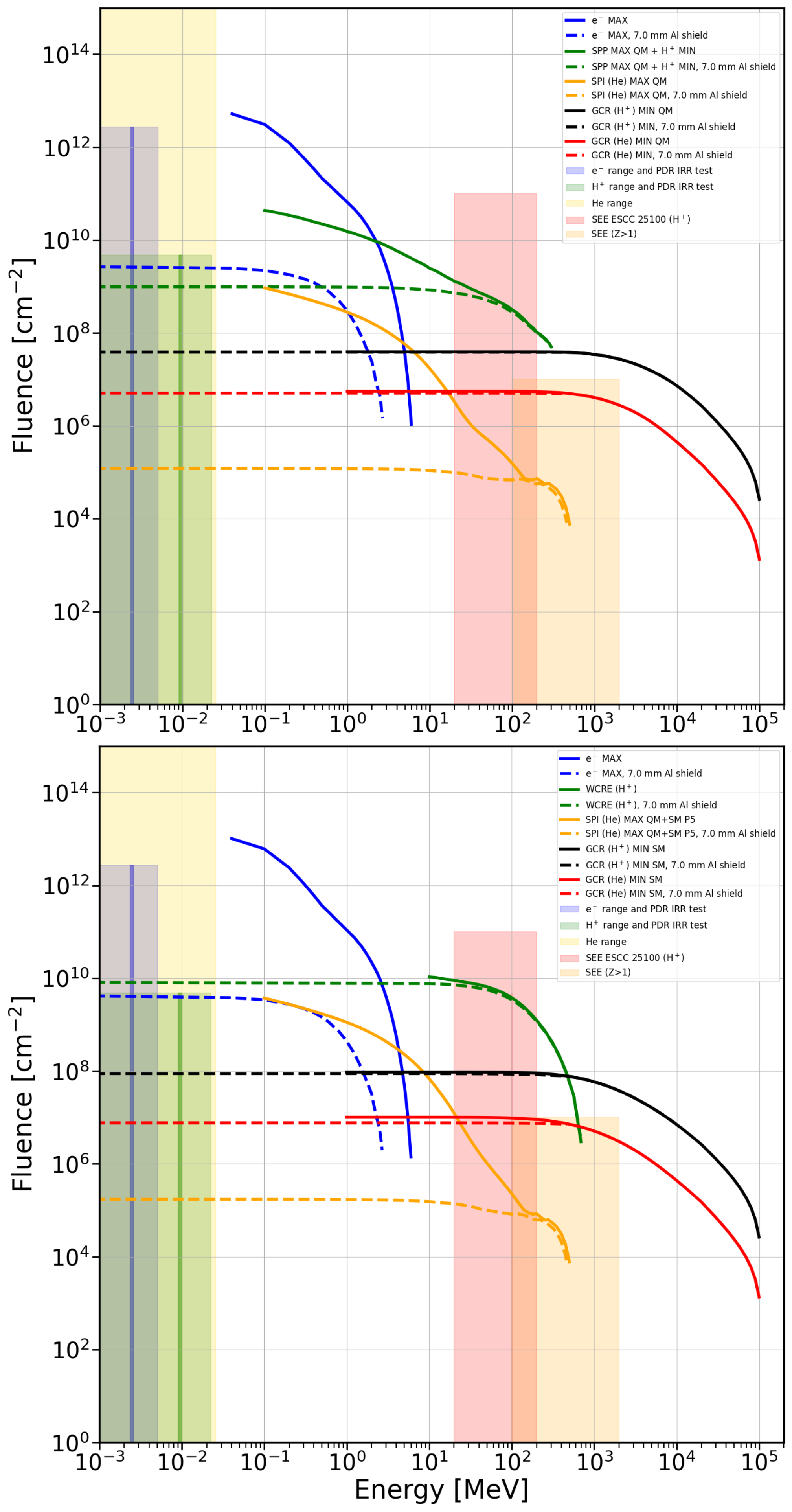}
  \end{center}
  \caption{Fluence spectra before and after passing 7 mm Al shield denoted by solids and dashed lines, respectively. Top figure corresponds to spectra of the NCRE, while the bottom figure to the WCRE environment.} \label{fig:transferred_fluence}
\end{figure}

Fluence of trapped electrons, protons, GCR, SPP, and SPI after passing 7~mm Al shield is presented in Figure \ref{fig:transferred_fluence}. Here, two {\color{red}{environments}} were considered, i.e. the NCRE (top figure) and the WCRE (bottom figure). Their magnitudes are stored in Table \ref{tab:compasso_transferred_fluence_magnitudes}. Transformed fluence spectra are used to estimate the magnitudes of deposited and transferred particle fluence within optical components, coatings, and their substrate materials. Electron and proton energies used during the irradiation test (see Section \ref{sec:irradiation_test}) are marked with thin blue and green bars. On the other hand, energy range for electrons, protons, and helium ions deposited within the coatings of the IC optical window are depicted as blue, green, and yellow bars. 

\begin{table}[!htbp]
	\scriptsize
	%\begin{center}
        \centering
        ~\clap{
	  \begin{tabular}{ c|c|c|l } 
	    {\color{red}{Species}} & Name & $\mathrm{F(E>10^{-9} MeV)}$ & Plot reference \\ \hline
            \multicolumn{4}{c}{Environment: NCRE}\\ \hline
            $\mathrm{e^{-}}$ & $\mathrm{e^{-}}$ MAX & 2.6759E+09 & Figure \ref{fig:transferred_fluence} (top) blue dashed line \\
            $\mathrm{H^{+}}$ & SPP MAX + $\mathrm{H^{+}}$ MIN & 9.8760E+08 & Figure \ref{fig:transferred_fluence} (top) green dashed line \\
            He ions & SPI (He) MAX & 1.2113E+05 & Figure \ref{fig:transferred_fluence} (top) orange dashed line \\
            $\mathrm{H^{+}}$ & GCR ($\mathrm{H^{+}}$) MIN & 3.8739E+07 & Figure \ref{fig:transferred_fluence} (top) black dashed line \\
            He ions & GCR (He) MIN & 5.0341E+06 & Figure \ref{fig:transferred_fluence} (top) red dashed line \\ \hline
            \multicolumn{4}{c}{Environment: WCRE}\\ \hline
            $\mathrm{e^{-}}$ & $\mathrm{e^{-}}$ MAX & 4.1284E+09 & Figure \ref{fig:transferred_fluence} (bottom) blue dashed line \\
            $\mathrm{H^{+}}$ & WCRE $\mathrm{H^{+}}$ & 8.1034E+09 & Figure \ref{fig:transferred_fluence} (bottom) green dashed line \\
            He ions & SPI (He) MAX & 1.7252E+05 & Figure \ref{fig:transferred_fluence} (bottom) orange dashed line \\
            $\mathrm{H^{+}}$ & GCR ($\mathrm{H^{+}}$) MIN & 8.7519E+07 & Figure \ref{fig:transferred_fluence} (bottom) black dashed line \\
            He ions & GCR (He) MIN & 7.6476E+06 & Figure \ref{fig:transferred_fluence} (bottom) {\color{red}{red}} dashed line \\ \hline
	  \hline
	  \end{tabular}
        }
	\caption{Fluence magnitudes $\mathrm{F(E>10^{-9} MeV)}$ of particles after passing 7 mm of Al shield.}
	\label{tab:compasso_transferred_fluence_magnitudes}
	%\end{center}
\end{table} 

%---------------------------------------------------------------------------
\section{Material Selection - Anti-reflective Coatings Designs}
\label{sec:coatings}
For radiation testing, eight series of anti-reflective (AR) coated samples were produced using the following combinations of coating materials, coating techniques, and coating designs. The sample substrates were made of fused silica, $15\times 15 \times 3$ $\mathrm{mm^3}$, polished and coated on both sides. 

Two coating techniques were used: electron beam evaporation (e-beam) and plasma ion-assisted deposition (PIAD). Both methods are similar in principle. They both use an electron beam to heat up and evaporate a coating material placed in a crucible. In the case of PIAD, an ion assist is added as a source to which argon gas is introduced to form plasma directed towards the samples using an electric field. This helps to create more compact coatings. Two different material combinations were used for the deposition, $\mathrm{TiO_2}$+$\mathrm{SiO_2}$ and $\mathrm{Ta_2O_5}$+$\mathrm{SiO_2}$. Corresponding refractive indices of the layer materials are {\color{red}{given}} in Table \ref{tab:refractive_index}. {\color{red}{These values are taken from the refractive index tables used by our coating systems.}} 

\begin{table}[!htbp]
	\scriptsize
	%\begin{center}
        \centering
        ~\clap{
	  \begin{tabular}{ c|c|c } 
	      Coating material & Refractive index @500 nm PIAD & Refractive index @500 nm E-BEAM  \\ \hline  
            $\mathrm{TiO_2}$ & 2.38 & 2.32 \\
            $\mathrm{Ta_2O_5}$ & 2.14 & 2.13 \\
            $\mathrm{SiO_2}$ & 1.48 & 1.48 \\ \hline
	  \hline
	  \end{tabular}
        }
	\caption{Refractive indices of used materials and deposited technologies.}
	\label{tab:refractive_index}
	%\end{center}
\end{table} 

\noindent The coating design {\color{red}{proposals are two-layered and four-layered schemes.}} {\color{red}{Table \ref{tab:coatings} summarizes these designs and the individual layer thicknesses, which were obtained by creating and optimizing the designs in Film Star software.}}

\begin{table}[!htbp]
	\scriptsize
	%\begin{center}
        \centering
        ~\clap{
	  \begin{tabular}{ c|l|c|l } 
	      Coating technique & Coating materials & Number of layers & Design  \\ \hline
            PIAD   & H=$\mathrm{TiO_2}$, L=$\mathrm{SiO_2}$   & 2 & 14.7H 122.4L \\
            PIAD   & H=$\mathrm{TiO_2}$, L=$\mathrm{SiO_2}$   & 4 & 22.3H 42.3L 43.4H 111.4L \\
            PIAD   & H=$\mathrm{Ta_2O_5}$, L=$\mathrm{SiO_2}$ & 2 & 22.2H 120.4L\\
            PIAD   & H=$\mathrm{Ta_2O_5}$, L=$\mathrm{SiO_2}$ & 4 & 27.2H 28.2L 77.3H 98.6L \\
            E-BEAM & H=$\mathrm{TiO_2}$, L=$\mathrm{SiO_2}$ & 2 & 13.9H 123.4L \\
            E-BEAM & H=$\mathrm{TiO_2}$, L=$\mathrm{SiO_2}$ & 4 & 21.3H 43.6L 41.3H 111.8L \\
            E-BEAM & H=$\mathrm{Ta_2O_5}$, L=$\mathrm{SiO_2}$ & 2 & 21.7H 119.4L \\
            E-BEAM & H=$\mathrm{Ta_2O_5}$, L=$\mathrm{SiO_2}$ & 4 & 25.3H 30.5L 76.8H 99.9L \\ \hline
	  \hline
	  \end{tabular}
        }
	\caption{Design of tested AR coatings combinations. Design denotes the thickness in nanometers of each layer starting from the substrate.}
	\label{tab:coatings}
	%\end{center}
\end{table} 

\noindent Using the above designs, ten samples per design were coated on both sides. Altogether, sixteen coating runs were made to produce eighty samples. After deposition, each sample was measured using a Cary 7000 spectrophotometer to obtain reflectance and transmittance spectra. Out of each ten sample series, five samples were annealed to 600$^{\circ}$C for e-beam coated samples and 700$^{\circ}$C for PIAD-coated samples. These temperatures were selected based on previous experimental work \cite{Oulehla2019}, because they effectively prevent the settling of the iodine molecules in the coating structure. After annealing, all the samples were subjected to iodine vapours to test their resistance to iodine molecules settling in the coating structure. The samples were put into the exsiccator. Its chamber was evacuated to $\sim$$1\times10^{-1}$ Pa level, and iodine vapour was introduced to the samples for about two days. After this procedure, the four best samples of each type, judging by the optical performance, were selected for radiation testing. After each step, every sample was measured again using the spectrophotometer.   

%---------------------------------------------------------------------------
\section{Results}
\subsection{The Complex Irradiation Facility}
\begin{figure}[!htbp] 
  \begin{center}
    \includegraphics[width=1.0\textwidth]{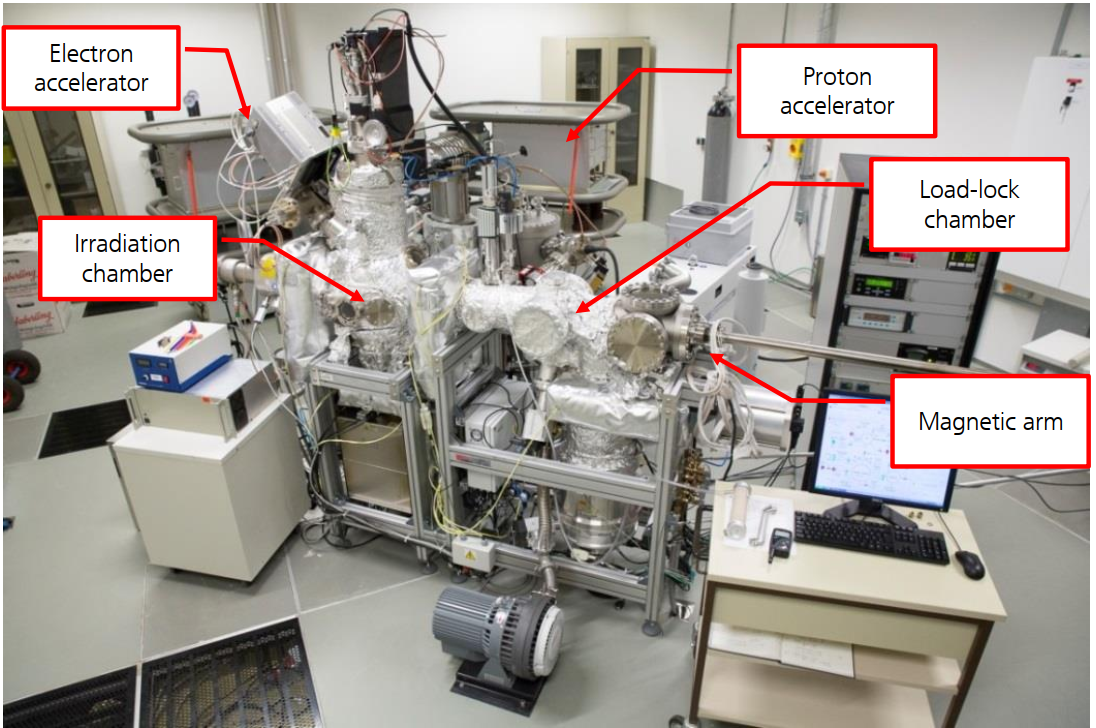}
  \end{center}
  \caption{The Complex Irradiation Facility at DLR-Bremen.} \label{fig:cif}
\end{figure}

An irradiation test was performed at DLR-Bremen in the so-called Complex Irradiation Facility (CIF). It is an Ultra High Vacuum facility equipped with two linear accelerators and three light sources to simulate the corpuscular and electromagnetic radiation of the Sun, see Figure \ref{fig:cif}. The light sources are an Argon-VUV source, a Deuterium lamp, and a Solar Simulator equipped with a Xenon lamp.

The facility produces monoenergetic protons and electrons ranging from 2 keV to 100 keV. A common beam line from both accelerator sections ends in the so-called irradiation chamber, where specimens under study are located. Accelerators work in two modes, i.e., low energetic and high energetic. In the first one, electrons and protons can be produced with an energy range from 2 keV to 10 keV with a current of up to 5 $\mathrm{\mu}$A. In the high energy mode, with energy ranging from 10 keV to 100 keV, particle current can reach its maximum value of 100 $\mathrm{\mu}$A. 

Electrons are produced by heating a $\mathrm{LaB_6}$ cathode with well-specified current parameters. The electrons are then accelerated to the required energy. Protons are generated by ionising $\mathrm{H_2}$ gas within a glass bulb by a radio frequency principle. The protons are then pushed by a voltage to the acceleration section and further to the beamline. The acceleration section ends with the so-called bending magnets, which bend the electron and the proton beams into the direction of the irradiation chamber, where the specimens are located. A detailed description of the facility can be found under \cite{renger}. {\color{red}{That description also mentions a built-in spectrometric complex for in-situ measurements in vacuum as a planned extension of the facility. It was not implemented at the time of the present campaign, in which the CIF served as a proton and electron source only. The optical {\color{red}{characterization}} of the specimens was performed ex-situ, see Section \ref{sec:optical_measurements}, and the outgassing behaviour of the coatings was not part of this investigation, so that the mass spectrometer of the facility was not used either.}} 

\subsection{Sample list}
ISI-Brno provided samples with eight different types of coating. Each type is subdivided into two {\color{red}{groups}}, which were or weren't baked out after the coating procedure to 600$^{\circ}$C (e-beam samples) and 700$^{\circ}$C (PIAD samples). Table \ref{tab:samples_and_coatings} gives a list of samples. Specimens with the symbol \textit{20} in their name denote the ones that were not baked out, and {\color{red}{those with}} the symbols \textit{600} and \textit{700} were baked out. On the other hand, the letter \textit{P} in the sample name means that a specimen substrate was coated using the PIAD method, and the letter \textit{E} indicates coating using the E-BEAM method. Note that four samples per coating type were delivered, while numbers varied from 1 to 5. The "missing sample" was used as reference material for measuring optical properties before the irradiation test. Sample layers denoted as 2 and 4 are the ones that are exposed to the radiation. 

\begin{table}[!htbp]
    \centering
    ~\clap{
    \begin{tabular}{ |l|l|l|l|l| } 
    \hline
    \multirow{2}{6em}{Sample name} & Layer 1 [nm] & Layer 2 [nm] & Layer 3 [nm] & Layer 4 [nm] \\ 
    \cline{2-5}
     & $\mathrm{TiO_2}$ or $\mathrm{Ta_2O_5}$ & $\mathrm{SiO_2}$ & $\mathrm{TiO_2}$ or $\mathrm{Ta_2O_5}$ & $\mathrm{SiO_2}$ \\ \hline
     Ti2P20-2, Ti2P700-1 & \multirow{4}{4em}{14.7} & \multirow{4}{4em}{122.4} & \multirow{4}{4em}{x} & \multirow{4}{4em}{x} \\
     Ti2P20-3, Ti2P700-3 & & & & \\
     Ti2P20-4, Ti2P700-4 & & & & \\
     Ti2P20-5, Ti2P700-5 & & & & \\ 
     \hline
     Ta2P20-1, Ta2P700-1 & \multirow{4}{4em}{22.2} & \multirow{4}{4em}{120.4} & \multirow{4}{4em}{x} & \multirow{4}{4em}{x} \\
     Ta2P20-2, Ta2P700-2 & & & & \\
     Ta2P20-3, Ta2P700-3 & & & & \\
     Ta2P20-4, Ta2P700-4 & & & & \\ 
     \hline
     Ti4P20-2, Ti4P700-1 & \multirow{4}{4em}{22.3} & \multirow{4}{4em}{42.3} & \multirow{4}{4em}{43.4} & \multirow{4}{4em}{111.4} \\
     Ti4P20-3, Ti4P700-2 & & & & \\
     Ti4P20-4, Ti4P700-3 & & & & \\
     Ti4P20-5, Ti4P700-4 & & & & \\ 
     \hline
     Ta4P20-1, Ta4P700-2 & \multirow{4}{4em}{27.2} & \multirow{4}{4em}{28.2} & \multirow{4}{4em}{77.3} & \multirow{4}{4em}{98.6} \\
     Ta4P20-2, Ta4P700-3 & & & & \\
     Ta4P20-4, Ta4P700-4 & & & & \\
     Ta4P20-5, Ta4P700-5 & & & & \\ 
     \hline
     Ti2E20-1, Ti2E600-1 & \multirow{4}{4em}{13.9} & \multirow{4}{4em}{123.4} & \multirow{4}{4em}{x} & \multirow{4}{4em}{x} \\
     Ti2E20-3, Ti2E600-3 & & & & \\
     Ti2E20-4, Ti2E600-4 & & & & \\
     Ti2E20-5, Ti2E600-5 & & & & \\ 
     \hline
     Ta2E20-1, Ta2E600-1 & \multirow{4}{4em}{21.7} & \multirow{4}{4em}{119.4} & \multirow{4}{4em}{x} & \multirow{4}{4em}{x} \\
     Ta2E20-3, Ta2E600-3 & & & & \\
     Ta2E20-4, Ta2E600-4 & & & & \\
     Ta2E20-5, Ta2E600-5 & & & & \\ 
     \hline
     Ti4E20-1, Ti4E600-1 & \multirow{4}{4em}{21.3} & \multirow{4}{4em}{43.6} & \multirow{4}{4em}{41.3} & \multirow{4}{4em}{111.8} \\
     Ti4E20-2, Ti4E600-2 & & & & \\
     Ti4E20-3, Ti4E600-3 & & & & \\
     Ti4E20-5, Ti4E600-4 & & & & \\ 
     \hline
     Ta4E20-1, Ta4E600-1 & \multirow{4}{4em}{25.3} & \multirow{4}{4em}{30.5} & \multirow{4}{4em}{76.8} & \multirow{4}{4em}{99.9} \\
     Ta4E20-3, Ta4E600-2 & & & & \\
     Ta4E20-4, Ta4E600-3 & & & & \\
     Ta4E20-5, Ta4E600-4 & & & & \\ 
     \hline
    \end{tabular}
    }
    \caption{Sample and coating system (type, thickness and layer) as prepared by ISI-Brno.}
    \label{tab:samples_and_coatings}
\end{table}

\subsection{Proton and electron irradiation test} \label{sec:irradiation_test}
\begin{figure}[!htbp] 
  \begin{center}
    \includegraphics[width=1.0\textwidth]{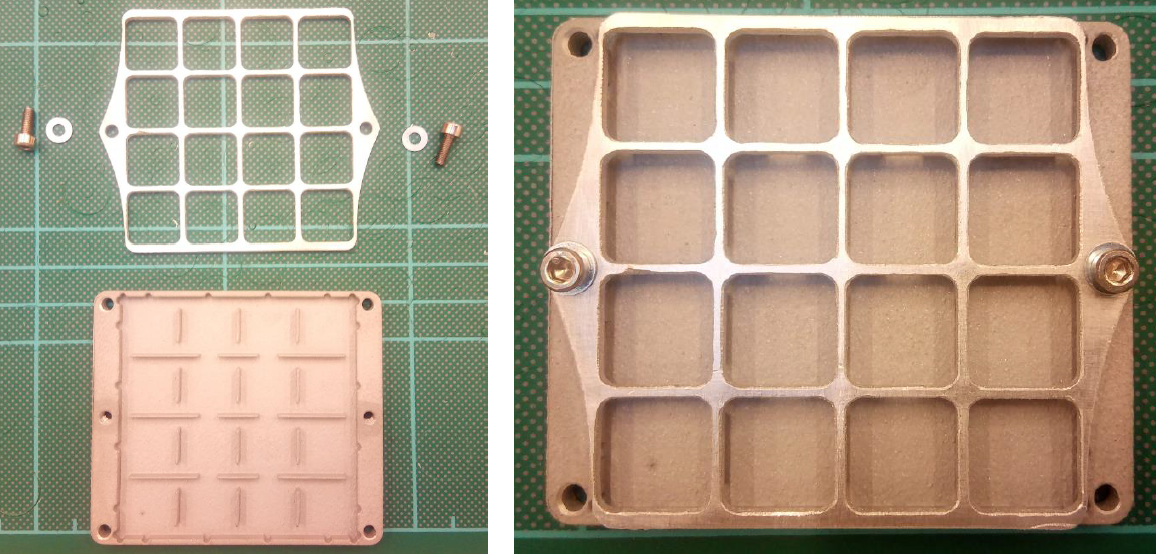}
  \end{center}
  \caption{A sample holder design at DLR-Bremen and used during the proton and electron irradiation test.} \label{fig:sample_holder}
\end{figure}

Samples were split into four sets (16 pcs per set). They were positioned in an aluminium sample holder designed and 3D-printed by DLR. An aluminium mask clamped all samples. Such a design is required since samples are exposed to radiation at the CIF in the vertical position. Figure \ref{fig:sample_holder} shows the holder before (left) and after (right) assembly. Tables from \ref{tab:set1} to \ref{tab:set4} provide a snapshot of specimen location within a sample holder. {\color{red}{Each set corresponds to one material family, and each row of the holder contains four nominally identical specimens of a single coating variant, defined by the deposition technology and the annealing temperature. The tables thus define the test configuration and provide the traceability between the specimen names used throughout Section \ref{sec:irradiation_test} and their position during the exposure.}}

\begin{table}[!htbp]
    \centering
    \begin{tabular}{ |c|l|l|l|l|l| } 
    \hline
    \multirow{4}{4em}{SET1} & Ti2P20-2 & Ti2P20-3 & Ti2P20-4 & Ti2P20-5 & Row 1 \\ 
     & Ti2P700-1 & Ti2P700-3 & Ti2P700-4 & Ti2P700-5 & Row 2 \\
     & Ti2E20-1  & Ti2E20-3  & Ti2E20-4  & Ti2E20-5  & Row 3 \\
     & Ti2E600-1 & Ti2E600-3 & Ti2E600-4 & Ti2E600-5 & Row 4 \\
    \hline
    \end{tabular}
    \caption{Location of samples within the sample holder for the SET1 during proton and electron irradiation test.}
    \label{tab:set1}
\end{table}

\begin{table}[!htbp]
    \centering
    \begin{tabular}{ |c|l|l|l|l|l| } 
    \hline
    \multirow{4}{4em}{SET2} & Ta2P20-1 & Ta2P20-2 & Ta2P20-3 & Ta2P20-4 & Row 1 \\ 
     & Ta2P700-1 & Ta2P700-2 & Ta2P700-3 & Ta2P700-4 & Row 2 \\
     & Ta2E20-1  & Ta2E20-3  & Ta2E20-4  & Ta2E20-5  & Row 3 \\
     & Ta2E600-1 & Ta2E600-3 & Ta2E600-4 & Ta2E600-5 & Row 4 \\
    \hline
    \end{tabular}
    \caption{Location of samples within the sample holder for the SET2 during proton and electron irradiation test.}
    \label{tab:set2}
\end{table}

\begin{table}[!htbp]
    \centering
    \begin{tabular}{ |c|l|l|l|l|l| } 
    \hline
    \multirow{4}{4em}{SET3} & Ti4P20-2 & Ti4P20-3 & Ti4P20-4 & Ti4P20-5 & Row 1 \\ 
     & Ti4P700-1 & Ti4P700-2 & Ti4P700-3 & Ti4P700-4 & Row 2 \\
     & Ti4E20-1  & Ti4E20-2  & Ti4E20-3  & Ti4E20-5  & Row 3 \\
     & Ti4E600-1 & Ti4E600-2 & Ti4E600-3 & Ti4E600-4 & Row 4 \\
    \hline
    \end{tabular}
    \caption{Location of samples within the sample holder for the SET3 during proton and electron irradiation test.}
    \label{tab:set3}
\end{table}

\begin{table}[!htbp]
    \centering
    \begin{tabular}{ |c|l|l|l|l|l| } 
    \hline
    \multirow{4}{4em}{SET4} & Ta4P20-1 & Ta4P20-2 & Ta4P20-4 & Ta4P20-5 & Row 1 \\ 
     & Ta4P700-2 & Ta4P700-3 & Ta4P700-4 & Ta4P700-5 & Row 2 \\
     & Ta4E20-1  & Ta4E20-3  & Ta4E20-4  & Ta4E20-5  & Row 3 \\
     & Ta4E600-1 & Ta4E600-2 & Ta4E600-3 & Ta4E600-4 & Row 4 \\
    \hline
    \end{tabular}
    \caption{Location of samples within the sample holder for the SET4 during proton and electron irradiation test.}
    \label{tab:set4}
\end{table}

All samples were exposed to protons (energy 10 keV) and electrons (2.5 keV), as {\color{red}{summarized}} in Table \ref{tab:fluence}. {\color{red}{No ion beam was applied during the campaign. Within the COMPASSO environment, the ion component is dominated by helium ions, whose fluence behind the payload shield is lower by two to three orders of magnitude than the electron and proton fluences, and which are stopped within the same coating layers as the 10 keV protons used here. The applied proton exposure therefore envelops the mission ion load. The quantitative justification is given in Section \ref{sec:optical_window_corp_radiation}. In addition, the CIF accelerators produce monoenergetic protons and electrons only, and an ion source covering the required helium energy range is not part of the facility.}}

\begin{table}[!htbp]
	%\scriptsize
	%\begin{center}
        \centering
        ~\clap{
	  \begin{tabular}{ l|c|c|l } 
	      Irradiation test & {\color{red}{Energy [keV]}} & {\color{red}{Flux [$\mathrm{cm^{-2}\,s^{-1}}$]}} & Particle fluence [$\mathrm{e^-}$ or $\mathrm{H^+}$ $\mathrm{cm^{-2}}$]  \\ \hline
            Proton irradiation test & {\color{red}{10}} & {\color{red}{9.70E+08}} & 4.85E9 $\pm$ 20.11\% \\ 
            Electron irradiation test & {\color{red}{2.5}} & {\color{red}{5.12E+10}} & 2.75E12 $\pm$ 2.62\% \\ \hline
	  \hline
	  \end{tabular}
        }
	\caption{Sample SET1, SET2, SET3, and SET4 received electron and proton fluence together with uncertainty{\color{red}{, and the corresponding beam energies and particle fluxes applied during the exposure}}.}
	\label{tab:fluence}
	%\end{center}
\end{table} 

{\color{red}{The magnitude of the applied particle flux requires justification, since an accelerated ground exposure may in principle activate degradation mechanisms that do not operate at the much lower fluxes encountered in orbit, in particular the accumulation of electrostatic charge within the dielectric coating and the associated risk of a breakdown discharge. The fluxes listed in Table \ref{tab:fluence} result from the calibration campaign of the facility, which provides the most stable and uniform beam parameters over the exposed area, and were verified against previous irradiation campaigns performed at the CIF. The standards applicable to the on-ground testing of optical coatings do not prescribe an explicit upper limit on the particle flux or on the dose rate. ISO 15856:2010 does, however, point out that a negative charge accumulating on the surface repels incident low-energy electrons and thereby reduces the net irradiance of the material, and recommends {\color{red}{neutralizing}} it either by proton irradiation or by earthing the specimens. In the present campaign all specimens were clamped in a metallic sample holder mounted on a grounded sample station, see Figure \ref{fig:sample_holder}, so that surface charge accumulation was effectively {\color{red}{neutralized}}. An estimate of the internal electric field reached at the end of the exposure, obtained by balancing the deposited current density against the ohmic leakage through the finite conductivity of each coating material, is reported in \ref{app:flux_safety}. The electron exposure is the dimensioning case. Even for it, the calculated fields stay below the reported dielectric strength of all three coating materials, by a factor of 4.6 for $\mathrm{SiO_2}$, 51 for $\mathrm{Ta_2O_5}$ and 132 for $\mathrm{TiO_2}$. The applied electron current density of 8.2E-09 $\mathrm{A\,cm^{-2}}$ is furthermore nearly eight orders of magnitude below the threshold of 0.3 to 0.6 $\mathrm{A\,cm^{-2}}$ above which the electron beam damage of a silicate glass becomes dose rate dependent \cite{JIA16}, and low-energy protons applied at fluxes of 2E+11 to 4E+12 $\mathrm{cm^{-2}\,s^{-1}}$, i.e. three orders of magnitude above the value used here, were reported to cause no flux dependent degradation of $\mathrm{TiO_2}$ and $\mathrm{ZrO_2}$ monolayers \cite{CPS24}. The applied test parameters are therefore considered to lie below the electrical breakdown threshold of the coatings, and no systematic bias caused by a dose rate effect is expected.}}

After irradiation, a visual inspection of all sample surfaces revealed no {\color{red}{color}} or structural changes. All samples were inspected under the Keyence VHX7000 confocal optical microscope. After reviewing microscope pictures of the samples, before and after the test, it was spotted that six specimens indicate negligible point-like morphological changes, i.e., {\color{red}{Ti2P20-5}} (SET1), Ta2E600-1 (SET2), Ta2P700-1 (SET2), Ta2P700-4 (SET2), Ti4P20-3 (SET3), Ta4P700-5 (SET4). {\color{red}{These changes were observed only after the proton exposure. The inspection carried out after the electron exposure revealed no surface modification on any of the specimens.}}

{\color{red}{The origin of these features deserves a comment. They are point-like and localised, and they produce no measurable loss of transmittance, which remains at the level of 99.8\% on the irradiated specimens. Sputtering of the coating material by the incident particles can be excluded on quantitative grounds. Sputtering yields for 10 keV protons were computed with SRIM \cite{ZZB10}, using 1E+04 incident ions per run, and are listed in Table \ref{tab:sputtering_yields} for each coating variant together with the corresponding sputtered areal densities. The applied fluence $F=$ 4.85E+09 $\mathrm{cm^{-2}}$ removes at most 7.6E+07 atoms $\mathrm{cm^{-2}}$. This is to be compared with the areal density of a single atomic monolayer, estimated as $N_{\mathrm{ML}}=n^{2/3}$, where $n=3\rho N_{\mathrm{A}}/M$ is the atomic density of $\mathrm{SiO_2}$, $\rho$ its mass density, $M$ its molar mass, $N_{\mathrm{A}}$ the Avogadro constant, and the factor 3 the number of atoms per $\mathrm{SiO_2}$ formula unit. With $\rho=2.2$ $\mathrm{g\,cm^{-3}}$ and $M=60.08$ $\mathrm{g\,mol^{-1}}$ one obtains $n$ of 6.6E+22 $\mathrm{cm^{-3}}$, a monolayer thickness $n^{-1/3}=0.25$ nm and $N_{\mathrm{ML}}$ of 1.6E+15 $\mathrm{cm^{-2}}$, so that the exposure removes about 5E-08 of a monolayer. This conclusion is not affected by a possible localisation of the erosion on a small fraction of the surface, since the yield required to erode a depth $d$ is $nd/F$, which is independent of the size of the affected area. The removal of a single monolayer would require 3.4E+05 atoms per ion. The features are consequently discussed here as the beam-assisted release of pre-existing localised growth defects of the coating, or as particulate contamination, the beam initiating the release rather than progressively removing material. Their appearance after the proton exposure only, and not after the electron exposure, is consistent with such a picture, since 2.5 keV electrons transfer a negligible momentum to the lattice, whereas a 10 keV proton can dislodge a weakly bound inclusion without eroding the surrounding film. It also argues against a charging-driven origin, which would be expected to be governed by the electron exposure. It should be noted that no measurement of their size, depth, areal density or of the associated surface roughness was performed. The available {\color{red}{characterization}} is limited to the comparison of the optical microscope images recorded before and after the exposure, which are reported in \ref{app:sample_pictures}, see Figures \ref{fig:irradiated_samples_1} and \ref{fig:irradiated_samples_2}.}}

\begin{table}[!htbp]
    \scriptsize
    \centering
    ~\clap{
    \begin{tabular}{ l|c|c|c|c }
    {\color{red}{Coating variant}} & {\color{red}{$Y_{\mathrm{O}}$ [atoms/ion]}} & {\color{red}{$Y_{\mathrm{Si}}$ [atoms/ion]}} & {\color{red}{Sputtered O [$\mathrm{cm^{-2}}$]}} & {\color{red}{Sputtered Si [$\mathrm{cm^{-2}}$]}} \\ \hline
    {\color{red}{Ti2P20, Ti2P700}} & {\color{red}{1.17E-02}} & {\color{red}{2.90E-03}} & {\color{red}{5.67E+07}} & {\color{red}{1.41E+07}} \\
    {\color{red}{Ta2P20, Ta2P700}} & {\color{red}{6.50E-03}} & {\color{red}{1.90E-03}} & {\color{red}{3.15E+07}} & {\color{red}{9.22E+06}} \\
    {\color{red}{Ti4P20, Ti4P700}} & {\color{red}{9.80E-03}} & {\color{red}{3.10E-03}} & {\color{red}{4.75E+07}} & {\color{red}{1.50E+07}} \\
    {\color{red}{Ta4P20, Ta4P700}} & {\color{red}{1.05E-02}} & {\color{red}{2.80E-03}} & {\color{red}{5.09E+07}} & {\color{red}{1.36E+07}} \\
    {\color{red}{Ti2E20, Ti2E600}} & {\color{red}{1.02E-02}} & {\color{red}{2.80E-03}} & {\color{red}{4.95E+07}} & {\color{red}{1.36E+07}} \\
    {\color{red}{Ta2E20, Ta2E600}} & {\color{red}{1.19E-02}} & {\color{red}{3.70E-03}} & {\color{red}{5.77E+07}} & {\color{red}{1.79E+07}} \\
    {\color{red}{Ti4E20, Ti4E600}} & {\color{red}{8.60E-03}} & {\color{red}{2.60E-03}} & {\color{red}{4.17E+07}} & {\color{red}{1.26E+07}} \\
    {\color{red}{Ta4E20, Ta4E600}} & {\color{red}{1.06E-02}} & {\color{red}{3.70E-03}} & {\color{red}{5.14E+07}} & {\color{red}{1.79E+07}} \\
    \hline
    \end{tabular}
    }
    \caption{{\color{red}{Sputtering yields of oxygen and silicon computed with SRIM for 10 keV protons incident on the coating stacks, and the corresponding sputtered areal densities for the applied proton fluence of 4.85E+09 $\mathrm{cm^{-2}}$.}}}
    \label{tab:sputtering_yields}
\end{table}

{\color{red}{The qualification criterion for the iodine cell optical window is the stability of its transmittance at the working wavelength, since it is this quantity that conditions the correct operation of the iodine frequency reference. A loss of interfacial integrity would necessarily manifest itself as a measurable change of this transmittance, which was monitored before and after the exposure. It should be noted that the post-irradiation assessment of the specimens was based on spectrophotometric and microscopic {\color{red}{characterization}}, and that no mechanical evaluation of the coating adhesion, such as a cross-cut or a scratch test, was performed. Such a {\color{red}{characterization}} becomes necessary for test campaigns approaching the fluences at which the accumulation of implanted species at the coating interfaces leads to bubble formation and interfacial delamination, i.e., of the order of 1E+15 $\mathrm{cm^{-2}}$ and above \cite{CPS24}. This threshold is consistent across independent experiments and particle species: interfacial inter-diffusion in $\mathrm{Mo/Si}$ multilayers under 4 keV alpha particles is already present at 2.6E+15 $\mathrm{cm^{-2}}$, the lowest dose applied in that work \cite{NZP13}, and, in $\mathrm{SnO_2}$--$\mathrm{TiO_2}$ nanocomposite films irradiated with 200 keV $\mathrm{Ar^+}$ ions over the range 5E+13 to 5E+16 $\mathrm{cm^{-2}}$, the reversal of the grain-size trend attributed by the authors to lattice defects and to the associated build-up of micro-strain occurs between 5E+15 and 5E+16 $\mathrm{cm^{-2}}$ \cite{KJG18}, albeit as an intragranular rather than an interfacial effect. The proton fluence applied in the present campaign is more than five orders of magnitude lower. In addition, the same study attributes the bubble and delamination mechanism specifically to the samples containing a metallic layer, namely $\mathrm{SiO_2/Al}$ and $\mathrm{SiO_2/Ag}$, in which the protons cross the dielectric protective layer and are implanted in the metal underneath, whereas for the purely dielectric $\mathrm{SiO_2/TiO_2}$ and $\mathrm{SiO_2/ZrO_2}$ bilayers no interfacial damage could be resolved by cross-sectional transmission electron microscopy (TEM) or grazing incidence X-ray diffraction (GIXRD), approximately half of the incident protons and most of the energy of the beam being deposited in the topmost $\mathrm{SiO_2}$ layer \cite{CPS24}. A single $\mathrm{SiO_2}$ layer was found to be stable at every energy and fluence investigated in that work, up to the highest applied value of 1E+17 $\mathrm{cm^{-2}}$. The coating stack considered here is likewise a purely dielectric system containing no metal--dielectric interface, although it should be noted that $\mathrm{Ta_2O_5}$ was not among the materials covered by \cite{CPS24}. That material is addressed by Di Sarcina et al. \cite{DGM14}, who exposed single layers of $\mathrm{HfO_2}$, $\mathrm{Ta_2O_5}$, $\mathrm{Nb_2O_5}$, $\mathrm{TiO_2}$ and $\mathrm{SiO_2}$, together with dielectric multilayers and fused silica substrates, to 60 keV protons at a fluence of 1E+13 $\mathrm{cm^{-2}}$, and report all of them except $\mathrm{SiO_2}$ to be unaffected. For the fused silica substrate, which is the substrate material used here as well, the transmittance decreases only below 300 nm, the effect in the visible and near infrared being negligible. That fluence is more than three orders of magnitude above the one applied in the present campaign, and the residual sensitivity of $\mathrm{SiO_2}$, ascribed to the formation of oxygen vacancies, acts in the ultraviolet, far from the 532 nm working wavelength. This degradation mechanism is therefore not expected to be activated under the conditions of this study.}}

\subsection{Optical Measurements of Irradiated Specimens} \label{sec:optical_measurements}%something like this

The irradiated samples were returned to ISI {\color{red}{Brno}}, and the final AR coating performance was retested using {\color{red}{the same}} Cary 7000 spectrophotometer {\color{red}{as at every preceding stage of the test sequence. All the optical measurements reported in this work, i.e. as deposited, after the annealing, after the exposure to iodine vapours and after the proton and electron irradiation, were performed ex-situ, in atmospheric environment and at room temperature. No optical measurement was carried out at the irradiation facility, neither in vacuum nor under the beam, so that the same instrument and the same measurement conditions apply to all the spectra compared in this section. The measured spectral graphs and comments on the results are listed in \ref{sec:spectras}}}. There are observable changes in the optical performance of the tested samples due to the annealing process. In the case of e-beam samples, there is a spectral shift towards the shorter wavelengths. This is caused by the annealing process, during which the originally porous coatings become more compact and dense, decreasing their thickness \cite{WOO}. In addition, the refractive index of the coating materials increases, resulting in an increase in reflectance \cite{MARTIN}.% These changes are summarized in Table \ref{tab:annealing} (transmittance level at 532 nm wavelength).

% \begin{table}[!htbp]
%     %\scriptsize
%         \centering
%         \begin{tabular}{c|c|c}
%             Sample & $\mathrm{T_{as\_deposited}} [\%]$ & $\mathrm{T_{annealed}} [\%]$ \\ \hline
%             Ti2E600 & 99.68 & 98.42 \\
%             Ta2E600 & 99.65 & 99.11 \\
%             Ti4E600 & 99.73 & 99.25 \\
%             Ta4E600 & 99.72 & 99.65 \\ \hline \hline
%         \end{tabular}
%         \caption{The influence of annealing on the transmittance of the e-beam samples at 532 nm.}
%         \label{tab:annealing}
% \end{table}

\begin{figure}[!htbp] 
  \begin{center}
    \includegraphics[width=1.0\textwidth]{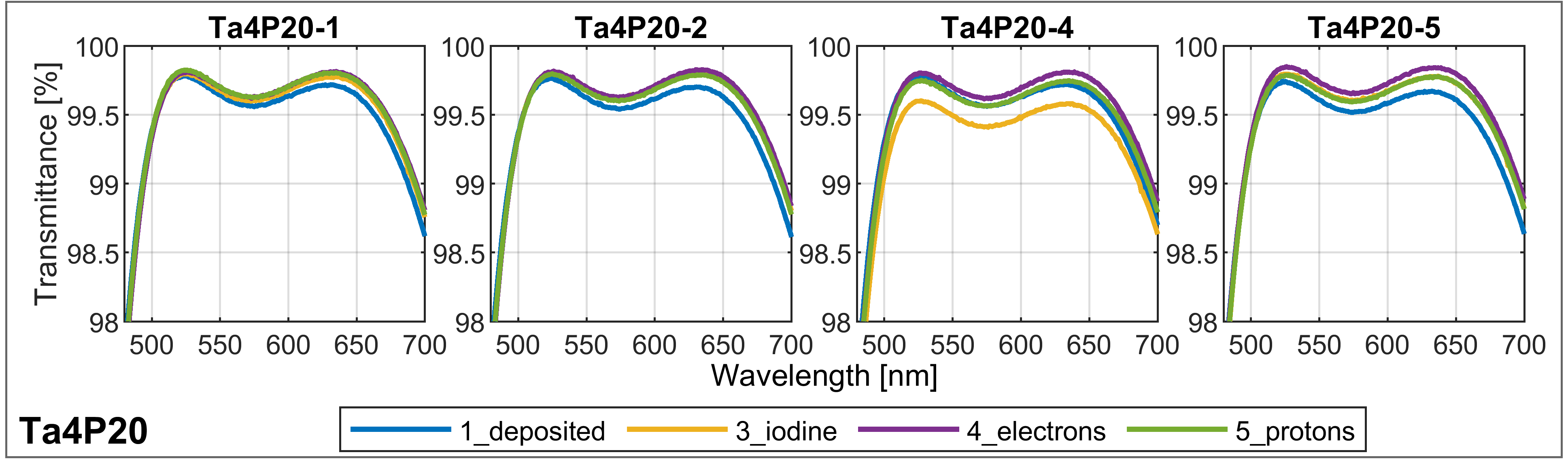}
  \end{center}
  \caption{{\color{red}{Transmittance of the selected four-layer PIAD coating deposited without bake-out, measured by spectrophotometry through the both-sides-coated specimens $\mathrm{Ta_2O_5}$. The curves correspond to the state of the specimens as deposited, after the exposure to iodine vapours, and after the proton and the electron irradiation.}}} \label{fig:final_coating}
\end{figure}

These spectral shifts towards the shorter wavelengths can be avoided by either using PIAD, where the position of the maxima in the transmittance spectrum does not
change with temperature, or by biasing the coating design towards longer wavelengths. Another effect that can be observed is the influence of annealing on the tantalum-based PIAD coatings. {\color{red}{When $\mathrm{Ta_2O_5}$ is annealed above a certain temperature, crystallization can occur in the initially amorphous films, which can cause an increase in scattering and subsequent decrease in transmittance. While annealing can initially improve transmittance by improving oxidation of $\mathrm{Ta_2O_5}$ \cite{Xu2009}, further heating causes crystallization \cite{Xu2008,Favaro2024}. This phenomenon has been reported across a range of deposition methods \cite{HE2009,MASSE2006,Ren2021}}}.
%Changes in transmittance measured by a spectrophotometer are summarized in Table \ref{tab:cryst}
% \begin{table}[!htbp]
%     %\scriptsize
%         \centering
%         \begin{tabular}{c|c|c}
%             Sample & $\mathrm{T_{as\_deposited}} [\%]$ & $\mathrm{T_{annealed}} [\%]$ \\ \hline
%             Ta2P700 & 99.85 & 99.16 \\
%             Ta4P700 & 99.72 & 98.50 \\ \hline \hline
%         \end{tabular}
%         \caption{The influence of annealing on the transmittance of the PIAD samples at 532 nm.}
%         \label{tab:cryst}
% \end{table}
% Furthermore, the results show no effect of either proton or electron radiation on the coatings' transmittance, confirming their suitability for the considered space operational conditions. 
{\color{red}{We chose the $\mathrm{Ta_2O_5/SiO_2}$ 4-layer design, PIAD, no bake-out as the best candidate for the cell coatings. This choice was made based on the fact, that the 4-layer design has a wider range of wavelengths around 532 nm, where the reflectance is very low, compensating for any inaccuracies during the coating process. PIAD was chosen, because it prevents any settling of iodine in the structure of the coating \cite{Oulehla2019}. Therefore no bake-out is necessary and there are no issues with crystallization induced increase in scattering losses in $\mathrm{Ta_2O_5}$, which was chosen for its very low scattering losses. The coatings prepared this way show}} no influence of applied irradiation and the maximum transmittance through the samples is~99.8\% at 532~nm. The corresponding spectrophotometric data {\color{red}{obtained in atmospheric environment}} are shown in Figure \ref{fig:final_coating}. {\color{red}{The figure reports the transmittance of the specimens from the series Ta4P20 at the successive stages of the test sequence, i.e. as deposited, after the exposure to iodine vapours, and after the proton and the electron irradiation. }} The final coating combination is as follows: Substrate (fused silica) - $\mathrm{Ta_2O_5}$ (27.2 nm) - $\mathrm{SiO_2}$ (28.2 nm) - $\mathrm{Ta_2O_5}$ (77.3 nm) - $\mathrm{SiO_2}$ (98.6 nm).   

%1

\FloatBarrier

\subsection{Iodine cell optical window exposed to corpuscular radiation} \label{sec:optical_window_corp_radiation}

\begin{figure}[!htbp] 
  \begin{center}
    \includegraphics[width=0.7\textwidth]{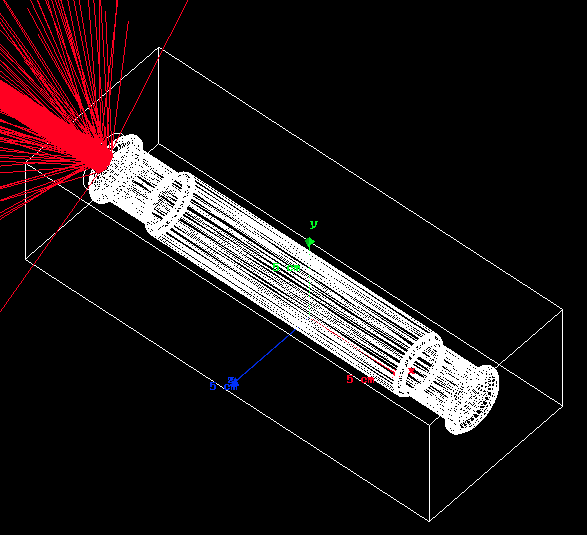}
  \end{center}
  \caption{A snapshot from the GRAS simulation. It shows the cell optical window hit by a beam of mono-energetic electrons, denoted by red lines. Also, a simulation was made for incident protons and helium ions.} \label{fig:gras_ic}
\end{figure}

The selected material geometry for the iodine cell window was then used to calculate radiation loads within its coatings. A Geant4 Radiation Analysis for Space (GRAS)\footnote{Version 05-02-01-r2153 of the GRAS software was used.} 3D Monte Carlo software \cite{SIE05} was used to perform a simulation to calculate particle energies, whose corresponding penetration depths are \textit{equal} to coating layer thicknesses. This information is required to calculate particle fluence magnitudes deposited and transmitted in the coating material. It was assumed that when a particle with a given kinetic energy is deposited within a following coating layer or a substrate with equal or less than 1\% of its initial fluence, then the penetration depth of such energy points to an interface between two material layers. Such a simulation was made for all coatings and the substrate of the iodine cell optical window. A snapshot from the simulation is shown in Figure \ref{fig:gras_ic}. There, the iodine cell is shown in the middle, and a beam of particles, depicted as red lines, strikes the optical window. {\color{red}{For the simulation a standard electromagnetic physics list, represented by the \texttt{G4EmStandardPhysics} constructor, was used. It includes the following processes: photoelectric effect, Compton scattering, gamma conversion (pair production), Rayleigh scattering, ionisation, bremsstrahlung, positron annihilation and multiple Coulomb scattering.}}

During a single simulation, 1E+04 mono-energetic particles were used. {\color{red}{This number refers to the auxiliary simulation described above, which serves to determine the energy boundaries of the coating layers, and not to the TID calculation reported later in this section. It should also be noted that the fluence magnitudes given below are not obtained by Monte Carlo counting, but by integrating the flux spectra transported through the shield between the energy boundaries derived here. In the low energy range relevant for the deposition within the coating layers, these integral spectra are essentially flat. Hence, a change of the boundary energies within this range leaves the integrated fluence magnitudes practically unchanged, and increasing the number of primary particles of the auxiliary simulation would refine the boundary energies themselves without modifying the fluence magnitudes reported in Table \ref{tab:compasso_IC_coatings_corp}.}} The first row {\textcolor{red}{of the table}} contains information about the material of the examined layer, while the second row holds information about its thickness. Next, there are three sets of rows. Each set corresponds to a different particle species, i.e., electrons, protons, and helium ions. Additionally, each row represents a set that contains information about the kinetic energies of particles, which correspond to those deposited within a given coating layer and those transmitted to underlying layers. Then, there are two blocks of rows that hold particle fluence magnitudes, which are deposited and transmitted to the underlying coating layers. Each block summarizes numbers from different environments under study. Table columns represent coating layers. The last one, with material named \textit{Vacuum}, holds fluence magnitudes $F(E>\mathrm{1E-9} \ \mathrm{MeV})$ for radiation environments within the spectroscopy board subunit and before hitting the iodine cell optical window. Next, the column named \textit{Coating B2}, holds particle fluence magnitudes for the first coating of the iodine cell optical window coating stack. The other columns follow that schema. Hence, the column order from right to left corresponds to the coating structure from its first layer down to the substrate. For all considered environments, it can be recognized that only a tiny fraction of the incident fluence is deposited within coating layers. Hence, most of the incident particle fluence is transmitted through coatings to the underlying layers and to the interior of the cell.        

\begin{table}[!htbp]
  \scriptsize
  %\footnotesize
  %\begin{center}
  \centering
  ~\clap{
	\begin{tabular}{ l|c|c|c|c|c|c } 
	   & Substrate & Coating A1 & Coating B1 & Coating A2 & Coating B2 & Environment  \\ \hline
          Material  & Fused silica & $\mathrm{Ta_2O_5}$ & $\mathrm{SiO_2}$ & $\mathrm{Ta_2O_5}$ & $\mathrm{SiO_2}$ & Vacuum \\
          Thickness & 3 mm & 27.2 nm & 28.2 nm & 77.3 nm & 98.6 nm & --- \\
          dep. $\mathrm{e^-}$ E [keV] & (5.36 ... 1.45E+03] & (4.65 ... 5.36] & (4.38 ... 4.65] & (1.74 ... 4.38] & (1E-06 ... 1.74] & --- \\
          tr. $\mathrm{e^-}$ E [keV] & (1.45E+03 ... MAX] & (5.36 ... MAX] & (4.65 ... MAX] & (4.38 ... MAX] & (1.74 ... MAX] & --- \\ 
          dep. $\mathrm{H^+}$ E [keV] & (22.8 ... 23.0E+03] & (17.85 ... 22.8] & (16.48 ... 17.85] & (4.96 ... 16.48] & (1E-06 ... 4.96] & --- \\
          tr. $\mathrm{H^+}$ E [keV] & (23.0E+03 ... MAX] & (22.8 ... MAX] & (17.85 ... MAX] & (16.48 ... MAX] & (4.96 ... MAX] & --- \\ 
          dep. $\mathrm{He}$ ions E [keV] & (25.85 ... 91.61E+03] & (20.0 ... 25.85] & (18.4 ... 20.0] & (5.34 ... 18.4] & (1E-06 ... 5.34] & --- \\
          tr. $\mathrm{He}$ ions E [keV] & (91.61E+03 ... MAX] & (25.85 ... MAX] & (20.0 ... MAX] & (18.4 ... MAX] & (5.34 ... MAX] & --- \\ \hline \hline
          \multicolumn{7}{c}{Environment: NCRE}\\ \hline
          dep. $\mathrm{e^-}$ fluence [$\mathrm{cm^{-2}}$] & 2.5497E+09 & 3.5888E+06 & 1.3648E+06 & 1.3344E+07  & 1.0414E+07 & 2.6759E+09 \\
          tr. $\mathrm{e^-}$ fluence [$\mathrm{cm^{-2}}$] & 9.7478E+07 & 2.6472E+09 & 2.6508E+09  & 2.6521E+09 & 2.6655E+09 & 2.6759E+09 \\
          dep. $\mathrm{H^+}$ fluence [$\mathrm{cm^{-2}}$] & 2.8719E+08 & 9.0823E+04 & 2.5137E+04 & 2.1137E+05  & 9.1006E+04  & 1.0263E+09  \\
          tr. $\mathrm{H^+}$ fluence [$\mathrm{cm^{-2}}$] & 7.3873E+08 & 1.0259E+09 & 1.0260E+09 & 1.0260E+09 & 1.0263E+09 & 1.0263E+09\\
          dep. $\mathrm{He}$ ions fluence [$\mathrm{cm^{-2}}$] & 7.3298E+04 & 8.5311E+00 & 2.3333E+00 & 1.9045E+01 & 7.7873E+00 & 5.1552E+06  \\ 
          tr. $\mathrm{He}$ ions fluence [$\mathrm{cm^{-2}}$] & 5.0819E+06 & 5.1552E+06 & 5.1552E+06 & 5.1552E+06 & 5.1552E+06 & 5.1552E+06 \\ \hline
          \multicolumn{7}{c}{Environment: WCRE}\\ \hline
          dep. $\mathrm{e^-}$ fluence [$\mathrm{cm^{-2}}$] & 3.9429E+09 & 5.6758E+06 & 2.1584E+06 & 2.1104E+07 & 1.6537E+07 & 4.1284E+09 \\
          tr. $\mathrm{e^-}$ fluence [$\mathrm{cm^{-2}}$] & 1.3997E+08 & 4.0829E+09 & 4.0886E+09 & 4.0907E+09 & 4.1118E+09 & 4.1284E+09 \\
          dep. $\mathrm{H^+}$ fluence [$\mathrm{cm^{-2}}$] & 1.0822E+09 & 2.2996E+05 & 6.3644E+04 & 5.3517E+05 & 2.2933E+05 & 8.1909E+09  \\
          tr. $\mathrm{H^+}$ fluence [$\mathrm{cm^{-2}}$] & 7.1077E+09 & 8.1899E+09 & 8.1901E+09 & 8.1902E+09 & 8.1907E+09 & 8.1909E+09 \\
          dep. $\mathrm{He}$ ions fluence [$\mathrm{cm^{-2}}$] & 1.8446E+05 & 1.7417E+01 & 4.7637E+00 & 3.8883E+01 & 1.5899E+01 & 7.8201E+06 \\
          tr. $\mathrm{He}$ ions fluence [$\mathrm{cm^{-2}}$] & 7.6356E+06 & 7.8200E+06 & 7.8200E+06 & 7.8200E+06 & 7.8201E+06 & 7.8201E+06 \\ \hline
	  \hline
	\end{tabular}
        }
	\caption{Deposited and transmitted particle fluence magnitudes for IC optical window coating stack. The numbers are summarized for two considered space environments, the NCRE and the WCRE.}
	\label{tab:compasso_IC_coatings_corp}
	%\end{center}
\end{table} 

Now, having particle fluence magnitudes deposited and transmitted within the iodine cell window coating structure from two considered environments and results from the irradiation test campaign, it is necessary to calculate an equivalent time in orbit for the irradiated sample material. For this purpose, a GRAS simulation was made to estimate how many 2.5 keV electrons and 10 keV protons will be deposited and transmitted within the iodine cell coating structure. It was found that 2.5 keV $\mathrm{e^-}$ are mostly deposited within iodine cell B2 coating (73.06\%) while the rest of the particles (26.94\%) are implanted in the A2 layer of the stack. On the other hand, 24.83\% of 10 keV $\mathrm{H^+}$ stuck within the B2 coating while the rest (75.17\%) are deposited within the A2 coating. Electron and proton energies used during the test are marked with thin blue and green bars, respectively, see Figure \ref{fig:transferred_fluence}. On the other hand, energy ranges for electrons, protons, and helium ions deposited within the coatings of the iodine cell optical window are depicted as light blue, green, and yellow bars. The height of the blue and green bars corresponds to electron and proton fluence magnitudes used during the test. Results are shown in Table \ref{tab:compasso_IC_coatings_corp_acc}. Columns in the table contain particle fluence magnitudes from the irradiation test, as well as a number of equivalent years in space (marked with bold font). This time is much longer than the mission time for deposited $\mathrm{e^-}$ and $\mathrm{H^+}$ in coatings B2 and A2. For transmitted particles, and only for WCRE environment, the time is approx. 325 days. However, deposited particles within the coating material have a degradation potential to irreversibly harm its lattice structure and, hence, have an impact on its physical properties. It is worth noticing that within the frame of considered environments, there is a particle fluence that is transferred through the iodine cell optical window, see Table \ref{tab:compasso_IC_coatings_corp} second column. Here, the radiation may partially ionize the iodine gas.  

{\color{red}{It should be emphasized that the irradiation test campaign was performed with electrons and protons only, hence Table \ref{tab:compasso_IC_coatings_corp_acc} contains no entries for helium ions. The equivalent time in space is, by its definition, a property of the fluence applied during the test. Hence, for a particle species that was not applied, it cannot be defined. The exclusion of ions from the test is justified by the magnitudes collected in Table \ref{tab:compasso_IC_coatings_corp}. Behind the 7~mm Al shield, the mission-integrated helium ion fluence within the spectroscopy subunit amounts to 5.1552E+06 $\mathrm{cm^{-2}}$ (NCRE) and 7.8201E+06 $\mathrm{cm^{-2}}$ (WCRE), which is 0.14\% and 0.06\% of the total incident charged particle fluence, respectively. More importantly, the helium fluence deposited within the whole coating stack equals 37.7 $\mathrm{cm^{-2}}$ (NCRE) and 77.0 $\mathrm{cm^{-2}}$ (WCRE), against 2.8711E+07 $\mathrm{cm^{-2}}$ (NCRE) and 4.5475E+07 $\mathrm{cm^{-2}}$ (WCRE) of deposited electrons, and 4.1834E+05 $\mathrm{cm^{-2}}$ (NCRE) and 1.0581E+06 $\mathrm{cm^{-2}}$ (WCRE) of deposited protons. Hence, helium ions constitute a fraction of the order of $10^{-4}$\,\% of the particle fluence deposited within the coatings. Furthermore, helium ions are stopped within the same two layers, B2 and A2, as the 10 keV protons used during the irradiation test, since their deposited energies range up to approx. 25.85 keV. The proton fluence applied during the test, 4.8500E+09 $\mathrm{cm^{-2}}$, exceeds the mission-integrated deposited helium fluence in those layers, i.e. 26.8 $\mathrm{cm^{-2}}$ (NCRE) and 54.8 $\mathrm{cm^{-2}}$ (WCRE), by approximately eight orders of magnitude. Even after accounting for the higher displacement damage and sputtering cross sections of He with respect to $\mathrm{H^+}$, the applied proton exposure envelops the helium-induced damage in the same layers by a very large margin. Finally, the helium ion contribution to the TID of the coating layers does not exceed 0.21 rad, i.e. at most approx. 1.6\% of the total dose in the directly irradiated layers, see Table \ref{tab:compasso_OCUS_IC_tid}. The ion component of the environment is therefore dimensioning for single event effects in EEE components, and not for the cumulative degradation of a passive dielectric coating stack.}}

\begin{table}[!htbp]
  \scriptsize
  %\footnotesize
	%\begin{center}
        \centering
        ~\clap{
         \begin{tabular}{ l|c|c|c|c|c|c } 
	  & Substrate & Coating A1 & Coating B1 & Coating A2 & Coating B2 & PDR Test  \\ \hline
          Material  & Fused silica & $\mathrm{Ta_2O_5}$ & $\mathrm{SiO_2}$ & $\mathrm{Ta_2O_5}$ & $\mathrm{SiO_2}$ & Vacuum \\
          Thickness & 3 mm & 27.2 nm & 28.2 nm & 77.3 nm & 98.6 nm & --- \\ \hline 
          \multicolumn{7}{c}{Environment: NCRE}\\ \hline
          dep. $\mathrm{e^-}$ fluence [$\mathrm{cm^{-2}}$] & --- & --- & --- & 7.4085E+11 & 2.0092E+12 & 2.7500E+12 \\
          dep. $\mathrm{e^-}$ eqv. time in space [ys] & --- & ---  & ---  & \textbf{1.1104E+05}  & \textbf{3.8586E+05}  & \textbf{2.0554E+03}  \\
          %dep. $\mathrm{e^-}$ eqv. time in space [ys], \textbf{\color{red}{RDM = 3}} & --- & ---  & ---  & \textbf{\color{red}{3.7013E+04}}  & \textbf{\color{red}{1.2862E+05}}  & \textbf{\color{red}{6.8513E+02}}  \\
          tr. $\mathrm{e^-}$ fluence [$\mathrm{cm^{-2}}$] & --- & --- & --- & --- & 7.4085E+11 & 2.7500E+12 \\
          tr. $\mathrm{e^-}$ eqv. time in space [ys] & --- & ---  & ---  & ---  & \textbf{5.5588E+02}  & \textbf{2.0554E+03}  \\ \hline
          %tr. $\mathrm{e^-}$ eqv. time in space [ys], \textbf{\color{red}{RDM = 3}} & --- & ---  & ---  & ---  & \textbf{\color{red}{1.8529E+02}}  & \textbf{\color{red}{6.8513E+02}}  \\ \hline
          dep. $\mathrm{H^+}$ fluence [$\mathrm{cm^{-2}}$] & --- & --- & --- & 3.6453E+09 & 1.2043E+09 & 4.8500E+09 \\
          dep. $\mathrm{H^+}$ eqv. time in space [ys] & --- & ---  & ---  & \textbf{3.4492E+04} & \textbf{2.6465E+04} & \textbf{9.4514E+00} \\
          %dep. $\mathrm{H^+}$ eqv. time in space [ys], \textbf{\color{red}{RDM = 3}} & --- & ---  & ---  & \textbf{\color{red}{1.1497E+04}} & \textbf{\color{red}{8.8218E+03}} & \textbf{\color{red}{3.1505E+00}} \\
          tr. $\mathrm{H^+}$ fluence [$\mathrm{cm^{-2}}$] & --- & --- & --- & --- & 3.6453E+09 & 4.8500E+09 \\
          tr. $\mathrm{H^+}$ eqv. time in space [ys] & --- & ---  & ---  & --- & \textbf{7.1058E+00} & \textbf{9.4514E+00} \\ \hline
          %tr. $\mathrm{H^+}$ eqv. time in space [ys], \textbf{\color{red}{RDM = 3}} & --- & ---  & ---  & --- & \textbf{\color{red}{2.3686E+00}} & \textbf{\color{red}{3.1505E+00}} \\ \hline
          \multicolumn{7}{c}{Environment: WCRE}\\ \hline
          dep. $\mathrm{e^-}$ fluence [$\mathrm{cm^{-2}}$] & --- & --- & --- & 7.4085E+11 & 2.0092E+12 & 2.7500E+12 \\
          dep. $\mathrm{e^-}$ eqv. time in space [ys] & --- & ---  & ---  & \textbf{7.0209E+04} & \textbf{2.4299E+05}  & \textbf{1.3322E+03}  \\
          tr. $\mathrm{e^-}$ fluence [$\mathrm{cm^{-2}}$] & --- & --- & --- & --- & 7.4085E+11 & 2.7500E+12 \\
          tr. $\mathrm{e^-}$ eqv. time in space [ys] & --- & ---  & ---  & ---  & \textbf{3.6035E+02} & \textbf{1.3322E+03}  \\ \hline
          dep. $\mathrm{H^+}$ fluence [$\mathrm{cm^{-2}}$] & --- & --- & --- & 3.6453E+09 & 1.2043E+09 & 4.8500E+09 \\
          dep. $\mathrm{H^+}$ eqv. time in space [ys] & --- & ---  & ---  & \textbf{1.3623E+04} & \textbf{1.0502E+04} & \textbf{1.1842E+00} \\
          tr. $\mathrm{H^+}$ fluence [$\mathrm{cm^{-2}}$] & --- & --- & --- & --- & 3.6453E+09 & 4.8500E+09 \\
          tr. $\mathrm{H^+}$ eqv. time in space [ys] & --- & ---  & ---  & --- & \textbf{8.9010E-01} & \textbf{1.1842E+00} \\
	  \hline
	\end{tabular}
         }
	 \caption{Electron and proton fluence magnitudes as well as equivalent time in space given in years for two considered environments. {\color{red}{Helium ions were not applied during the irradiation test campaign. Consequently, no test fluence and no equivalent time in space can be defined for this species, and the table holds no entries for it. The justification for excluding ions from the test is given in Section \ref{sec:optical_window_corp_radiation}.}}}
	\label{tab:compasso_IC_coatings_corp_acc}
	%\end{center}
\end{table} 

The optical window of the iodine cell was also analyzed for ionizing radiation dose (TID) levels in its coatings. Results are obtained from a simulation using the GRAS software. {\color{red}{This time, the simulation used a total of 2E+09 primary particles, with spectra taken from Table \ref{tab:ncre_spectra_recap} and Table \ref{tab:wcre_spectra_recap}.}} TID magnitudes from the cell left \textit{L} and right \textit{R} hand side of the spectroscopy unit are given in Table \ref{tab:compasso_OCUS_IC_tid}. The first column contains a component symbol, which consists of the side from which the incident radiation hits the unit. Next, an abbreviation of the environment is given. Then, the main component name \textit{(1)} is given. In the case of the optical window coatings \textit{(1C)}, there are two extra symbols. First, gives its name, and the second, gives its location with respect to the irradiation side, e.g., \textit{L\_NCRE\_1C\_B2\_R} means that the radiation of the NCRE environment strikes the component from the \textit{L} side. The component was the \textit{B2} coating of the IC optical window \textit{1C} located on the \textit{R} hand side of the unit. The second, third, and fourth columns of the table contain TID values calculated from incident electron, proton, and helium ion radiation, respectively. A TID record consists of two numbers, e.g., 1.0E+00 $\pm$ 1.0E-02 $\mathrm{[rad]}$. The first is the TID magnitude, while the second is its standard deviation. The fifth column represents a sum of TID from all particle species. From the propagation of uncertainty formula

\begin{equation}
    \Delta y = \sqrt{\sum_{i=1}^{N} \left( \frac{\partial y}{\partial x_i} \Delta x_i \right)^2},
\end{equation}

\noindent $\Delta \mathrm{TID}$ values are calculated for the $\sum\mathrm{TID}$ values. Resulting magnitudes for the iodine cell component indicate similar TID values for coatings and the cell \textit{(1)} itself with respect to the direction of the incident radiation (\textit{L} and \textit{R}). TID values computed for the NCRE environment are smaller than the magnitudes for the WCRE environment. The glass of the cell, as well as the optical coatings, received radiation doses of a maximum of approx. 6 [rad] for NCRE environment and approx. 50 [rad] for WCRE environment. Calculated magnitudes don't affect the coating layers physical properties \cite{TPK75,WWR92}.   

{\color{red}{The origin of the standard deviations reported in Table \ref{tab:compasso_OCUS_IC_tid} deserves an explanation. The dose scorer accumulates the energy deposited within a scoring volume on a per-event basis and reports the standard error of the mean over all primary histories. Here, a history denotes a single primary particle together with all of its secondaries, i.e., one event in the Geant4 nomenclature. Histories that deposit no energy within the volume contribute zeros to the accumulated sums. Hence, if $n$ primary histories deposit a comparable amount of energy within the volume, the relative standard deviation of the scored dose reduces to $\Delta \mathrm{TID} / \mathrm{TID} \simeq 1/\sqrt{n}$, where $\mathrm{TID}$ and $\Delta \mathrm{TID}$ denote the dose value and its standard deviation reported in Table \ref{tab:compasso_OCUS_IC_tid}, i.e., the first and the second number of each record, and $n$ is the number of histories depositing energy within the scoring volume. It is therefore governed by the number of histories that reach and deposit energy within the scoring volume, and not by the number of primaries launched. Since the scoring volumes are coating layers of 27.2 nm to 98.6 nm thickness located behind the 7 mm Al shield, only a few histories deposit energy within an individual layer. A relative standard deviation of 100\% thus corresponds to a single scoring event, and the entries reported as 0.00E+00 $\pm$ 0.00E+00 correspond to layers in which no history of a given species deposited energy at all. The quantity used for the assessment is the total dose $\sum$TID, which is dominated by the well sampled proton contribution, i.e. from 58\% to 99\% of the dose in the directly irradiated layers. Hence, although the electron entries carry a larger relative standard deviation than the proton ones, their absolute contribution to the dose is smaller, so that their uncertainty propagates into $\sum$TID with a correspondingly smaller weight. For the dimensioning WCRE environment, its relative standard deviation ranges from 0.9\% to 4.6\% in all directly irradiated layers but one, for which it amounts to 12.1\%.}} 

\begin{table}[!htbp]
  \footnotesize
  \centering
  ~\clap{
    \begin{tabular}{ l|c|c|c|c } 
      Component symbol & TID $\mathrm{e^-}$ [rad] & TID $\mathrm{H^+}$ [rad] & TID $\mathrm{He}$ [rad] & $\sum$ TID [rad] \\ \hline
      \multicolumn{5}{c}{Environment: NCRE}\\ \hline
      L\_NCRE\_1 & 1.27E-02 $\pm$ 3.00E-04 & 1.18E+00 $\pm$ 9.46E-03 & 2.27E-02 $\pm$ 5.04E-05 & 1.22E+00 $\pm$ 9.47E-03 \\
      L\_NCRE\_1C\_B2\_L & 1.09E-01 $\pm$ 4.46E-02 & 4.45E+00 $\pm$ 1.58E-01 & 6.23E-02 $\pm$ 7.70E-04 & 4.62E+00 $\pm$ 1.64E-01  \\
      L\_NCRE\_1C\_A2\_L & 1.84E+00 $\pm$ 1.01E+00 & 3.19E+00 $\pm$ 1.18E-01 & 4.51E-02 $\pm$ 5.24E-04 & 5.08E+00 $\pm$ 1.02E+00  \\
      L\_NCRE\_1C\_B1\_L & 2.29E-01 $\pm$ 1.07E-01 & 6.13E+00 $\pm$ 1.18E+00 & 6.28E-02 $\pm$ 1.54E-03 & 6.42E+00 $\pm$ 1.18E+00  \\
      L\_NCRE\_1C\_A1\_L & 2.20E+00 $\pm$ 1.98E+00 & 3.05E+00 $\pm$ 1.22E-01 & 1.17E-03 $\pm$ 9.60E-04 & 5.25E+00 $\pm$ 1.99E+00  \\
      L\_NCRE\_1C\_B2\_R & 2.10E-03 $\pm$ 2.10E-03 & 2.33E-01 $\pm$ 2.32E-02 & 6.66E-03 $\pm$ 1.73E-04 & 2.41E-01 $\pm$ 2.33E-02  \\
      L\_NCRE\_1C\_A2\_R & 9.63E-03 $\pm$ 8.77E-03 & 1.93E-01 $\pm$ 3.03E-02 & 4.86E-03 $\pm$ 4.82E-03 & 2.08E-01 $\pm$ 3.19E-02  \\
      L\_NCRE\_1C\_B1\_R & 9.92E-03 $\pm$ 9.92E-03 & 2.18E-01 $\pm$ 3.60E-02 & 6.42E-03 $\pm$ 2.45E-04 & 2.34E-01 $\pm$ 3.73E-02  \\
      L\_NCRE\_1C\_A1\_R & 1.08E-02 $\pm$ 1.08E-02 & 1.52E-01 $\pm$ 2.08E-02 & 4.99E-03 $\pm$ 1.62E-04 & 1.69E-01 $\pm$ 2.34E-02  \\ \hline
      %L\_NCRE\_1\_G1 & 1.26E-02 $\pm$ 6.50E-03 & 1.89E+00 $\pm$ 1.61E-01 & 3.44E-02 $\pm$ 6.61E-04 & 1.94E+00 $\pm$ 1.61E-01  \\
      %L\_NCRE\_1\_G2 & 5.79E-03 $\pm$ 4.22E-03 & 4.92E-01 $\pm$ 8.42E-02 & 1.22E-02 $\pm$ 4.42E-04 & 5.10E-01 $\pm$ 8.43E-02  \\ \hline
      R\_NCRE\_1 & 1.55E-02 $\pm$ 4.28E-04 & 1.49E+00 $\pm$ 1.04E-02 & 2.79E-02 $\pm$ 5.53E-05 & 1.53E+00 $\pm$ 1.04E-02  \\
      R\_NCRE\_1C\_B2\_R & 5.62E-01 $\pm$ 3.75E-01 & 5.55E+00 $\pm$ 1.80E-01 & 8.23E-02 $\pm$ 1.00E-03 & 6.19E+00 $\pm$ 4.16E-01  \\
      R\_NCRE\_1C\_A2\_R & 7.55E-01 $\pm$ 6.97E-01 & 4.00E+00 $\pm$ 2.71E-01 & 5.96E-02 $\pm$ 6.28E-04 & 4.81E+00 $\pm$ 7.47E-01  \\
      R\_NCRE\_1C\_B1\_R & 4.06E-01 $\pm$ 2.69E-01 & 5.93E+00 $\pm$ 4.70E-01 & 8.47E-02 $\pm$ 1.94E-03 & 6.42E+00 $\pm$ 5.42E-01  \\
      R\_NCRE\_1C\_A1\_R & 1.08E-01 $\pm$ 5.08E-02 & 3.58E+00 $\pm$ 1.34E-01 & 5.82E-02 $\pm$ 7.90E-04 & 3.74E+00 $\pm$ 1.43E-01  \\
      R\_NCRE\_1C\_B2\_L & 2.82E-03 $\pm$ 2.82E-03 & 2.74E-01 $\pm$ 4.34E-02 & 7.99E-03 $\pm$ 2.95E-04 & 2.85E-01 $\pm$ 4.34E-02  \\
      R\_NCRE\_1C\_A2\_L & 0.00E+00 $\pm$ 0.00E+00 & 1.57E-01 $\pm$ 1.18E-02 & 5.49E-03 $\pm$ 1.53E-04 & 1.62E-01 $\pm$ 1.18E-02  \\
      R\_NCRE\_1C\_B1\_L & 6.69E-03 $\pm$ 5.30E-03 & 2.39E-01 $\pm$ 3.56E-02 & 7.15E-03 $\pm$ 3.34E-04 & 2.53E-01 $\pm$ 3.60E-02  \\
      R\_NCRE\_1C\_A1\_L & 0.00E+00 $\pm$ 0.00E+00 & 1.76E-01 $\pm$ 2.79E-02 & 5.80E-03 $\pm$ 2.82E-04 & 1.81E-01 $\pm$ 2.79E-02  \\
      %R\_NCRE\_1\_G1 & 1.73E-02 $\pm$ 1.05E-02 & 5.61E-01 $\pm$ 1.01E-01 & 1.37E-02 $\pm$ 3.55E-04 & 5.92E-01 $\pm$ 1.02E-01  \\
      %R\_NCRE\_1\_G2 & 5.06E-03 $\pm$ 4.94E-03 & 2.52E+00 $\pm$ 1.79E-01 & 4.36E-02 $\pm$ 7.47E-04 & 2.57E+00 $\pm$ 1.79E-01  \\
      \hline
      \multicolumn{5}{c}{Environment: WCRE}\\ \hline
      L\_WCRE\_1         & 2.22E-02 $\pm$ 7.34E-04 & 1.16E+01 $\pm$ 1.45E-02 & 4.93E-02 $\pm$ 1.74E-04 & 1.17E+01 $\pm$ 1.45E-02  \\
      L\_WCRE\_1C\_B2\_L & 3.90E-01 $\pm$ 1.95E-01 & 4.26E+01 $\pm$ 4.04E-01 & 1.72E-01 $\pm$ 2.84E-03 & 4.32E+01 $\pm$ 4.48E-01  \\
      L\_WCRE\_1C\_A2\_L & 1.74E+00 $\pm$ 1.36E+00 & 2.81E+01 $\pm$ 1.77E-01 & 1.16E-01 $\pm$ 1.84E-01 & 3.00E+01 $\pm$ 1.39E+00  \\
      L\_WCRE\_1C\_B1\_L & 2.55E-01 $\pm$ 1.60E-01 & 4.26E+01 $\pm$ 7.11E-01 & 1.68E-01 $\pm$ 4.14E-03 & 4.30E+01 $\pm$ 7.29E-01  \\
      L\_WCRE\_1C\_A1\_L & 3.92E+00 $\pm$ 3.85E+00 & 2.80E+01 $\pm$ 2.71E-01 & 1.15E-01 $\pm$ 2.62E-03 & 3.20E+01 $\pm$ 3.86E+00  \\
      L\_WCRE\_1C\_B2\_R & 1.14E-02 $\pm$ 1.14E-02 & 2.63E+00 $\pm$ 1.19E-01 & 1.33E-02 $\pm$ 4.34E-04 & 2.65E+00 $\pm$ 1.19E-01  \\
      L\_WCRE\_1C\_A2\_R & 0.00E+00 $\pm$ 0.00E+00 & 1.78E+00 $\pm$ 5.11E-02 & 9.36E-03 $\pm$ 2.88E-04 & 1.79E+00 $\pm$ 5.11E-02  \\
      L\_WCRE\_1C\_B1\_R & 0.00E+00 $\pm$ 0.00E+00 & 2.48E+00 $\pm$ 7.71E-02 & 1.32E-02 $\pm$ 6.65E-04 & 2.49E+00 $\pm$ 7.71E-02  \\
      L\_WCRE\_1C\_A1\_R & 9.06E-04 $\pm$ 9.06E-04 & 1.83E+00 $\pm$ 7.72E-02 & 9.41E-03 $\pm$ 3.93E-04 & 1.84E+00 $\pm$ 7.74E-02  \\ \hline
      %L\_WCRE\_1\_G1     & 1.62E-02 $\pm$ 8.85E-03 & 1.79E+01 $\pm$ 2.16E-01 & 7.74E-02 $\pm$ 3.36E-03 & 1.80E+01 $\pm$ 2.17E-01  \\
      %L\_WCRE\_1\_G2     & 3.77E-03 $\pm$ 3.77E-03 & 5.07E+00 $\pm$ 1.31E-01 & 2.62E-02 $\pm$ 3.28E-03 & 5.10E+00 $\pm$ 1.31E-01  \\ \hline
      R\_WCRE\_1         & 2.62E-02 $\pm$ 7.47E-04 & 1.46E+01 $\pm$ 1.58E-02 & 6.12E-02 $\pm$ 1.94E-04 & 1.47E+01 $\pm$ 1.58E-02  \\
      R\_WCRE\_1C\_B2\_R & 5.15E-01 $\pm$ 2.27E-01 & 5.31E+01 $\pm$ 4.37E-01 & 2.12E-01 $\pm$ 3.24E-03 & 5.39E+01 $\pm$ 4.93E-01  \\
      R\_WCRE\_1C\_A2\_R & 7.24E-01 $\pm$ 3.13E-01 & 3.59E+01 $\pm$ 2.22E-01 & 1.46E-01 $\pm$ 1.73E-03 & 3.67E+01 $\pm$ 3.84E-01  \\
      R\_WCRE\_1C\_B1\_R & 2.24E-01 $\pm$ 1.12E-01 & 5.28E+01 $\pm$ 7.42E-01 & 2.09E-01 $\pm$ 4.52E-03 & 5.33E+01 $\pm$ 7.51E-01  \\
      R\_WCRE\_1C\_A1\_R & 3.89E-01 $\pm$ 1.87E-01 & 3.56E+01 $\pm$ 4.19E-01 & 1.45E-01 $\pm$ 2.72E-03 & 3.61E+01 $\pm$ 4.59E-01  \\
      R\_WCRE\_1C\_B2\_L & 0.00E+00 $\pm$ 0.00E+00 & 2.97E+00 $\pm$ 8.40E-02 & 1.49E-02 $\pm$ 4.71E-04 & 2.99E+00 $\pm$ 8.40E-02  \\
      R\_WCRE\_1C\_A2\_L & 0.00E+00 $\pm$ 0.00E+00 & 2.05E+00 $\pm$ 6.39E-02 & 1.07E-02 $\pm$ 3.60E-04 & 2.06E+00 $\pm$ 6.39E-02  \\
      R\_WCRE\_1C\_B1\_L & 0.00E+00 $\pm$ 0.00E+00 & 2.85E+00 $\pm$ 1.28E-01 & 1.46E-02 $\pm$ 9.12E-04 & 2.86E+00 $\pm$ 1.28E-01  \\
      R\_WCRE\_1C\_A1\_L & 0.00E+00 $\pm$ 0.00E+00 & 1.93E+00 $\pm$ 6.90E-02 & 1.36E-02 $\pm$ 7.48E-04 & 1.95E+00 $\pm$ 6.90E-02  \\
      %R\_WCRE\_1\_G1     & 0.00E+00 $\pm$ 0.00E+00 & 5.28E+00 $\pm$ 1.29E-01 & 2.80E-02 $\pm$ 1.42E-03 & 5.31E+00 $\pm$ 1.29E-01  \\
      %R\_WCRE\_1\_G2     & 1.22E-02 $\pm$ 6.90E-03 & 2.36E+01 $\pm$ 2.50E-01 & 9.88E-02 $\pm$ 3.91E-03 & 2.37E+01 $\pm$ 2.50E-01  \\
      \hline \hline
    \end{tabular}
    }
  \caption{TID values for IC and its optical window coatings. Columns hold component symbol, TID values from $\mathrm{e^{-}}$, $\mathrm{H^{+}}$ and helium ions for the second, the third and the fourth column, respectively. Next, the fifth column holds TID values from all three particle species.}
  \label{tab:compasso_OCUS_IC_tid}
\end{table}

%---------------------------------------------------------------------------
\section{Conclusions}

The irradiation test campaign and subsequent radiation analysis demonstrate that proton and electron exposure brings no significant impact on the optical performance of the tested AR coatings. However, some samples Ti2P20-5 (SET1), Ta2E600-1 (SET2), Ta2P700-1 (SET2), Ta2P700-4 (SET2), Ti4P20-3 (SET3), and Ta4P700-5 (SET4) indicate minor, point-like defects on their surface {\color{red}{after the proton exposure, while no surface modification was observed after the electron exposure.}} {\color{red}{Sputtering of the coating material by the incident particles is excluded as their origin by the SRIM calculation of Table \ref{tab:sputtering_yields}, and they are tentatively attributed to the beam-assisted release of pre-existing localised growth defects of the coating, or to particulate contamination, see Section \ref{sec:irradiation_test}.}} Consequently, a four-layer $\mathrm{Ta_2O_5/SiO_2}$ stack produced via PIAD (without bake-out) was selected for the iodine cell optical window, achieving a maximal transmittance of 99.8\% at 532 nm.

Furthermore, radiation modelling and 3D Monte Carlo simulations provide the following assurances for the COMPASSO mission. The calculated TID doses of approximately 6 [rad] (nominal) and 50 [rad] (worst-case) are well below the threshold required to affect the physical properties of coating layers \cite{TPK75,WWR92}. The selected coatings are expected to remain unaffected by corpuscular radiation for several years in the ISS radiation environment. Analysis confirms that ATOX and electromagnetic radiation are effectively blocked by the payload's aluminium shell, {\color{red}{posing no degradation risk}} to the internal spectroscopy subunit. {\color{red}{It is also emphasized that the lifetime assessment of the coatings is based on the electron and proton loads. The ion component of the environment, dominated by helium ions, contributes a fraction of the order of $10^{-4}$\,\% of the particle fluence deposited within the coating stack and at most 1.6\% of its total ionizing dose. It is therefore a negligible contributor to the cumulative degradation of the selected coating and is enveloped by the proton exposure applied during the irradiation test.}} Hence, optical components of the COMPASSO mission are highly resilient to radiation and maintain their structural and spectral integrity within the LEO radiation environment.

%---------------------------------------------------------------------------
\section{Data Availability Statement}

The original contributions presented in the study are included in the article. Further inquiries can be directed to the corresponding author.

\section{Acknowledgements}
This work was conducted at the German Aerospace Center (DLR) research and development branch, and partially supported (M. Sznajder) by a program of the Polish Ministry of Science under the title \textit{Regional Excellence Initiative}, project No. RID/SP/0050/2024/1. The AR coatings development and spectral measurements {\color{red}{were}} supported by the Ministry of Education, Youth and Sport of the Czech Republic as a part of the QUEENTEC project, reg. nr. CZ.02.01.01/00/22\_008/0004649 Quantum Engineering and Nanotechnology, by EURAMET project TOCK, reg. n. 22IEM01 and by the Czech Academy of Sciences (projects RVO:68081731 and SAV21-VP31).

%\clearpage
\appendix
\section{\color{red}{On the choice of the AE8/AP8 models for the trapped radiation environment}}

{\color{red}{The purpose of the following discussion is to justify the choice of the AE8/AP8 models, rather than the more recent AE9/AP9 (IRENE) models, for computing the trapped electron and proton spectra used throughout this work. The presented RHA results of the COMPASSO iodine reference cell cover only a fraction of the whole COMPASSO payload system. Since the payload was qualified for compliance with the ECSS standards, the RHA activity followed the corresponding ECSS guidelines. In particular, the ECSS-E-ST-10-04C standard still recommends the AE8/AP8 models for describing the trapped electron and proton environment, which motivates their usage.}}

\begin{figure}[!htbp]
  \centering
  \includegraphics[width=0.9\textwidth]{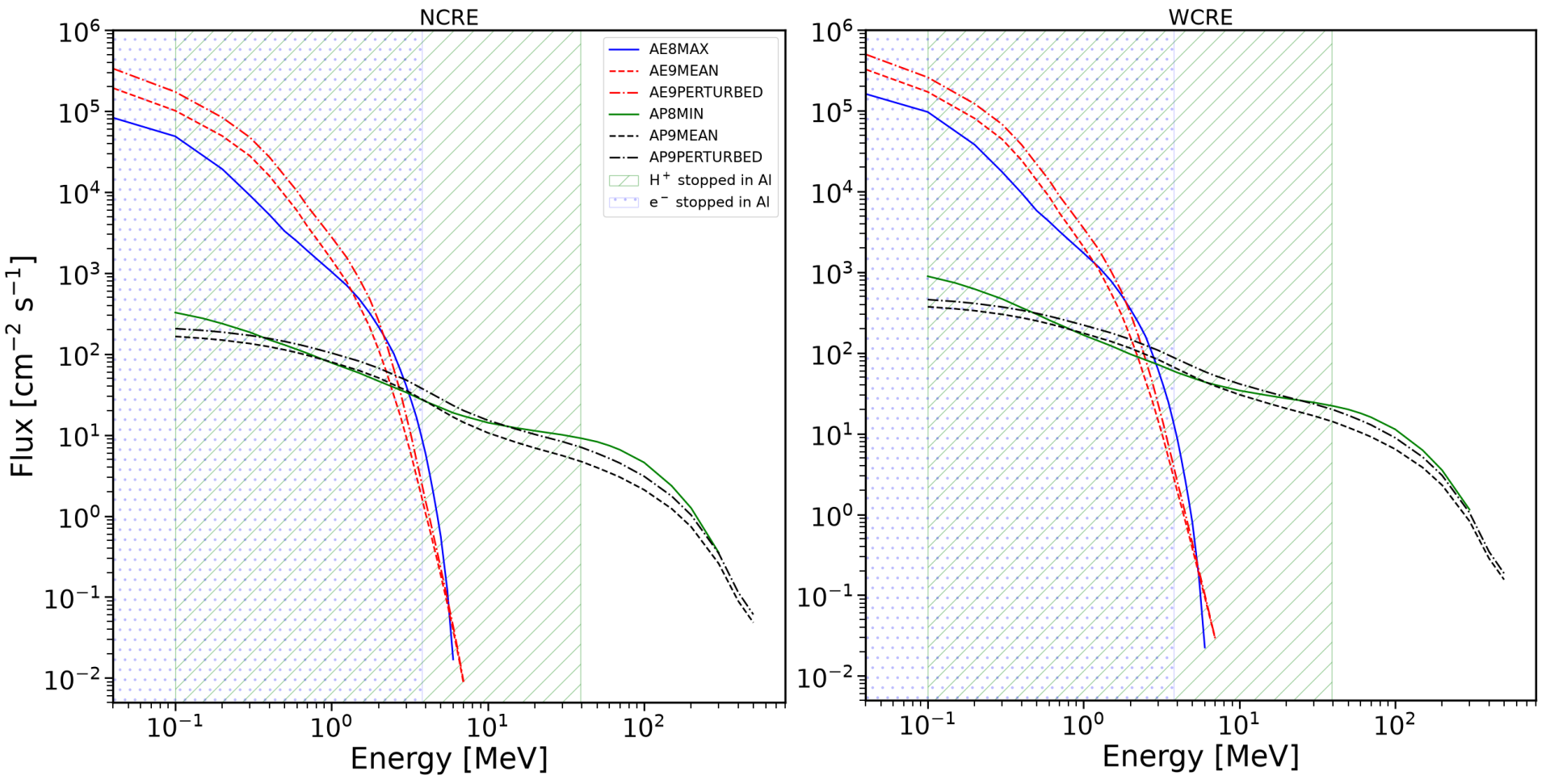}
  \caption{\color{red}{Comparison of the electron and proton integral flux spectra for the NCRE and WCRE environments calculated with the AE8/AP8 and AE9/AP9 (IRENE) models.}}
  \label{fig:model_comparison}
\end{figure}

{\color{red}{To confirm that this choice does not compromise the analysis, the AE8/AP8 and AE9/AP9 (IRENE) models were compared directly for the NCRE and WCRE environments in terms of the integral flux, as shown in Figure \ref{fig:model_comparison}. The AE8/AP8 models are found to be more conservative, i.e., over the energy ranges that penetrate the shield and contribute to the dose, they predict a flux equal to or higher than that of the AE9/AP9 (IRENE) models. Consequently, using the AE8/AP8 models does not lead to an underestimation of the total cumulative dose. In Figure \ref{fig:model_comparison}, the blue dotted and the green areas denote the electron and proton kinetic energy ranges stopped by the COMPASSO shield. The electron boundary kinetic energy above which electrons pass the shield is approximately 3.8~MeV, obtained from the work of Katz and Penfold \cite{KPA52}, while the corresponding proton boundary energy is 39~MeV, derived from the NIST PSTAR database \cite{NIST92}. The spectra were computed with the AE9/AP9 (IRENE) models (v1.50) in both the \textit{mean} and \textit{perturbed} run modes. The \textit{mean} mode provides the model's best-estimate (average) trapped-particle flux in each energy and spatial bin, representing the nominal, deterministic radiation environment. The \textit{perturbed} mode builds on this estimate by incorporating the statistical uncertainties associated with measurement error and gap-filling in the underlying data, thereby providing a more conservative flux estimate that reflects the confidence bounds of the model itself, independent of any time-dependent space-weather variability. For the AE8/AP8 models, a 50\% confidence level was used for the trapped particles, which likewise yields an average flux estimate. Data for both models were obtained from the ESA Space Environment Information System (SPENVIS) \cite{spenvis}.}}

%---------------------------------------------------------------------------
\section{\color{red}{Optical microscopy of the irradiated specimens}}
\label{app:sample_pictures}

{\color{red}{Point-like morphological changes were identified on six of the exposed specimens after the proton exposure. The corresponding optical micrographs, recorded with the Keyence VHX7000 confocal optical microscope, cover the whole surface of each specimen. The left column shows the surface state before and the right column after the exposure. The difference in color between the specimens follows from the different anti-reflective coating designs listed in Table \ref{tab:samples_and_coatings}, whereas the bright rim along the edges is an artefact of the illumination at the specimen border and is not related to the coating. The regions in which a change was identified are marked with red frames.}}

\begin{figure}[!htbp]
  \centering
  \includegraphics[width=0.7\textwidth]{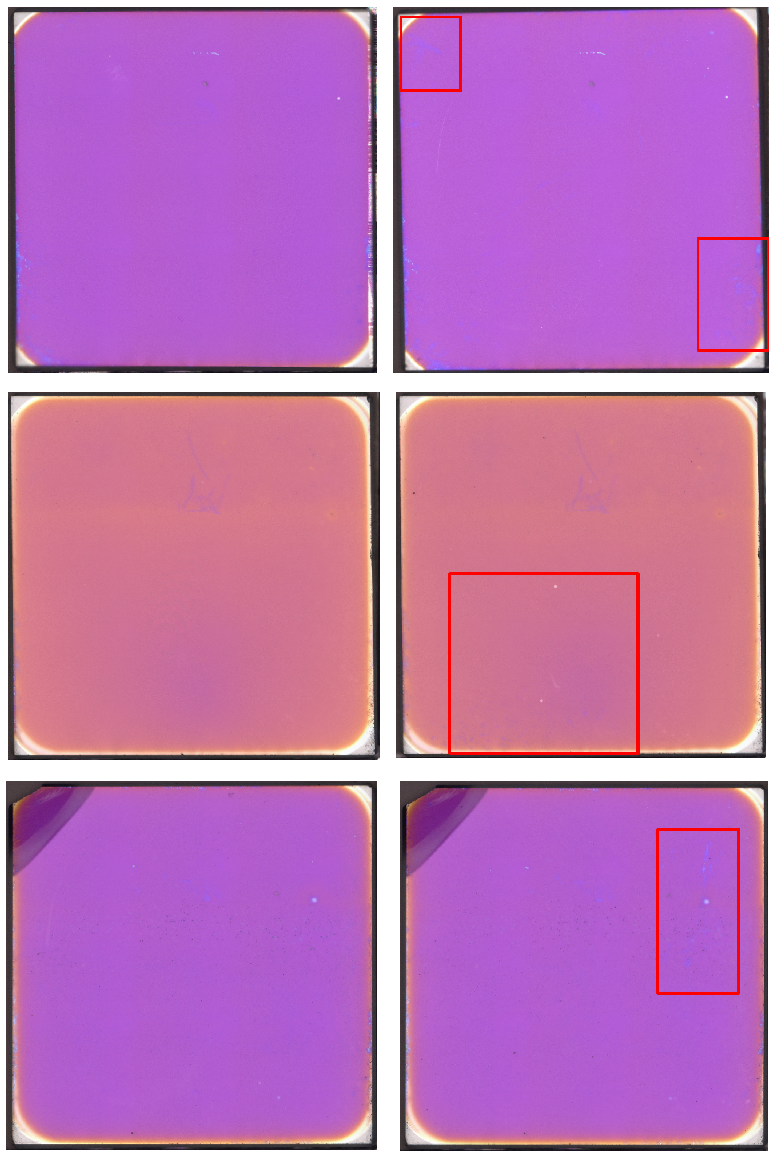}
  \caption{\color{red}{From top to bottom: specimens Ti2P20-5 (SET1), Ta2E600-1 (SET2) and Ta2P700-1 (SET2). Surface state before (left) and after (right) the proton exposure.}}
  \label{fig:irradiated_samples_1}
\end{figure}

\begin{figure}[!htbp]
  \centering
  \includegraphics[width=0.7\textwidth]{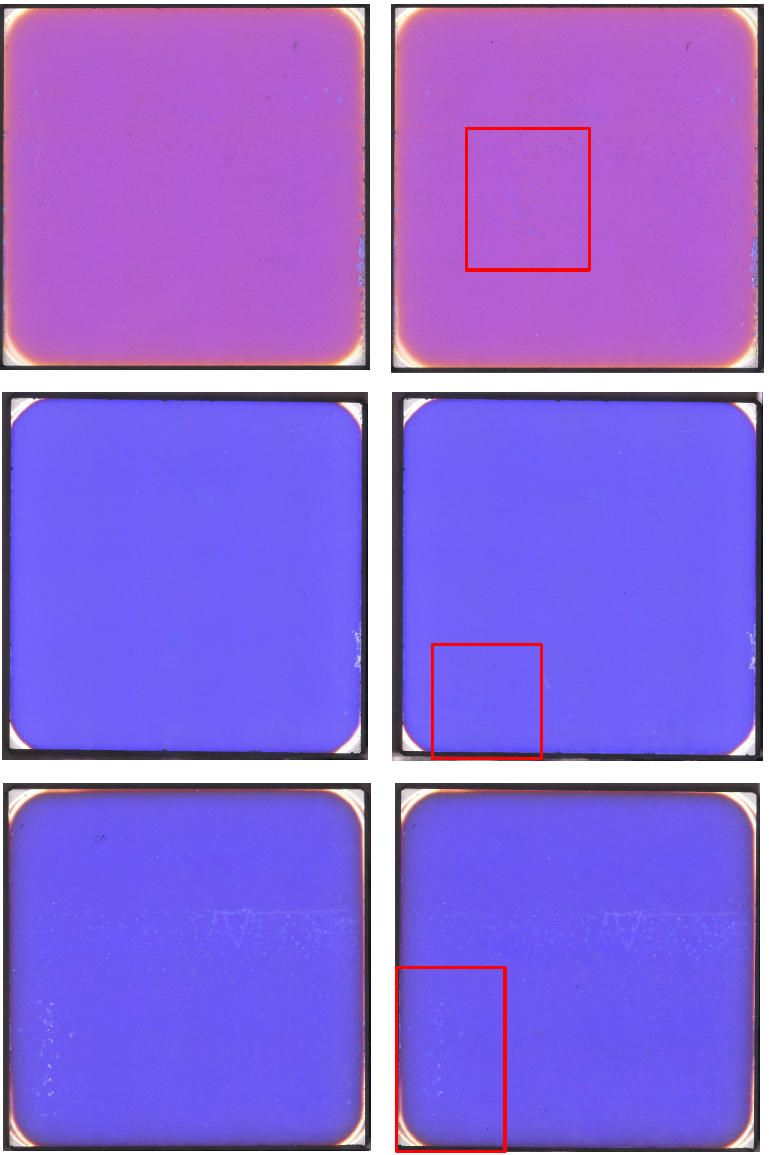}
  \caption{\color{red}{From top to bottom: specimens Ta2P700-4 (SET2), Ti4P20-3 (SET3) and Ta4P700-5 (SET4). Surface state before (left) and after (right) the proton exposure.}}
  \label{fig:irradiated_samples_2}
\end{figure}

{\color{red}{In Figure \ref{fig:irradiated_samples_1}, specimen Ti2P20-5 (SET1) shows two such regions, in the upper left and in the lower right part of the surface, specimen Ta2E600-1 (SET2) shows a group of bright point-like features in the lower left quadrant, and specimen Ta2P700-1 (SET2) shows a single bright feature on the right-hand side. In Figure \ref{fig:irradiated_samples_2}, specimen Ta2P700-4 (SET2) shows a faint modification in the central part of the surface, while specimens Ti4P20-3 (SET3) and Ta4P700-5 (SET4) show point-like features in the lower left quadrant.}}

{\color{red}{Several features are present in both columns and are therefore not related to the exposure. The dark region in the upper left corner of Ta2P700-1 is a chip of the specimen edge, which was already present when the specimen was received for the test. The diagonal streak in the upper part of Ta2E600-1 and the textured band crossing the middle of Ta4P700-5 originate from the deposition process or from the handling of the specimens before the exposure. The remaining fifty-eight specimens of the campaign show no difference between the two states, and no specimen shows any modification after the electron exposure.}}

{\color{red}{It should be emphasized that these are overview micrographs of the whole specimen. The identified features are close to the resolution limit at this magnification, so that their lateral size, their depth and their areal density cannot be extracted from them, and no dedicated metrology of the features was performed.}}

%---------------------------------------------------------------------------
\section{\color{red}{Assessment of the particle flux applied during the irradiation test}}
\label{app:flux_safety}

{\color{red}{Two rate-dependent limitations are relevant for a dielectric coating stack exposed to charged particles: the dielectric strength of the coating materials, which bounds the internal electric field that the deposited charge may build up, and the current density above which the radiation damage in oxides is reported to become dose rate dependent. The particle fluxes applied during the irradiation test, listed in Table \ref{tab:fluence}, are assessed below against both.}}

\paragraph{\color{red}{Standards perspective}} \hfill \break

{\color{red}{ISO 15856:2010 \cite{iso-15856} specifies no maximum proton or electron flux for the on-ground testing of optical coatings. Its section 6.3.3.8 states, regarding irradiation with low-energy electrons, that \textit{an increase of a negative charge on the surface of the material will repel incident low-energy electrons, thereby reducing the net irradiance of the material}, and that \textit{this effect can be reduced by neutralizing the charge with proton irradiation or earthing of the material at on-ground testing}. The same standard defines low-energy electrons, i.e. those causing a degradation of the surface properties of a material, as electrons with a kinetic energy below 50 keV. All specimens of the present campaign were clamped in a metallic sample holder mounted on a grounded sample station, so that the charge accumulated on the specimen surface was effectively neutralized.}}

{\color{red}{ECSS-E-ST-20-06C Rev.1 \cite{ecss-e-st-20-06C} addresses spacecraft charging. Clause 9 of the standard covers internal parts and materials and lists optical elements among them, so its provisions apply to the coated window considered here. Requirement 9.2.3a refers back to requirement 6.2.1a, which sets a maximum electric field within a dielectric of $10^{7}$ $\mathrm{V\,m^{-1}}$. That limit holds unless \textit{it is demonstrated that the material, in the situation in which it is used, is capable to sustain a higher bulk field without ESD}, and the standard notes that the default value carries no safety margin of its own. The dielectric strength of the individual coating materials is therefore taken here as the reference field.}}

{\color{red}{ECSS-Q-ST-70-06C \cite{ecss-q-st-70-06C}, section 5.1.4.2, requires that the maximum dose rate, or the energy flux on a material surface, be determined both by the allowable temperature increase of the sample and by the admissible acceleration factor. The temperature of the sample holder was monitored during the exposure and no increase was observed, the specimens remaining at room temperature, so that the first criterion is satisfied. The acceleration factor was set on the basis of previous irradiation campaigns performed at the CIF and of the dielectric strength assessment reported below. Neither ECSS-E-HB-10-12A \cite{ecss-e-hb-10-12A} nor ECSS-E-ST-10-12C \cite{ecss-e-st-10-12C} provides guidance on the particle flux to be used during an irradiation test of optical components, nor on accelerated testing.}}

\paragraph{\color{red}{Dielectric strength}} \hfill \break

{\color{red}{A convenient quantity for assessing whether the applied test parameters were safe for the dielectric materials under examination is the dielectric strength. A simple model of the electric field built up within a dielectric under charged particle irradiation follows from combining the charge continuity equation with Ohm's law and Gauss's law. It describes the temporal build-up of the internal field when a deposited current density $j_{\mathrm{dep}}$ is balanced against the ohmic leakage through the finite conductivity $\sigma$ of the material,}}

\begin{equation} \label{eq:electric_field}
{\color{red}{
\mathcal{E}(t) = \frac{j_{\mathrm{dep}}}{\sigma} \left( 1 - e^{-t/\tau}\right), \qquad \tau = \frac{\epsilon_r \epsilon_0}{\sigma},
}}
\end{equation} 

{\color{red}{\noindent where $\epsilon_r$ is the relative permittivity of the material, $\epsilon_0$ the vacuum permittivity and $\tau$ the dielectric relaxation time. Expression \ref{eq:electric_field} is also recommended to use by the ECSS-E-ST-20-06C standard (section C.8.5) \cite{ecss-e-st-20-06C}. The latter is the quantity given in \cite{JIA16} as $\tau = \epsilon_0 \epsilon_r \rho_R$, with $\rho_R = 1/\sigma$ the resistivity of the material, and is obtained there from the same combination of local charge conservation, Ohm's law and Gauss's law used above.}}

{\color{red}{In this description $\sigma$ is treated as a constant material property, that is, the conductivity measured in the absence of irradiation. In reality the conductivity of a dielectric does not remain constant under exposure, since the ionisation produced by the incident particles generates free charge carriers and gives rise to an additional, radiation-induced contribution which depends on the dose rate and raises the conductivity above the value measured without irradiation \cite{SRR00}. A higher conductivity implies both a shorter relaxation time $\tau$ and a lower steady-state field $j_{\mathrm{dep}}/\sigma$, so that neglecting this contribution can only overestimate the field reached during the exposure. The values derived below are therefore to be understood as conservative upper bounds.}}

{\color{red}{For an irradiation test it is more convenient to express the deposited current density through the particle flux $f$, so that $j_{\mathrm{dep}} \equiv f\,q\,\eta$, where $q$ denotes the charge of the incident particle and $\eta$ is a dimensionless parameter stating which fraction of the incident particles is deposited within the material. For a worst case assessment it is set to $\eta=1$. With the fluxes of Table \ref{tab:fluence} this gives a deposited current density of 8.2E-05 $\mathrm{A\,m^{-2}}$ for the electrons and of 1.55E-06 $\mathrm{A\,m^{-2}}$ for the protons. Table \ref{tab:dielectric_strength} collects the material parameters of the three coating materials used in this work, the resulting relaxation time, the electric field reached at the end of the electron and of the proton exposure, and the dielectric strength taken from the literature.}}

\begin{table}[!htbp]
    \scriptsize
    \centering
    ~\clap{
    \begin{tabular}{c|c|c|c|c|c|c}
    {\color{red}{Material}} & {\color{red}{$\sigma$ [$\mathrm{S\,m^{-1}}$]}} & {\color{red}{$\epsilon_r$ [--]}} & {\color{red}{$\tau$ [s]}} & {\color{red}{$\mathcal{E}_{\mathrm{e}}$ [$\mathrm{V\,m^{-1}}$]}} & {\color{red}{$\mathcal{E}_{\mathrm{p}}$ [$\mathrm{V\,m^{-1}}$]}} & {\color{red}{$\mathcal{E}_{\mathrm{max}}$ [$\mathrm{V\,m^{-1}}$]}} \\
    \hline
    {\color{red}{$\mathrm{SiO_2}$}}   & {\color{red}{1E-14 \cite{SZE07}}} & {\color{red}{3.9 \cite{ROB04}}} & {\color{red}{3.45E+03}} & {\color{red}{1.27E+08}} & {\color{red}{2.25E+05}} & {\color{red}{5.85E+08 \cite{KG66}, 1E+09 \cite{SZE07}}} \\
    {\color{red}{$\mathrm{TiO_2}$}}   & {\color{red}{1E-13 \cite{MMV26}}} & {\color{red}{80 \cite{ROB04}}}  & {\color{red}{7.08E+03}} & {\color{red}{6.20E+06}} & {\color{red}{1.10E+04}} & {\color{red}{8.15E+08 to 9.8E+08 \cite{DLP14}}} \\
    {\color{red}{$\mathrm{Ta_2O_5}$}} & {\color{red}{1E-11 \cite{CNB25}}} & {\color{red}{22 \cite{ROB04}}}  & {\color{red}{1.95E+01}} & {\color{red}{7.68E+06}} & {\color{red}{3.52E+04}} & {\color{red}{3.9E+08 \cite{CNB25}}} \\
    \hline
    \end{tabular}
    }
    \caption{{\color{red}{Electrical conductivity $\sigma$, relative permittivity $\epsilon_r$, relaxation time $\tau$, electric field reached at the end of the electron exposure $\mathcal{E}_{\mathrm{e}}$ and of the proton exposure $\mathcal{E}_{\mathrm{p}}$, and dielectric strength $\mathcal{E}_{\mathrm{max}}$ of the three coating materials. The conductivity is the reciprocal of the resistivity quoted in the cited sources.}}}
    \label{tab:dielectric_strength}
\end{table}

{\color{red}{For all three materials the field reached at the end of the exposure remains below the reported dielectric strength. The smallest margin is obtained for $\mathrm{SiO_2}$, where the electron exposure produces 1.27E+08 $\mathrm{V\,m^{-1}}$ against a dielectric strength of 5.85E+08 $\mathrm{V\,m^{-1}}$, that is, a factor of 4.6. For $\mathrm{Ta_2O_5}$ and $\mathrm{TiO_2}$ the corresponding margins amount to 51 and 132. The proton exposure is not dimensioning, the associated fields being two to three orders of magnitude lower. The model was kept intentionally conservative. All the incident charge is assumed to stay in the coating ($\eta=1$), the radiation-induced conductivity is ignored, and no charge is allowed to leave, either as secondary electrons or through the grounding described above. Each of these assumptions raises the calculated field, so the real fields are lower. The applied test parameters therefore stay below the electrical breakdown threshold of all three coating materials.}}

\paragraph{\color{red}{Literature perspective}} \hfill \break

{\color{red}{A further indication is provided by the work of Corso et al. \cite{CPS24}, which shows that low-energy protons of 1, 16 and 50 keV, applied at fluxes from 2E+11 up to 4E+12 $\mathrm{H^+\,cm^{-2}\,s^{-1}}$, caused no flux dependent degradation of $\mathrm{TiO_2}$ and $\mathrm{ZrO_2}$ monolayers. The review of Jiang \cite{JIA16} on electron beam damage in oxides reports a threshold current density above which the damage becomes dose rate dependent. For a silicate glass this threshold lies between 0.3 and 0.6 $\mathrm{A\,cm^{-2}}$, i.e. 1.87E+18 to 3.75E+18 $\mathrm{cm^{-2}\,s^{-1}}$. The same total dose changed the oxygen K edge above this threshold, but not below it. The electron exposure applied here corresponds to a current density of 8.2E-09 $\mathrm{A\,cm^{-2}}$, nearly eight orders of magnitude below that threshold. In pure $\mathrm{SiO_2}$ glasses no such defect signature was detected at any of the current densities investigated in that review, which span 1E-02 to 1E+02 $\mathrm{A\,cm^{-2}}$, i.e. 6.24E+16 to 6.24E+20 $\mathrm{cm^{-2}\,s^{-1}}$.}}

\paragraph{\color{red}{Conclusions}} \hfill \break

{\color{red}{The particle fluxes applied during the irradiation test are within the safety threshold with respect to the electrical breakdown of the coating materials, and lie several orders of magnitude below the current densities at which a dose rate dependence of the damage has been reported. No systematic bias caused by a dose rate effect is therefore expected in the results presented in Section \ref{sec:irradiation_test}.}}

\newpage

\section{\textcolor{red}{Transmittance spectra}}
\label{sec:spectras}

{\color{red}The following graphs show the transmittance spectra of the samples collected directly after deposition and then after each experimental step. The samples were transported between the facilities in a dust-protective box, and surface cleaning was carried out prior to every spectral measurement as a standard part of the handling protocol.

Generally, there is no evidence of iodine exposure or of proton and electron irradiation affecting the coatings' transmittance, confirming their suitability for the desired application.

The samples labelled Ta2E20 and Ta2E600 show no influence of the exposure to iodine or the radiation. For the Ta2E600 series, the effect of annealing is evident in the graphs (originally porous coatings become more compact and dense, decreasing their thickness), shifting the maximum transmittance towards shorter wavelengths. Similarly, for the Ta2P20 and Ta2P700 samples, no evident changes are induced by iodine or radiation. There is a decrease in transmittance after annealing of Ta2P700 due to {\color{red}{crystallization}} in the tantalum oxide layers.

The results show that four-layer coatings (Ta4E600) exhibit better tolerance to spectral shifts induced by annealing than two-layer systems (Ta2E600). The spectral shifts are still present, but because of the shape of the spectral curve, they do not limit the coating's optical performance as much as in the two-layer design. We observe the same decrease in transmittance due to {\color{red}{crystallization}} in the Ta4P700 four-layer series.

The same conclusions can be drawn for the titanium-based sample series. One sample, Ti2P20-3, deviates from the rest and shows a reduced transmittance level after the iodine exposure step. Since none of the remaining samples of the same series and batch reproduces this behaviour, the deviation cannot be ascribed to the iodine exposure and is instead attributed to residual surface contamination acquired during handling. This interpretation is supported by the subsequent measurement: after the proton and electron irradiation step and the associated cleaning, the transmittance maximum recovered essentially to its as-deposited value, which would not be expected for a change intrinsic to the coating.

A similar behavior is observed for the Ti2P700-1 and Ti4P700-1 samples. In the former, transmittance decreases after annealing, a change absent in the other three samples of the series, while in the latter, the decrease is more pronounced than in the remaining samples. As both deviations occur in single samples rather than across a series, they are likewise assigned to residual surface contamination associated with the annealing step rather than to a property of the coating itself. 
}

\begin{figure}[p]
  \centering
  \includegraphics[width=\textwidth]{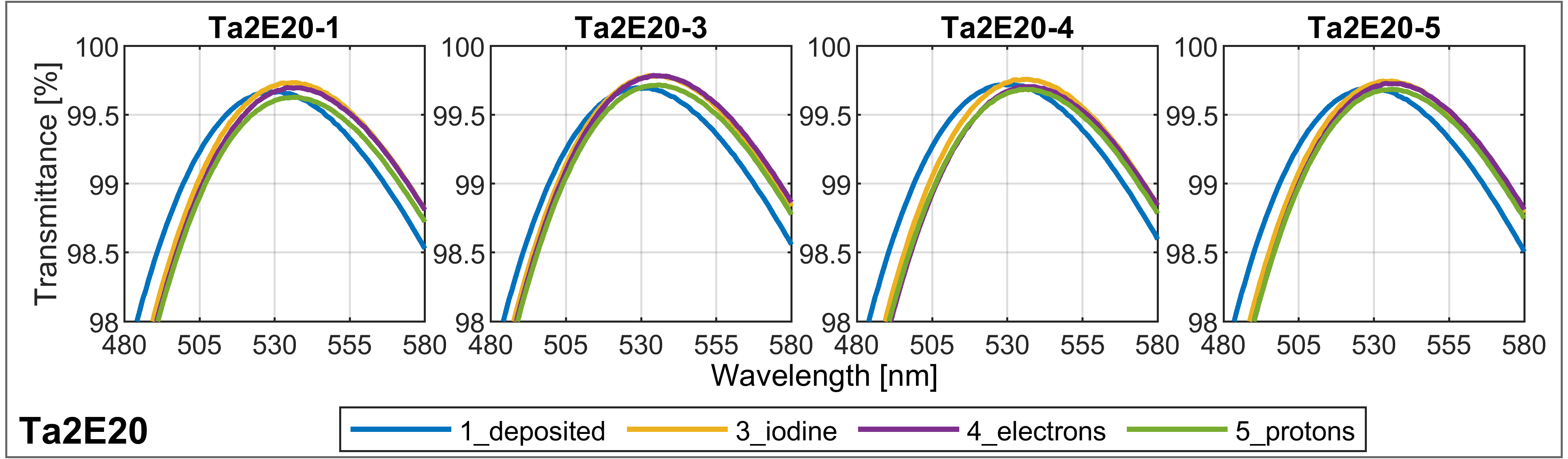}\\[4pt]
  \includegraphics[width=\textwidth]{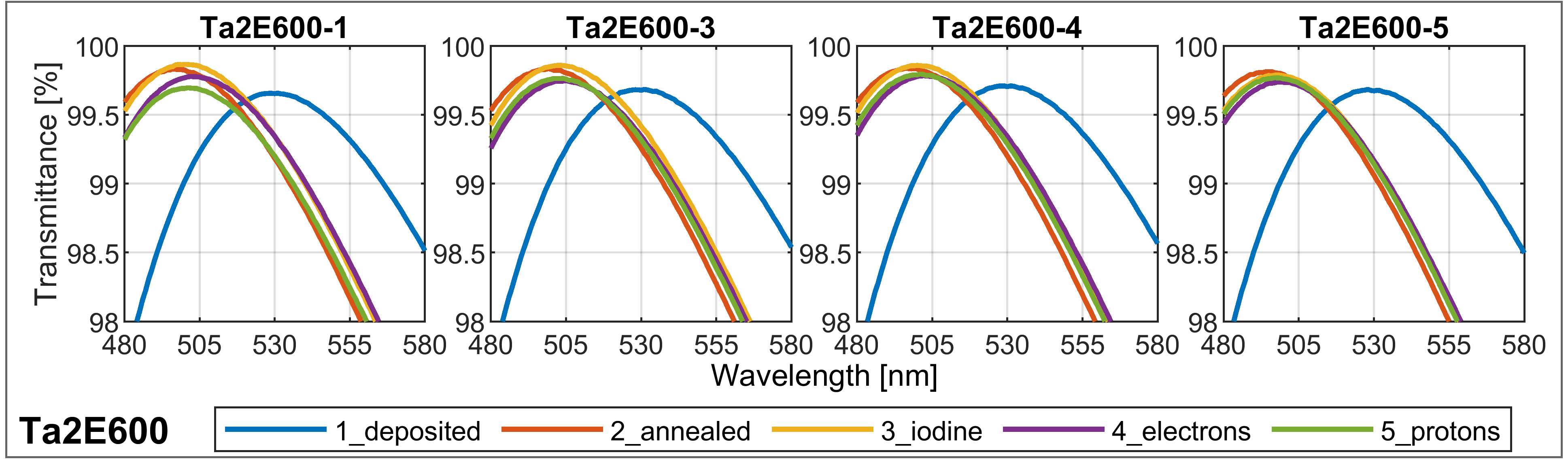}\\[4pt]
  \includegraphics[width=\textwidth]{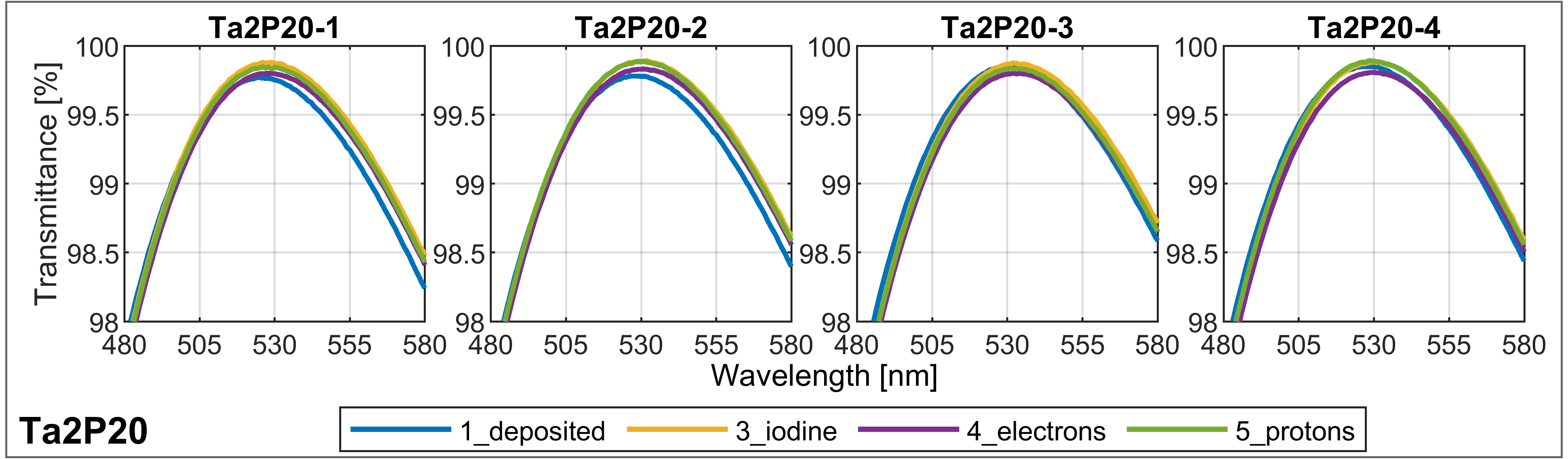}\\[4pt]
  \includegraphics[width=\textwidth]{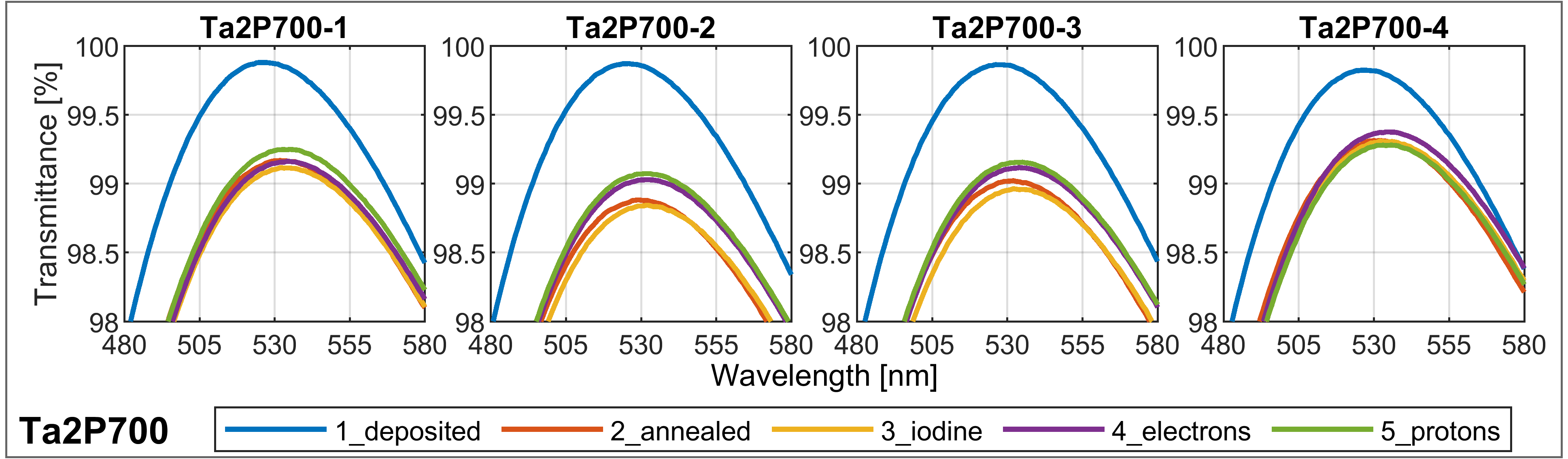}
  \caption{Transmittance spectra of the coated samples - part 1/4.}
  \label{fig:spektra1}
\end{figure}

\begin{figure}[p]
  \centering
  \includegraphics[width=\textwidth]{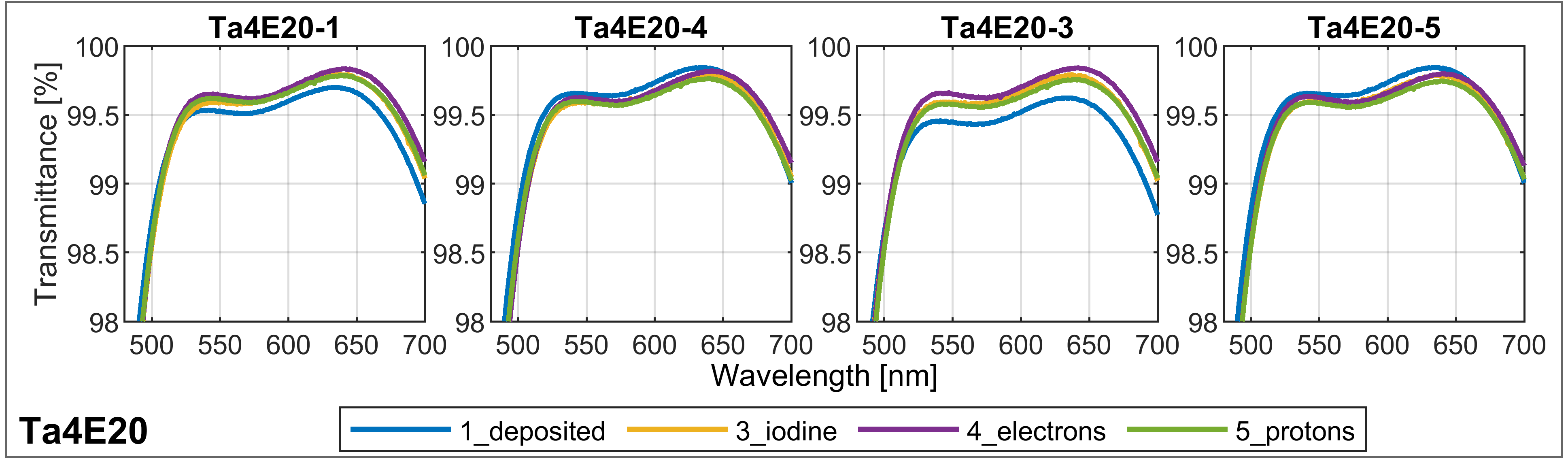}\\[4pt]
  \includegraphics[width=\textwidth]{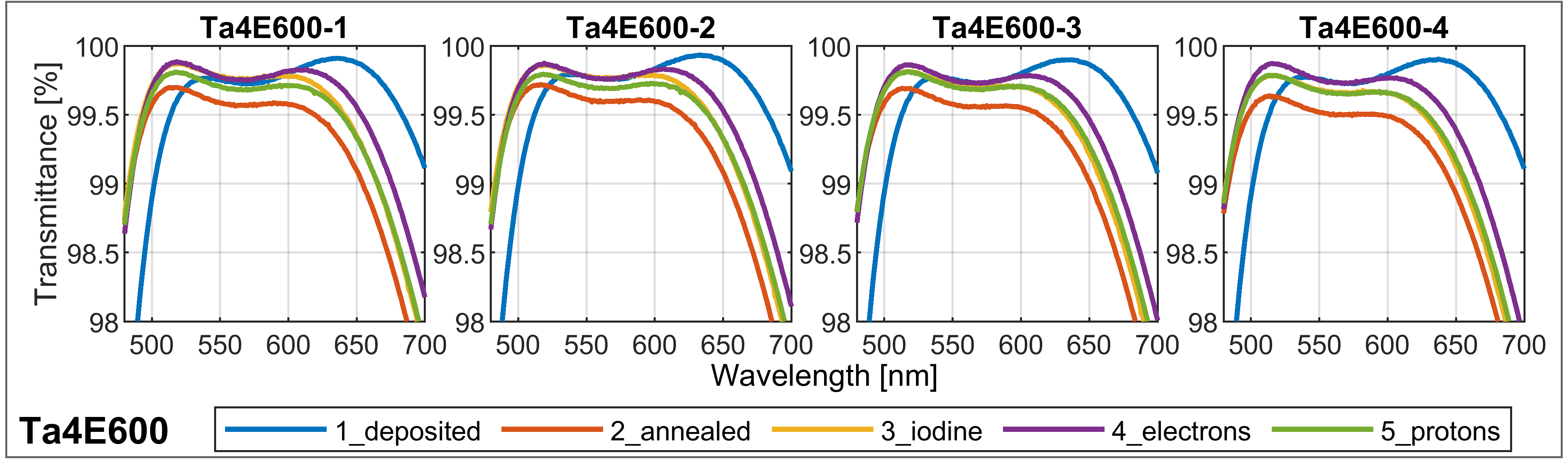}\\[4pt]
  \includegraphics[width=\textwidth]{Ta4P20.png}\\[4pt]
  \includegraphics[width=\textwidth]{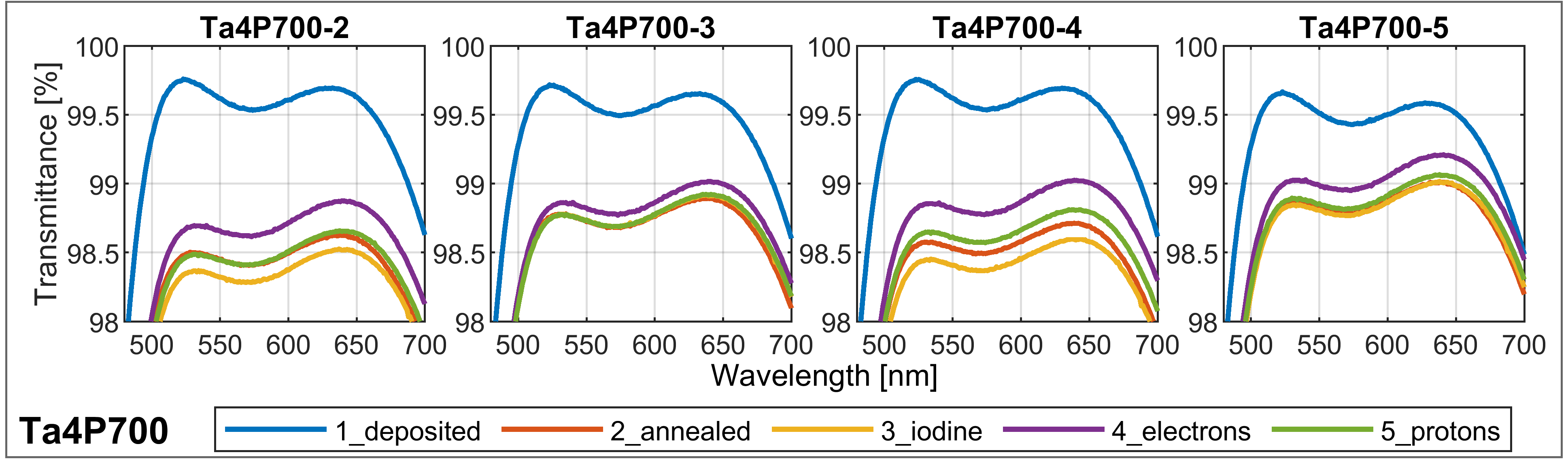}
  \caption{Transmittance spectra of the coated samples - part 2/4.}
  \label{fig:spektra2}
\end{figure}

\begin{figure}[p]
  \centering
  \includegraphics[width=\textwidth]{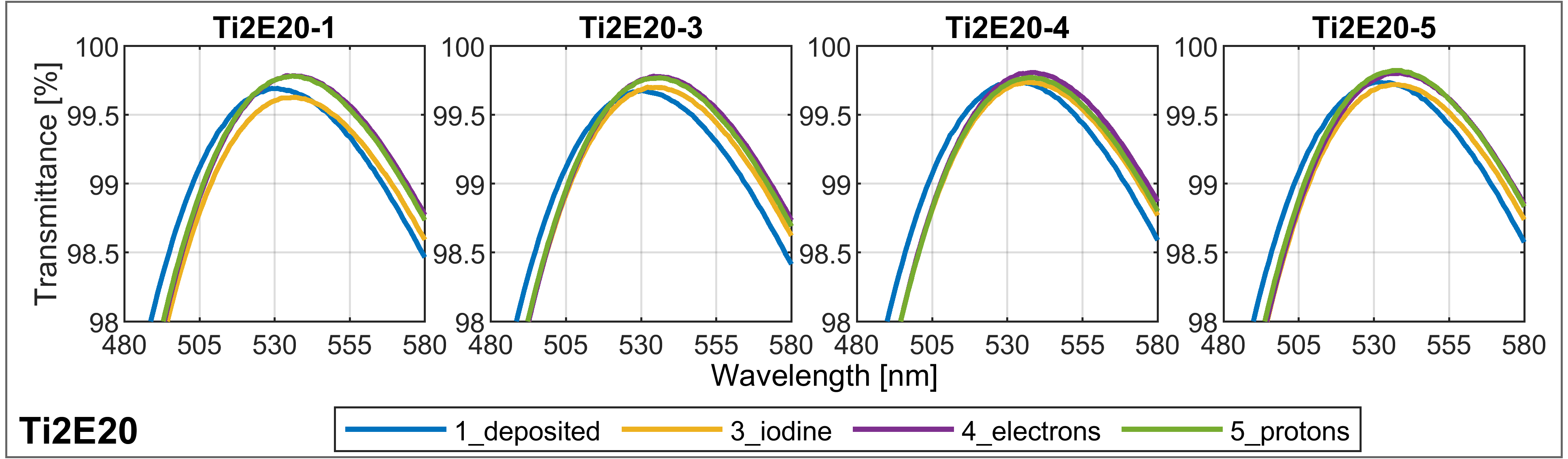}\\[4pt]
  \includegraphics[width=\textwidth]{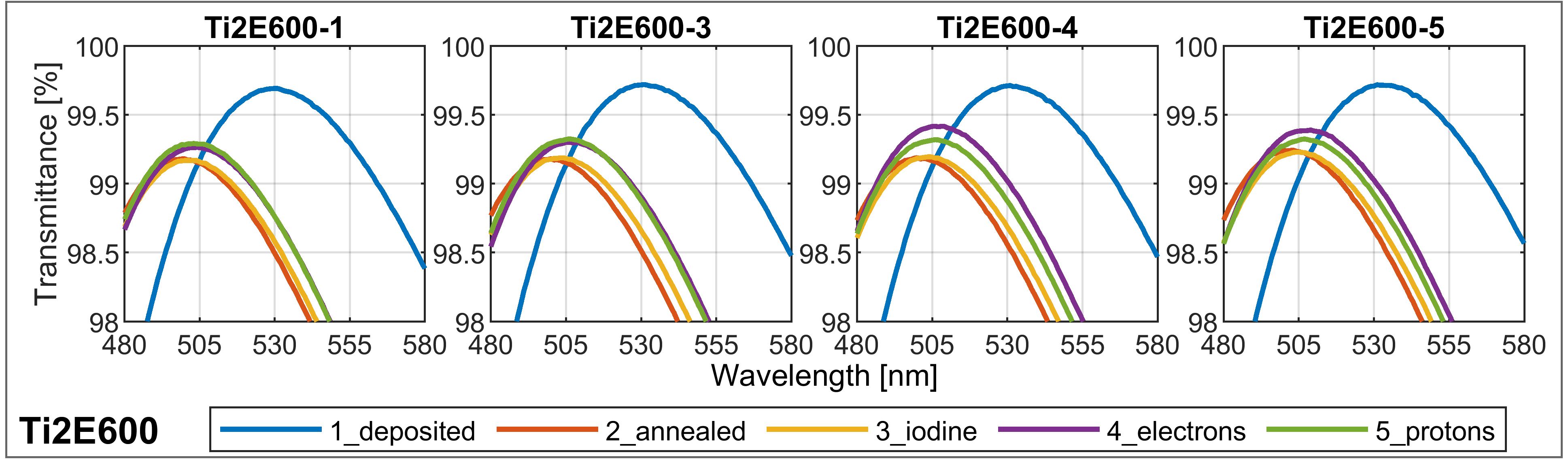}\\[4pt]
  \includegraphics[width=\textwidth]{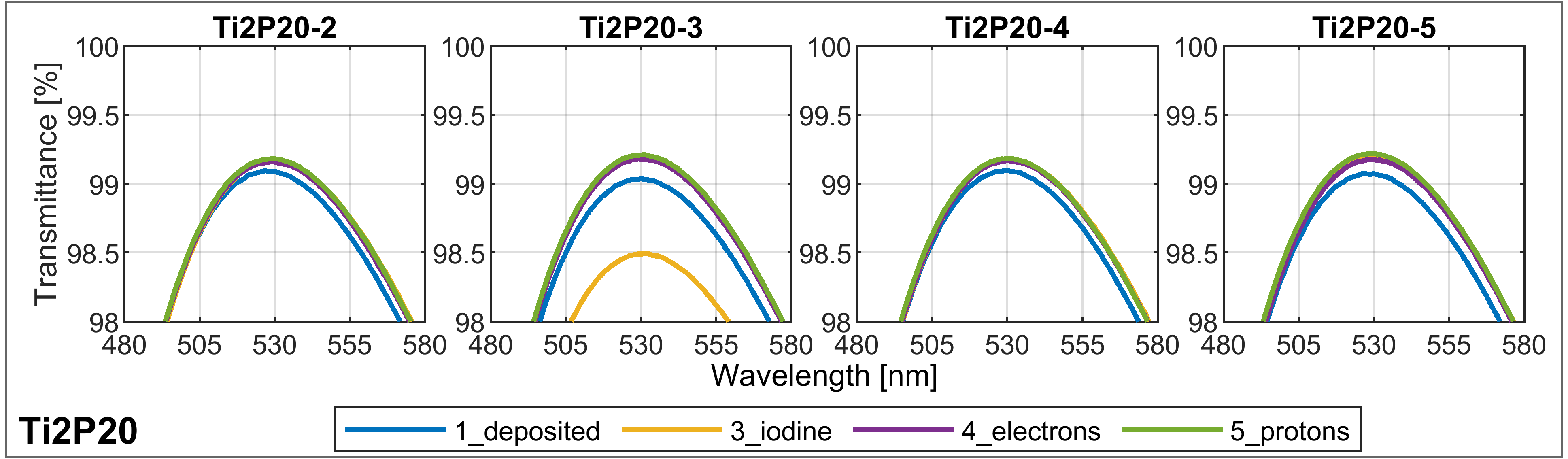}\\[4pt]
  \includegraphics[width=\textwidth]{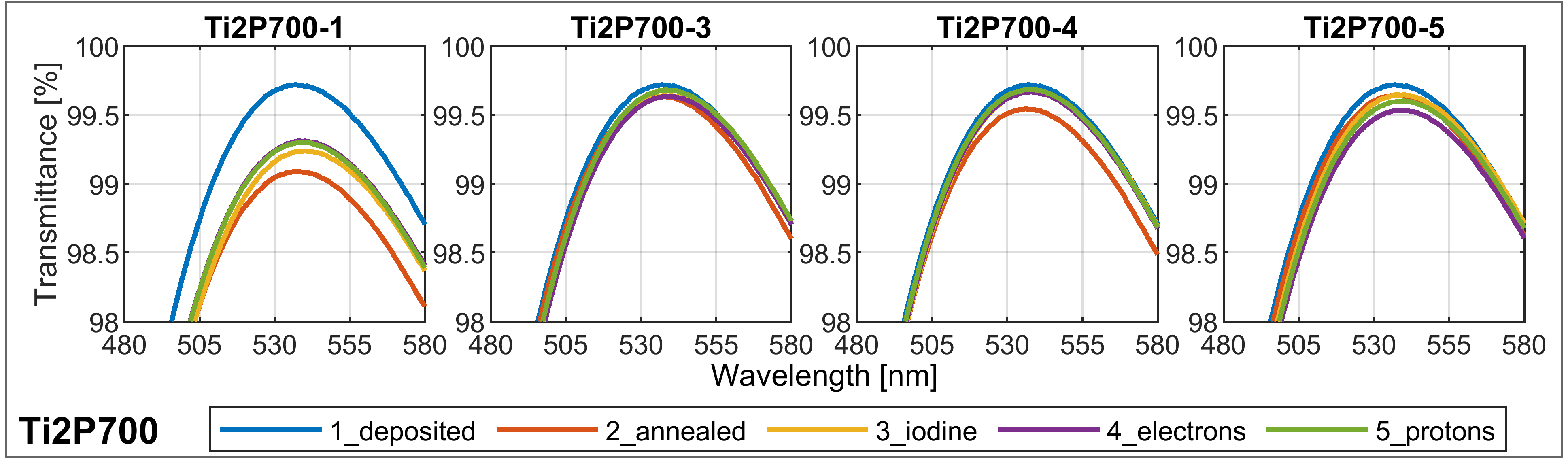}
  \caption{Transmittance spectra of the coated samples - part 3/4.}
  \label{fig:spektra3}
\end{figure}

\begin{figure}[p]
  \centering
  \includegraphics[width=\textwidth]{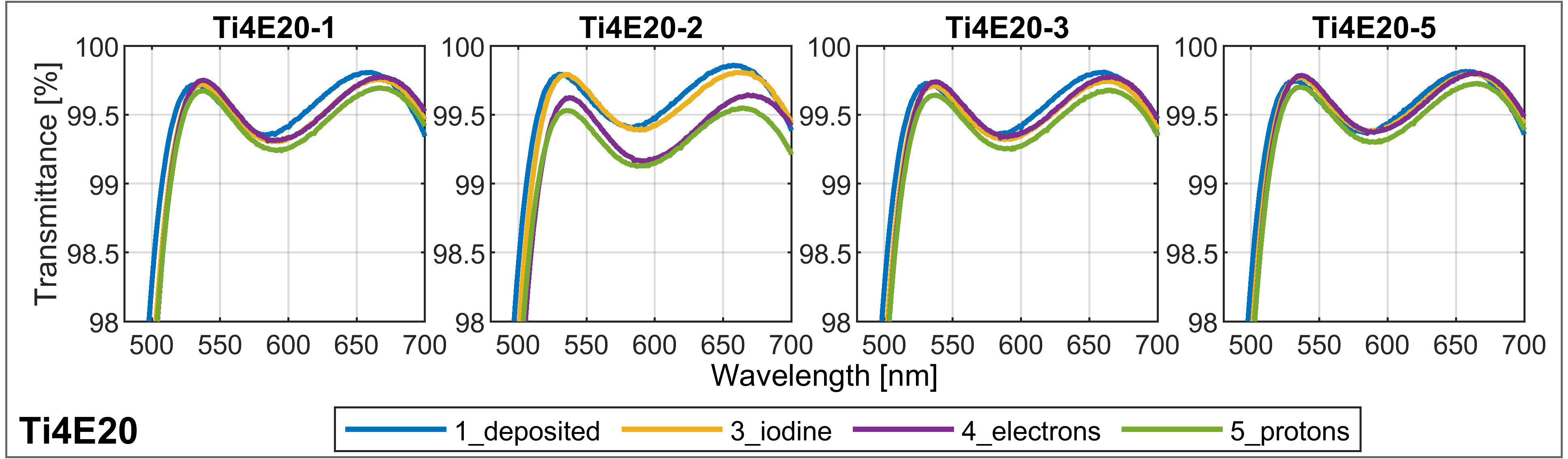}\\[4pt]
  \includegraphics[width=\textwidth]{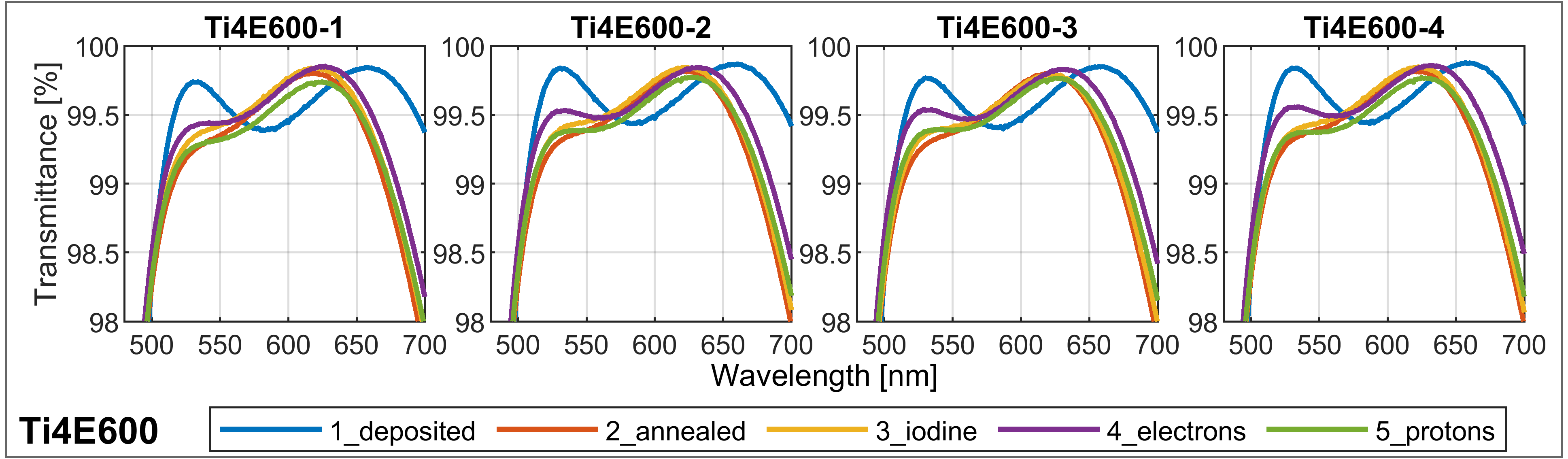}\\[4pt]
  \includegraphics[width=\textwidth]{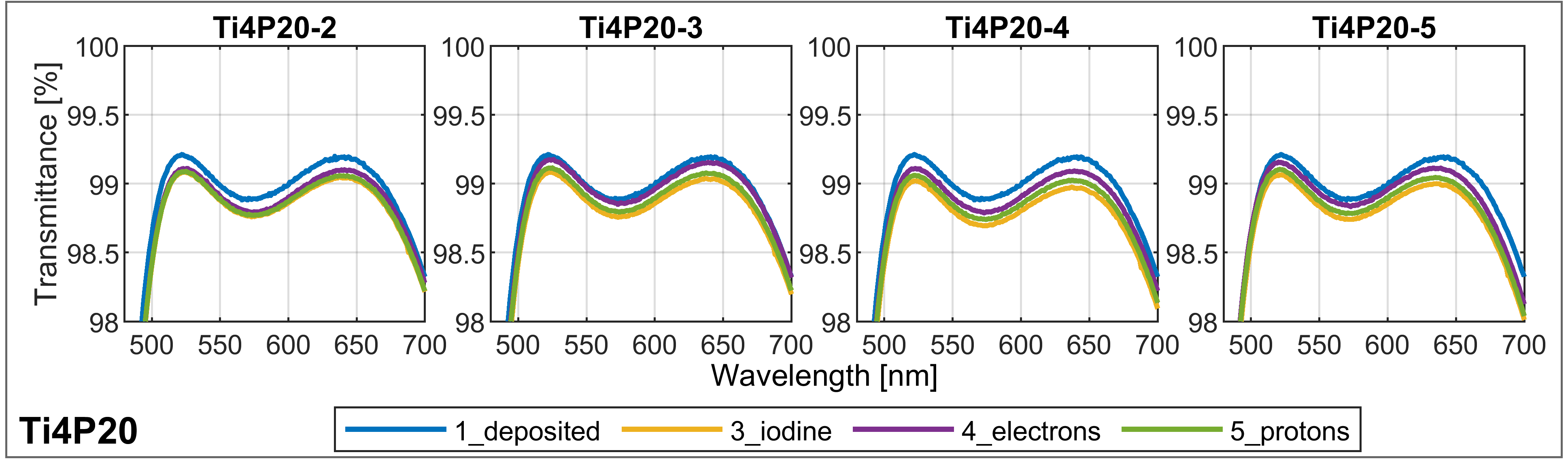}\\[4pt]
  \includegraphics[width=\textwidth]{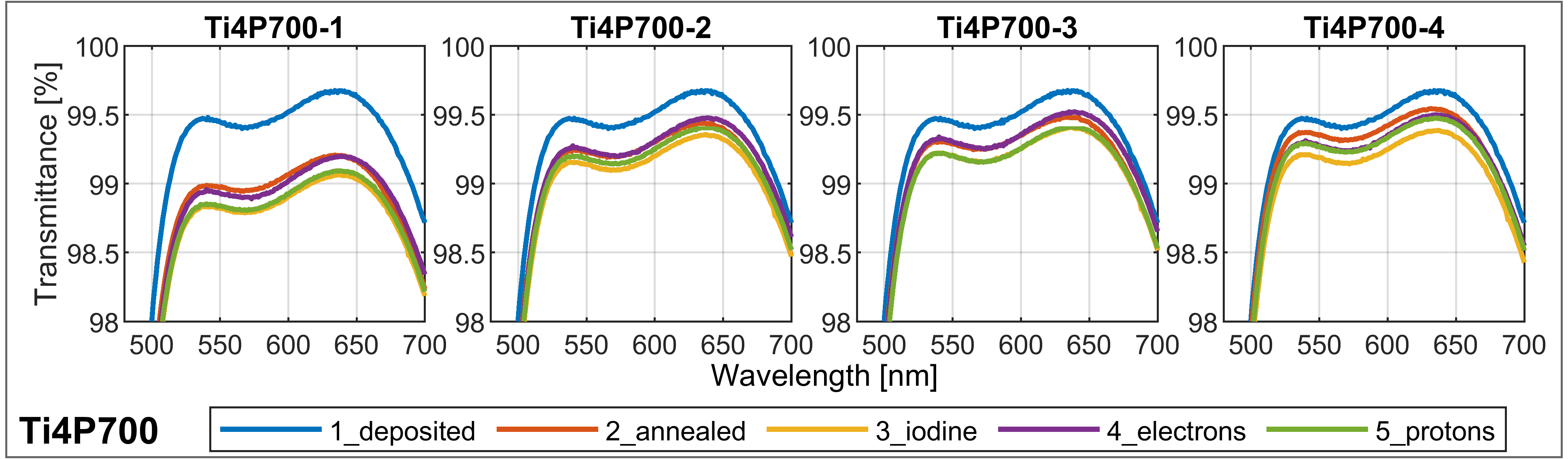}
  \caption{Transmittance spectra of the coated samples - part 4/4.}
  \label{fig:spektra4}
\end{figure}

\newpage

%---------------------------------------------------------------------------
\section{\textcolor{red}{Nomenclature}}
\label{sec:nomenclature}

\textcolor{red}{
The symbols, abbreviations, and acronyms used throughout this manuscript are listed below in alphabetical order. Table \ref{tab:nomenclature_symbols} collects the physical and mathematical symbols, while Table \ref{tab:nomenclature_abbreviations} collects the abbreviations and acronyms. Entries that are dimensionless or that denote a name rather than a physical quantity are marked with \textit{---} in the unit column.
}

\subsection{\textcolor{red}{Symbols}}

\begingroup
\leftcaptions
\begin{longtable}{@{}p{0.14\textwidth}|p{0.60\textwidth}|p{0.20\textwidth}@{}}
  Symbol & Description & Unit \\
  \midrule
  \endfirsthead
  \multicolumn{3}{@{}l}{\textit{Table \ref{tab:nomenclature_symbols} continued from previous page.}} \\
  Symbol & Description & Unit \\
  \midrule
  \endhead
  \midrule
  \multicolumn{3}{r@{}}{\textit{Continued on next page.}} \\
  \endfoot
  \hline \hline
  \caption{Physical and mathematical symbols used in the manuscript.}
  \label{tab:nomenclature_symbols} \\
  \endlastfoot
  $\alpha$             & Absorption (attenuation) coefficient of a material & $\mathrm{nm^{-1}}$ \\
  $\Delta \mathrm{TID}$ & Standard deviation of a total ionizing dose value & rad \\
  $\Delta x_{i}$        & Uncertainty of the input quantity $x_{i}$ in the propagation of uncertainty formula & unit of $x_{i}$ \\
  $\Delta y$            & Combined standard uncertainty of the derived quantity $y$ & unit of $y$ \\
  $d$                   & Eroded depth of the coating material & nm \\
  $E$                   & Kinetic energy of an incident particle & keV, MeV \\
  $\mathcal{E}$         & Electric field built up within a dielectric coating under charged particle irradiation & $\mathrm{V\,m^{-1}}$ \\
  $\mathcal{E}_{\mathrm{max}}$ & Dielectric strength of a coating material & $\mathrm{V\,m^{-1}}$ \\
  $\epsilon_{0}$        & Vacuum permittivity & $8.854\times10^{-12}$ $\mathrm{F\,m^{-1}}$ \\
  $\epsilon_{r}$        & Relative permittivity of a coating material & --- \\
  $\eta$                & Fraction of the incident particles deposited within the coating material & --- \\
  $f$                   & Particle flux applied during the irradiation test & $\mathrm{cm^{-2}\,s^{-1}}$ \\
  $F$                   & Particle fluence, i.e. particle flux integrated over the mission time & $\mathrm{cm^{-2}}$ \\
  H                     & High-refractive-index layer of an anti-reflective coating design & --- \\
  $I$                   & Light intensity at a penetration depth $x$ within a target material & $\mathrm{W\ m^{-2}}$ \\
  $I_{\mathrm{0}}$      & Light intensity incident on the target material surface & $\mathrm{W\ m^{-2}}$ \\
  $j_{\mathrm{dep}}$    & Current density deposited within the coating material & $\mathrm{A\,m^{-2}}$ \\
  $k$                   & Extinction coefficient of a material & --- \\
  L                     & Low-refractive-index layer of an anti-reflective coating design & --- \\
  $\lambda$             & Wavelength of electromagnetic radiation & nm \\
  $M$                   & Molar mass of the coating material & $\mathrm{g\,mol^{-1}}$ \\
  $n$                   & Atomic density of the coating material & $\mathrm{cm^{-3}}$ \\
  $N$                   & Number of input quantities in the propagation of uncertainty formula & --- \\
  $N_{\mathrm{A}}$       & Avogadro constant & $6.022\times10^{23}$ $\mathrm{mol^{-1}}$ \\
  $N_{\mathrm{h}}$       & Number of primary histories depositing energy within a scoring volume of the Monte Carlo simulation & --- \\
  $N_{\mathrm{ML}}$      & Areal density of an atomic monolayer of the coating material & $\mathrm{cm^{-2}}$ \\
  $q$                   & Electric charge of an incident particle & C \\
  $R_{\mathrm{E}}$      & Earth radius, used as a unit of radial distance in the equatorial plane & 1 $R_{\mathrm{E}}$ = 6371 km \\
  $\rho$                & Mass density of the coating material & $\mathrm{g\,cm^{-3}}$ \\
  $\rho_{R}$            & Electrical resistivity of a coating material, i.e. the reciprocal of $\sigma$ & $\mathrm{\Omega\,m}$ \\
  $\sigma$              & Electrical conductivity of a coating material & $\mathrm{S\,m^{-1}}$ \\
  $\sum \mathrm{TID}$   & Total ionizing dose summed over all considered particle species & rad \\
  $t$                   & Time elapsed since the beginning of the exposure & s \\
  $T$                   & Transmittance of a coated sample at a wavelength of 532 nm & \% \\
  $\tau$                & Dielectric relaxation time of a coating material & s \\
  $x$                   & Penetration depth of radiation within a target material & nm, mm \\
  $Y$                   & Sputtering yield of a coating constituent & atoms/ion \\
  $Z$                   & Atomic number of an ion species & --- \\
\end{longtable}
\endgroup

\subsection{\textcolor{red}{Abbreviations and Acronyms}}

\begingroup
\leftcaptions
\begin{longtable}{@{}p{0.14\textwidth}|p{0.60\textwidth}|p{0.20\textwidth}@{}}
  Abbreviation & Description & Unit \\
  \midrule
  \endfirsthead
  \multicolumn{3}{@{}l}{\textit{Table \ref{tab:nomenclature_abbreviations} continued from previous page.}} \\
  Abbreviation & Description & Unit \\
  \midrule
  \endhead
  \midrule
  \multicolumn{3}{r@{}}{\textit{Continued on next page.}} \\
  \endfoot
  \hline \hline
  \caption{Abbreviations and acronyms used in the manuscript.}
  \label{tab:nomenclature_abbreviations} \\
  \endlastfoot
  AE8, AE9 & NASA models of the trapped electron radiation environment; AE9 is part of the IRENE model suite & --- \\
  AM0       & Air Mass Zero, i.e. the solar spectrum outside the Earth's atmosphere & --- \\
  AP8, AP9  & NASA models of the trapped proton radiation environment; AP9 is part of the IRENE model suite & --- \\
  AR        & Anti-Reflective (coating) & --- \\
  ASTM      & American Society for Testing and Materials & --- \\
  ATOX      & Atomic Oxygen & --- \\
  AU        & Astronomical Unit, i.e. the mean Sun--Earth distance & 1 AU = $1.496\times10^{8}$ km \\
  CAD       & Computer-Aided Design & --- \\
  CIF       & Complex Irradiation Facility at DLR-Bremen & --- \\
  COMPASSO & DLR in-orbit demonstration mission for compact optical clock technologies, hosted on the ISS Bartolomeo platform & --- \\
  CREME96   & Cosmic Ray Effects on Micro-Electronics, 1996 revision of the model & --- \\
  DLR       & Deutsches Zentrum f\"ur Luft- und Raumfahrt (German Aerospace Center) & --- \\
  e-beam    & Electron beam evaporation deposition technique & --- \\
  ECSS      & European Cooperation for Space Standardization & --- \\
  EEE       & Electrical, Electronic and Electromechanical (components) & --- \\
  EMR       & Electromagnetic Radiation & --- \\
  ESA       & European Space Agency & --- \\
  ESCC      & European Space Components Coordination & --- \\
  ESP       & Emission of Solar Protons model & --- \\
  EUV       & Extreme Ultraviolet range of the electromagnetic spectrum (10 nm to 121 nm) & --- \\
  EY        & Erosion Yield, i.e. the target material volume removed by a single atomic oxygen atom & $\mathrm{cm^{3}}$ ATOX atom$^{-1}$ \\
  GCR       & Galactic Cosmic Rays & --- \\
  GIXRD     & Grazing Incidence X-Ray Diffraction & --- \\
  GNSS      & Global Navigation Satellite System & --- \\
  GRAS      & Geant4 Radiation Analysis for Space, a 3D Monte Carlo simulation software & --- \\
  IC        & Iodine Cell & --- \\
  IR        & Iodine Reference, i.e. the iodine-based optical frequency reference subsystem & --- \\
  IRENE     & International Radiation Environment Near Earth, the AE9/AP9 model suite & --- \\
  ISI       & Institute of Scientific Instruments of the Czech Academy of Sciences, Brno & --- \\
  ISO       & International Organization for Standardization & --- \\
  ISS       & International Space Station & --- \\
  LEO       & Low Earth Orbit & --- \\
  LET       & Linear Energy Transfer & $\mathrm{MeV\ cm^{2}\ mg^{-1}}$ \\
  MAX       & Solar maximum activity state & --- \\
  MC        & Monte Carlo (simulation method) & --- \\
  MIN       & Solar minimum activity state & --- \\
  MISSE     & Materials International Space Station Experiment & --- \\
  NASA      & National Aeronautics and Space Administration & --- \\
  NCRE      & Nominal Corpuscular Radiation Environment & --- \\
  NIST      & National Institute of Standards and Technology & --- \\
  P5        & Peak 5-min average particle flux of the CREME96 model & --- \\
  PDR       & Preliminary Design Review & --- \\
  PIAD      & Plasma Ion-Assisted Deposition & --- \\
  PSTAR     & NIST database of proton stopping power and range in matter & --- \\
  QM        & Quiet state of the Earth's magnetosphere & --- \\
  RA        & Radiation Analysis & --- \\
  RAM       & Spacecraft side facing the direction of its orbital motion & --- \\
  RHA       & Radiation Hardness Assurance & --- \\
  SAA       & South Atlantic Anomaly & --- \\
  SC        & Solar Constant, i.e. the total solar irradiance at 1 AU & 1 SC = 1362 $\mathrm{W\ m^{-2}}$ \\
  SEE       & Single Event Effect & --- \\
  SM        & Stormy state of the Earth's magnetosphere & --- \\
  SPENVIS   & Space Environment Information System of ESA & --- \\
  SPI       & Solar Peak Ions & --- \\
  SPP       & Solar Peak Protons & --- \\
  SR        & System Requirements of the COMPASSO project & --- \\
  SRIM      & Stopping and Range of Ions in Matter software & --- \\
  TEM       & Transmission Electron Microscopy & --- \\
  TID       & Total Ionizing Dose & rad \\
  TNID      & Total Non-Ionizing Dose & $\mathrm{MeV\ g^{-1}}$ \\
  TSI       & Total Solar Irradiance & $\mathrm{W\ m^{-2}}$ \\
  UHV       & Ultra High Vacuum & --- \\
  UV        & Ultraviolet range of the electromagnetic spectrum (100 nm to 400 nm per ECSS) & --- \\
  VUV       & Vacuum Ultraviolet & --- \\
  WCRE      & Worst Case Corpuscular Radiation Environment & --- \\
  WD        & Worst day average particle flux of the CREME96 model & --- \\
  XUV       & Soft X-ray and extreme ultraviolet range (0.1 nm to 10 nm) & --- \\
\end{longtable}
\endgroup

\clearpage
%\bibliographystyle{elsarticle-num-names} 
%\bibliography{refs}

\begin{thebibliography}{67}
\expandafter\ifx\csname natexlab\endcsname\relax\def\natexlab#1{#1}\fi
\providecommand{\url}[1]{\texttt{#1}}
\providecommand{\href}[2]{#2}
\providecommand{\path}[1]{#1}
\providecommand{\DOIprefix}{doi:}
\providecommand{\ArXivprefix}{arXiv:}
\providecommand{\URLprefix}{URL: }
\providecommand{\Pubmedprefix}{pmid:}
\providecommand{\doi}[1]{\href{http://dx.doi.org/#1}{\path{#1}}}
\providecommand{\Pubmed}[1]{\href{pmid:#1}{\path{#1}}}
\providecommand{\bibinfo}[2]{#2}
\ifx\xfnm\relax \def\xfnm[#1]{\unskip,\space#1}\fi
%Type = Incollection
\bibitem[{Wernham and Piegari(2018)}]{WER18}
\bibinfo{author}{D.~Wernham}, \bibinfo{author}{A.~Piegari},
\newblock \bibinfo{title}{Chapter 22 - optical coatings in the space environment},
\newblock in: \bibinfo{editor}{A.~Piegari}, \bibinfo{editor}{F.~Flory} (Eds.), \bibinfo{booktitle}{Optical Thin Films and Coatings (Second Edition)}, Woodhead Publishing Series in Electronic and Optical Materials, \bibinfo{edition}{second edition} ed., \bibinfo{publisher}{Woodhead Publishing}, \bibinfo{year}{2018}, pp. \bibinfo{pages}{789--811}. \DOIprefix\doi{10.1016/B978-0-08-102073-9.00022-9}.
%Type = Inproceedings
\bibitem[{Kuhnhenn et~al.(2019)Kuhnhenn, Steffens, and Baum}]{KUH18}
\bibinfo{author}{J.~Kuhnhenn}, \bibinfo{author}{M.~Steffens}, \bibinfo{author}{M.~Baum},
\newblock \bibinfo{title}{{Radiation testing of optical coatings: better testing with simulations}},
\newblock in: \bibinfo{editor}{Z.~Sodnik}, \bibinfo{editor}{N.~Karafolas}, \bibinfo{editor}{B.~Cugny} (Eds.), \bibinfo{booktitle}{International Conference on Space Optics — ICSO 2018}, volume \bibinfo{volume}{11180}, \bibinfo{organization}{International Society for Optics and Photonics}, \bibinfo{publisher}{SPIE}, \bibinfo{year}{2019}, p. \bibinfo{pages}{111804L}. \DOIprefix\doi{10.1117/12.2536084}.
%Type = Article
\bibitem[{Giorgi et~al.(2019)Giorgi, Schmidt, Trainotti, Mata-Calvo, Fuchs, Hoque, Berdermann, Furthner, Günther, Schuldt, Sanjuan, Gohlke, Oswald, Braxmaier, Balidakis, Dick, Flechtner, Ge, Glaser, König, Michalak, Murböck, Semmling, and Schuh}]{GIO19}
\bibinfo{author}{G.~Giorgi}, \bibinfo{author}{T.~Schmidt}, \bibinfo{author}{C.~Trainotti}, \bibinfo{author}{R.~Mata-Calvo}, \bibinfo{author}{C.~Fuchs}, \bibinfo{author}{M.~Hoque}, \bibinfo{author}{J.~Berdermann}, \bibinfo{author}{J.~Furthner}, \bibinfo{author}{C.~Günther}, \bibinfo{author}{T.~Schuldt}, \bibinfo{author}{J.~Sanjuan}, \bibinfo{author}{M.~Gohlke}, \bibinfo{author}{M.~Oswald}, \bibinfo{author}{C.~Braxmaier}, \bibinfo{author}{K.~Balidakis}, \bibinfo{author}{G.~Dick}, \bibinfo{author}{F.~Flechtner}, \bibinfo{author}{M.~Ge}, \bibinfo{author}{S.~Glaser}, \bibinfo{author}{R.~König}, \bibinfo{author}{G.~Michalak}, \bibinfo{author}{M.~Murböck}, \bibinfo{author}{M.~Semmling}, \bibinfo{author}{H.~Schuh},
\newblock \bibinfo{title}{Advanced technologies for satellite navigation and geodesy},
\newblock \bibinfo{journal}{Advances in Space Research} \bibinfo{volume}{64} (\bibinfo{year}{2019}) \bibinfo{pages}{1256--1273}. \DOIprefix\doi{10.1016/j.asr.2019.06.010}.
%Type = Article
\bibitem[{{Schuldt} et~al.(2021){Schuldt}, {Gohlke}, {Oswald}, {W{\"u}st}, {Blomberg}, {D{\"o}ringshoff}, {Bawamia}, {Wicht}, {Lezius}, {Voss}, {Krutzik}, {Herrmann}, {Kovalchuk}, {Peters}, and {Braxmaier}}]{SCH21}
\bibinfo{author}{T.~{Schuldt}}, \bibinfo{author}{M.~{Gohlke}}, \bibinfo{author}{M.~{Oswald}}, \bibinfo{author}{J.~{W{\"u}st}}, \bibinfo{author}{T.~{Blomberg}}, \bibinfo{author}{K.~{D{\"o}ringshoff}}, \bibinfo{author}{A.~{Bawamia}}, \bibinfo{author}{A.~{Wicht}}, \bibinfo{author}{M.~{Lezius}}, \bibinfo{author}{K.~{Voss}}, \bibinfo{author}{M.~{Krutzik}}, \bibinfo{author}{S.~{Herrmann}}, \bibinfo{author}{E.~{Kovalchuk}}, \bibinfo{author}{A.~{Peters}}, \bibinfo{author}{C.~{Braxmaier}},
\newblock \bibinfo{title}{{Optical clock technologies for global navigation satellite systems}},
\newblock \bibinfo{journal}{GPS Solutions} \bibinfo{volume}{25} (\bibinfo{year}{2021}) \bibinfo{pages}{83}. \DOIprefix\doi{10.1007/s10291-021-01113-2}.
%Type = Inproceedings
\bibitem[{Steimle et~al.(2019)Steimle, Walz, Fuchs, Pedersen, and Lombardi}]{STE19}
\bibinfo{author}{C.~Steimle}, \bibinfo{author}{C.~Walz}, \bibinfo{author}{C.~Fuchs}, \bibinfo{author}{D.~Pedersen}, \bibinfo{author}{C.~Lombardi},
\newblock \bibinfo{title}{Bartolomeo external platform entering into commercial services},
\newblock \bibinfo{year}{2019}.
%Type = Article
\bibitem[{{Kuschewski} et~al.(2024){Kuschewski}, {W{\"u}st}, {Oswald}, {Blomberg}, {Gohlke}, {Bischof}, {Boac}, {Alam}, {Bußmeier}, {Abich}, {Röder}, {D{\"o}ringshoff}, {Holá}, {Oulehla}, {Schuldt}, and {Braxmaier}}]{KUS24}
\bibinfo{author}{F.~{Kuschewski}}, \bibinfo{author}{J.~{W{\"u}st}}, \bibinfo{author}{M.~{Oswald}}, \bibinfo{author}{T.~{Blomberg}}, \bibinfo{author}{M.~{Gohlke}}, \bibinfo{author}{J.~{Bischof}}, \bibinfo{author}{A.~{Boac}}, \bibinfo{author}{T.~{Alam}}, \bibinfo{author}{A.~{Bußmeier}}, \bibinfo{author}{K.~{Abich}}, \bibinfo{author}{N.~{Röder}}, \bibinfo{author}{K.~{D{\"o}ringshoff}}, \bibinfo{autor}{J.~{Hrabina}}, \bibinfo{author}{M.~{Holá}}, \bibinfo{author}{J.~{Oulehla}}, \bibinfo{author}{T.~{Schuldt}}, \bibinfo{author}{C.~{Braxmaier}},
\newblock \bibinfo{title}{{COMPASSO mission and its iodine clock: outline of the clock design}},
\newblock \bibinfo{journal}{GPS Solutions} \bibinfo{volume}{28} (\bibinfo{year}{2024}) \bibinfo{pages}{10}. \DOIprefix\doi{10.1007/s10291-023-01551-0}.
%Type = Article
\bibitem[{Hrabina et~al.(2017)Hrabina, Zucco, Philippe, Pham, Holá, Acef, Lazar, and Číp}]{Hrabina2017}
\bibinfo{author}{J.~Hrabina}, \bibinfo{author}{M.~Zucco}, \bibinfo{author}{C.~Philippe}, \bibinfo{author}{M.~Pham}, \bibinfo{author}{M.~Holá}, \bibinfo{author}{O.~Acef}, \bibinfo{author}{J.~Lazar}, \bibinfo{author}{O.~Číp},
\newblock \bibinfo{title}{Iodine absorption cells purity testing},
\newblock \bibinfo{journal}{Sensors} \bibinfo{volume}{17} (\bibinfo{year}{2017}) \bibinfo{pages}{1--13}. \DOIprefix\doi{10.3390/s17010102}.
%Type = Article
\bibitem[{Hrabina et~al.(2014)Hrabina, Šarbort, Acef, Du~Burck, Chiodo, Holá, Číp, and Lazar}]{Hrabina2014}
\bibinfo{author}{J.~Hrabina}, \bibinfo{author}{M.~Šarbort}, \bibinfo{author}{O.~Acef}, \bibinfo{author}{F.~Du~Burck}, \bibinfo{author}{N.~Chiodo}, \bibinfo{author}{M.~Holá}, \bibinfo{author}{O.~Číp}, \bibinfo{author}{J.~Lazar},
\newblock \bibinfo{title}{Spectral properties of molecular iodine in absorption cells filled to specified saturation pressure},
\newblock \bibinfo{journal}{Applied Optics} \bibinfo{volume}{53} (\bibinfo{year}{2014}) \bibinfo{pages}{7435--7441}. \DOIprefix\doi{10.1364/AO.53.007435}.
%Type = Article
\bibitem[{{Freden}(1969)}]{FSC69}
\bibinfo{author}{S.~C. {Freden}},
\newblock \bibinfo{title}{{Inner-Belt Van Allen Radiation}},
\newblock \bibinfo{journal}{Space Science Reviews} \bibinfo{volume}{9} (\bibinfo{year}{1969}) \bibinfo{pages}{198--242}. \DOIprefix\doi{10.1007/BF00215633}.
%Type = Article
\bibitem[{{Li} and {Hudson}(2019)}]{LH19}
\bibinfo{author}{W.~{Li}}, \bibinfo{author}{M.~K. {Hudson}},
\newblock \bibinfo{title}{{Earth's Van Allen Radiation Belts: From Discovery to the Van Allen Probes Era}},
\newblock \bibinfo{journal}{Journal of Geophysical Research (Space Physics)} \bibinfo{volume}{124} (\bibinfo{year}{2019}) \bibinfo{pages}{8319--8351}.
%Type = Article
\bibitem[{{Ripoll} et~al.(2020){Ripoll}, {Claudepierre}, {Ukhorskiy}, {Colpitts}, {Li}, {Fennell}, and {Crabtree}}]{RCU20}
\bibinfo{author}{J.~F. {Ripoll}}, \bibinfo{author}{S.~G. {Claudepierre}}, \bibinfo{author}{A.~Y. {Ukhorskiy}}, \bibinfo{author}{C.~{Colpitts}}, \bibinfo{author}{X.~{Li}}, \bibinfo{author}{J.~F. {Fennell}}, \bibinfo{author}{C.~{Crabtree}},
\newblock \bibinfo{title}{{Particle Dynamics in the Earth's Radiation Belts: Review of Current Research and Open Questions}},
\newblock \bibinfo{journal}{Journal of Geophysical Research (Space Physics)} \bibinfo{volume}{125} (\bibinfo{year}{2020}) \bibinfo{pages}{e26735}.
%Type = Book
\bibitem[{{Holmes-Siedle} and {Adams}(2012)}]{SA12}
\bibinfo{author}{A.~{Holmes-Siedle}}, \bibinfo{author}{L.~{Adams}}, \bibinfo{title}{Handbook of Radiation Effects}, \bibinfo{publisher}{Oxford University Press}, \bibinfo{address}{New York}, \bibinfo{year}{2012}.
%Type = Article
\bibitem[{Baker et~al.(2013)Baker, Kanekal, Hoxie, Henderson, Li, Spence, Elkington, Friedel, Goldstein, Hudson, Reeves, Thorne, Kletzing, and Claudepierre}]{BKH13}
\bibinfo{author}{D.~N. Baker}, \bibinfo{author}{S.~G. Kanekal}, \bibinfo{author}{V.~C. Hoxie}, \bibinfo{author}{M.~G. Henderson}, \bibinfo{author}{X.~Li}, \bibinfo{author}{H.~E. Spence}, \bibinfo{author}{S.~R. Elkington}, \bibinfo{author}{R.~H.~W. Friedel}, \bibinfo{author}{J.~Goldstein}, \bibinfo{author}{M.~K. Hudson}, \bibinfo{author}{G.~D. Reeves}, \bibinfo{author}{R.~M. Thorne}, \bibinfo{author}{C.~A. Kletzing}, \bibinfo{author}{S.~G. Claudepierre},
\newblock \bibinfo{title}{A long-lived relativistic electron storage ring embedded in earth{\textquoteright}s outer van allen belt},
\newblock \bibinfo{journal}{Science} \bibinfo{volume}{340} (\bibinfo{year}{2013}) \bibinfo{pages}{186--190}.
%Type = Article
\bibitem[{NASA(2010)}]{iss_2010}
\bibinfo{author}{NASA},
\newblock \bibinfo{title}{Reference guide to the international space station. – assembly complete edition.},
\newblock \bibinfo{journal}{NP-2010-09-682-HQ}  (\bibinfo{year}{2010}).
%Type = Article
\bibitem[{NASA(2015)}]{iss_2015}
\bibinfo{author}{NASA},
\newblock \bibinfo{title}{Reference guide to the international space station. – utlization edition.},
\newblock \bibinfo{journal}{NP-2015-05-022-JSC}  (\bibinfo{year}{2015}).
%Type = Article
\bibitem[{ECSS(2020)}]{ecss-04C}
\bibinfo{author}{ECSS},
\newblock \bibinfo{title}{Space engineering, space environment},
\newblock \bibinfo{journal}{ECSS-E-ST-10-04C Rev.1}  (\bibinfo{year}{2020}).
%Type = Misc
\bibitem[{Univesity(2023)}]{ap8min_ap8max}
\bibinfo{author}{V.~Univesity}, \bibinfo{title}{Ap8min and ap8max trapped proton models}, \bibinfo{howpublished}{https://creme.isde.vanderbilt.edu/CREME-MC/help/ap8min-and-ap8max-trapped-proton-models}, \bibinfo{year}{2023}.
%Type = Misc
\bibitem[{(ESCC)(2014)}]{escc-25100}
\bibinfo{author}{E.~S. C.~C. (ESCC)}, \bibinfo{title}{Single event effects test method and guidelines}, \bibinfo{year}{2014}.
%Type = Article
\bibitem[{{Tylka} et~al.(1997){Tylka}, {Adams}, {Boberg}, {Brownstein}, {Dietrich}, {Flueckiger}, {Petersen}, {Shea}, {Smart}, and {Smith}}]{TAB97}
\bibinfo{author}{A.~J. {Tylka}}, \bibinfo{author}{J.~H. {Adams}}, \bibinfo{author}{P.~R. {Boberg}}, \bibinfo{author}{B.~{Brownstein}}, \bibinfo{author}{W.~F. {Dietrich}}, \bibinfo{author}{E.~O. {Flueckiger}}, \bibinfo{author}{E.~L. {Petersen}}, \bibinfo{author}{M.~A. {Shea}}, \bibinfo{author}{D.~F. {Smart}}, \bibinfo{author}{E.~C. {Smith}},
\newblock \bibinfo{title}{{CREME96: a revision of the C\_osmic R\_ay E\_ffects on M\_icro-E\_lectronics code}},
\newblock \bibinfo{journal}{IEEE Transactions on Nuclear Science} \bibinfo{volume}{44} (\bibinfo{year}{1997}) \bibinfo{pages}{2150--2160}. \DOIprefix\doi{10.1109/23.659030}.
%Type = Techreport
\bibitem[{Simonsen et~al.(1992)Simonsen, Atwell, Nealy, and Cucinotta}]{SAN92}
\bibinfo{author}{L.~C. Simonsen}, \bibinfo{author}{W.~Atwell}, \bibinfo{author}{J.~E. Nealy}, \bibinfo{author}{F.~A. Cucinotta}, \bibinfo{title}{Radiation Dose to Critical Body Organs for October 1989 Proton Event}, \bibinfo{type}{Technical Report} \bibinfo{number}{NASA-TP-3237}, NASA, \bibinfo{year}{1992}. \URLprefix \url{https://ntrs.nasa.gov/api/citations/19930003688/downloads/19930003688.pdf}.
%Type = Article
\bibitem[{{St{\o}rmer}(1930)}]{STO30}
\bibinfo{author}{C.~{St{\o}rmer}},
\newblock \bibinfo{journal}{Astrophysics} \bibinfo{volume}{1} (\bibinfo{year}{1930}) \bibinfo{pages}{237}.
%Type = Article
\bibitem[{{Adams} et~al.(1981){Adams}, {Silberberg}, and {Tsao}}]{AST81}
\bibinfo{author}{J.~{Adams}, J.~H.}, \bibinfo{author}{R.~{Silberberg}}, \bibinfo{author}{C.~H. {Tsao}},
\newblock \bibinfo{title}{{Cosmic ray effects on microelectronics. Part 1: The near-Earth particle environment}},
\newblock \bibinfo{journal}{NAVAL RESEARCH LAB WASHINGTON DC}  (\bibinfo{year}{1981}) \bibinfo{pages}{34134}.
%Type = Article
\bibitem[{{de Rooij}(2010)}]{ADR10}
\bibinfo{author}{A.~{de Rooij}},
\newblock \bibinfo{title}{{Corrosion in Space}},
\newblock \bibinfo{journal}{Encyclopedia of Aerospace Engineering}  (\bibinfo{year}{2010}) \bibinfo{pages}{1--10}. \DOIprefix\doi{10.1002/9780470686652.eae242}.
%Type = Article
\bibitem[{NASA(1994)}]{NAS94}
\bibinfo{author}{NASA},
\newblock \bibinfo{title}{{Space Station Ionizing Radiation Design Environment. International Space Station Alpha.}},
\newblock \bibinfo{journal}{SSP 30512 Revision C}  (\bibinfo{year}{1994}).
%Type = Article
\bibitem[{ASTM(2022)}]{ASTM}
\bibinfo{author}{ASTM},
\newblock \bibinfo{title}{{Standard solar constant and zero air mass solar spectral irradiance tables}},
\newblock \bibinfo{journal}{ASTM-E490}  (\bibinfo{year}{2022}).
%Type = Article
\bibitem[{Gueymard(2004)}]{GUE04}
\bibinfo{author}{C.~A. Gueymard},
\newblock \bibinfo{title}{The sun’s total and spectral irradiance for solar energy applications and solar radiation models},
\newblock \bibinfo{journal}{Solar Energy} \bibinfo{volume}{76} (\bibinfo{year}{2004}) \bibinfo{pages}{423--453}. \DOIprefix\doi{j.solener.2003.08.039}.
%Type = Article
\bibitem[{{Nicoletta} and {Eubanks}(1972)}]{NEU72}
\bibinfo{author}{C.~A. {Nicoletta}}, \bibinfo{author}{A.~G. {Eubanks}},
\newblock \bibinfo{title}{{Effect of simulated space radiation on selected optical materials}},
\newblock \bibinfo{journal}{Applied Optics} \bibinfo{volume}{11} (\bibinfo{year}{1972}) \bibinfo{pages}{1365}. \DOIprefix\doi{10.1364/AO.11.001365}.
%Type = Misc
\bibitem[{{Silverman}(1995)}]{SEM95}
\bibinfo{author}{E.~M. {Silverman}}, \bibinfo{title}{{Space environmental effects on spacecraft: LEO materials selection guide, part 2}}, \bibinfo{howpublished}{Progress Report, Apr. 1993 - Mar. 1995 TRW, Inc., Redondo Beach, CA. Space and Electronics Group.}, \bibinfo{year}{1995}.
%Type = Article
\bibitem[{{Sigmund}(1969)}]{SIGP69}
\bibinfo{author}{P.~{Sigmund}},
\newblock \bibinfo{title}{{Theory of Sputtering. I. Sputtering Yield of Amorphous and Polycrystalline Targets}},
\newblock \bibinfo{journal}{Physical Review} \bibinfo{volume}{184} (\bibinfo{year}{1969}) \bibinfo{pages}{383--416}. \DOIprefix\doi{10.1103/PhysRev.184.383}.
%Type = Article
\bibitem[{{MacDonald}(1970)}]{MACD70}
\bibinfo{author}{R.~J. {MacDonald}},
\newblock \bibinfo{title}{{The ejection of atomic particles from ion bombarded solids}},
\newblock \bibinfo{journal}{Advances in Physics} \bibinfo{volume}{19} (\bibinfo{year}{1970}) \bibinfo{pages}{457--524}. \DOIprefix\doi{10.1080/00018737000101161}.
%Type = Article
\bibitem[{{de Rooij}(1989)}]{ADR89}
\bibinfo{author}{A.~{de Rooij}},
\newblock \bibinfo{title}{{The Oxidation of Silver by Atomic Oxygen}},
\newblock \bibinfo{journal}{ESA Journal} \bibinfo{volume}{13} (\bibinfo{year}{1989}) \bibinfo{pages}{363--382}.
%Type = Article
\bibitem[{Silverman(1995)}]{ldef1}
\bibinfo{author}{E.~M. Silverman},
\newblock \bibinfo{title}{Space environmental effects on spacecraft: Leo materials selection guide},
\newblock \bibinfo{journal}{NASA Contractor Report} \bibinfo{volume}{4661} (\bibinfo{year}{1995}).
%Type = Techreport
\bibitem[{{de Groh} and Banks(2019)}]{DGB19}
\bibinfo{author}{K.~K. {de Groh}}, \bibinfo{author}{B.~A. Banks}, \bibinfo{title}{{Atomic Oxygen Erosion Data From the MISSE 2--8 Missions}}, \bibinfo{type}{Technical Report} \bibinfo{number}{NASA/TM--2019-219982}, NASA Glenn Research Center, \bibinfo{year}{2019}.
%Type = Article
\bibitem[{Ziegler et~al.(2010)Ziegler, Ziegler, and Biersack}]{ZZB10}
\bibinfo{author}{J.~F. Ziegler}, \bibinfo{author}{M.~Ziegler}, \bibinfo{author}{J.~Biersack},
\newblock \bibinfo{title}{Srim – the stopping and range of ions in matter (2010)},
\newblock \bibinfo{journal}{Nuclear Instruments and Methods in Physics Research Section B: Beam Interactions with Materials and Atoms} \bibinfo{volume}{268} (\bibinfo{year}{2010}) \bibinfo{pages}{1818--1823}. \DOIprefix\doi{https://doi.org/10.1016/j.nimb.2010.02.091}, \bibinfo{note}{19th International Conference on Ion Beam Analysis}.
%Type = Article
\bibitem[{Oulehla et~al.(2019)Oulehla, Pokorný, Hrabina, Holá, Číp, and Lazar}]{Oulehla2019}
\bibinfo{author}{J.~Oulehla}, \bibinfo{author}{P.~Pokorný}, \bibinfo{author}{J.~Hrabina}, \bibinfo{author}{M.~Holá}, \bibinfo{author}{O.~Číp}, \bibinfo{author}{J.~Lazar},
\newblock \bibinfo{title}{Influence of coating technology and thermal annealing on the optical performance of ar coatings in iodine-filled absorption cells},
\newblock \bibinfo{journal}{Optics Express} \bibinfo{volume}{27} (\bibinfo{year}{2019}) \bibinfo{pages}{9361--9371}. \DOIprefix\doi{10.1364/OE.27.009361}.
%Type = Article
\bibitem[{Renger et~al.(2014)Renger, Sznajder, Witzke, and Geppert}]{renger}
\bibinfo{author}{T.~Renger}, \bibinfo{author}{M.~Sznajder}, \bibinfo{author}{A.~Witzke}, \bibinfo{author}{U.~Geppert},
\newblock \bibinfo{title}{The complex irradiation facility at dlr-bremen},
\newblock \bibinfo{journal}{Journal of Materials Science and Engineering A} \bibinfo{volume}{4} (\bibinfo{year}{2014}) \bibinfo{pages}{1}.
%Type = Article
\bibitem[{Jiang(2015)}]{JIA16}
\bibinfo{author}{N.~Jiang},
\newblock \bibinfo{title}{Electron beam damage in oxides: a review},
\newblock \bibinfo{journal}{Reports on Progress in Physics} \bibinfo{volume}{79} (\bibinfo{year}{2015}) \bibinfo{pages}{016501}. \DOIprefix\doi{10.1088/0034-4885/79/1/016501}.
%Type = Article
\bibitem[{Corso et~al.(2024)Corso, Padovani, Santi, H{\"u}bner, Kentsch, Bazzan, and Pelizzo}]{CPS24}
\bibinfo{author}{A.~J. Corso}, \bibinfo{author}{M.~Padovani}, \bibinfo{author}{G.~Santi}, \bibinfo{author}{R.~H{\"u}bner}, \bibinfo{author}{U.~Kentsch}, \bibinfo{author}{M.~Bazzan}, \bibinfo{author}{M.~G. Pelizzo},
\newblock \bibinfo{title}{Optical thin films in space environment: Investigation of proton irradiation damage},
\newblock \bibinfo{journal}{ACS Applied Materials \& Interfaces} \bibinfo{volume}{16} (\bibinfo{year}{2024}) \bibinfo{pages}{38645--38657}. \DOIprefix\doi{10.1021/acsami.4c03362}.
%Type = Article
\bibitem[{Nardello et~al.(2013)Nardello, Zuppella, Polito, Corso, Zuccon, and Pelizzo}]{NZP13}
\bibinfo{author}{M.~Nardello}, \bibinfo{author}{P.~Zuppella}, \bibinfo{author}{V.~Polito}, \bibinfo{author}{A.~J. Corso}, \bibinfo{author}{S.~Zuccon}, \bibinfo{author}{M.~G. Pelizzo},
\newblock \bibinfo{title}{Stability of euv multilayer coatings to low energy alpha particles bombardment},
\newblock \bibinfo{journal}{Optics Express} \bibinfo{volume}{21} (\bibinfo{year}{2013}) \bibinfo{pages}{28334--28343}. \DOIprefix\doi{10.1364/OE.21.028334}.
%Type = Article
\bibitem[{Kumar et~al.(2018)Kumar, Jaiswal, Gupta, Ram, Sulania, Ojha, Sun, Koratkar, and Kumar}]{KJG18}
\bibinfo{author}{V.~Kumar}, \bibinfo{author}{M.~K. Jaiswal}, \bibinfo{author}{R.~Gupta}, \bibinfo{author}{J.~Ram}, \bibinfo{author}{I.~Sulania}, \bibinfo{author}{S.~Ojha}, \bibinfo{author}{X.~Sun}, \bibinfo{author}{N.~Koratkar}, \bibinfo{author}{R.~Kumar},
\newblock \bibinfo{title}{Effect of low energy (kev) ion irradiation on structural, optical and morphological properties of sno$_2$--tio$_2$ nanocomposite thin films},
\newblock \bibinfo{journal}{Journal of Materials Science: Materials in Electronics} \bibinfo{volume}{29} (\bibinfo{year}{2018}) \bibinfo{pages}{13328--13336}. \DOIprefix\doi{10.1007/s10854-018-9457-6}.
%Type = Article
\bibitem[{Sarcina et~al.(2014)Sarcina, Grilli, Menchini, Piegari, Scaglione, Sytchkova, and Zola}]{DGM14}
\bibinfo{author}{I.~D. Sarcina}, \bibinfo{author}{M.~L. Grilli}, \bibinfo{author}{F.~Menchini}, \bibinfo{author}{A.~Piegari}, \bibinfo{author}{S.~Scaglione}, \bibinfo{author}{A.~Sytchkova}, \bibinfo{author}{D.~Zola},
\newblock \bibinfo{title}{Behavior of optical thin-film materials and coatings under proton and gamma irradiation},
\newblock \bibinfo{journal}{Applied Optics} \bibinfo{volume}{53} (\bibinfo{year}{2014}) \bibinfo{pages}{A314--A320}. \DOIprefix\doi{10.1364/AO.53.00A314}.
%Type = Article
\bibitem[{Woo and Hwangbo(2006)}]{WOO}
\bibinfo{author}{S.-H. Woo}, \bibinfo{author}{C.~K. Hwangbo},
\newblock \bibinfo{title}{Effects of annealing on the optical, structural, and chemical properties of tio2 and mgf2 thin films prepared by plasma ion-assisted deposition},
\newblock \bibinfo{journal}{Applied Optics} \bibinfo{volume}{45} (\bibinfo{year}{2006}) \bibinfo{pages}{1447--1455}. \DOIprefix\doi{10.1364/AO.45.001447}.
%Type = Article
\bibitem[{Martin et~al.(1997)Martin, Rousselot, Rondot, Palmino, and Mercier}]{MARTIN}
\bibinfo{author}{N.~Martin}, \bibinfo{author}{C.~Rousselot}, \bibinfo{author}{D.~Rondot}, \bibinfo{author}{F.~Palmino}, \bibinfo{author}{R.~Mercier},
\newblock \bibinfo{title}{Microstructure modification of amorphous titanium oxide thin films during annealing treatment},
\newblock \bibinfo{journal}{Thin Solid Films} \bibinfo{volume}{300} (\bibinfo{year}{1997}) \bibinfo{pages}{113--121}. \DOIprefix\doi{https://doi.org/10.1016/S0040-6090(96)09510-7}.
%Type = Article
\bibitem[{Xu et~al.(2009)Xu, Dong, Yuan, He, Shao, and Fan}]{Xu2009}
\bibinfo{author}{C.~Xu}, \bibinfo{author}{H.~Dong}, \bibinfo{author}{L.~Yuan}, \bibinfo{author}{H.~He}, \bibinfo{author}{J.~Shao}, \bibinfo{author}{Z.~Fan},
\newblock \bibinfo{title}{Investigation of annealing effects on the laser-induced damage threshold of amorphous ta2o5 films},
\newblock \bibinfo{journal}{Optics \& Laser Technology} \bibinfo{volume}{41} (\bibinfo{year}{2009}) \bibinfo{pages}{258--263}. \URLprefix \url{https://www.sciencedirect.com/science/article/pii/S003039920800131X}. \DOIprefix\doi{https://doi.org/10.1016/j.optlastec.2008.06.009}.
%Type = Article
\bibitem[{Xu et~al.(2008)Xu, Xiao, Ma, Jin, Shao, and Fan}]{Xu2008}
\bibinfo{author}{C.~Xu}, \bibinfo{author}{Q.~Xiao}, \bibinfo{author}{J.~Ma}, \bibinfo{author}{Y.~Jin}, \bibinfo{author}{J.~Shao}, \bibinfo{author}{Z.~Fan},
\newblock \bibinfo{title}{High temperature annealing effect on structure, optical property and laser-induced damage threshold of ta2o5 films},
\newblock \bibinfo{journal}{Applied Surface Science} \bibinfo{volume}{254} (\bibinfo{year}{2008}) \bibinfo{pages}{6554--6559}. \URLprefix \url{https://www.sciencedirect.com/science/article/pii/S0169433208007265}. \DOIprefix\doi{https://doi.org/10.1016/j.apsusc.2008.04.034}.
%Type = Article
\bibitem[{Favaro et~al.(2024)Favaro, Milotti, Alonso Diaz~Riega, Busdon, Bazzan, Granata, Hofman, Michel, Pinard, Conti, Capaccioli, Shcheblanov, Lema{\^i}tre, Martinez, Cagnoli, Piergiovanni, Fabrizi, Pelizzo, and Corso}]{Favaro2024}
\bibinfo{author}{G.~Favaro}, \bibinfo{author}{V.~Milotti}, \bibinfo{author}{D.~Alonso Diaz~Riega}, \bibinfo{author}{N.~Busdon}, \bibinfo{author}{M.~Bazzan}, \bibinfo{author}{M.~Granata}, \bibinfo{author}{D.~Hofman}, \bibinfo{author}{C.~Michel}, \bibinfo{author}{L.~Pinard}, \bibinfo{author}{L.~Conti}, \bibinfo{author}{S.~Capaccioli}, \bibinfo{author}{N.~S. Shcheblanov}, \bibinfo{author}{A.~Lema{\^i}tre}, \bibinfo{author}{V.~Martinez}, \bibinfo{author}{G.~Cagnoli}, \bibinfo{author}{F.~Piergiovanni}, \bibinfo{author}{F.~Fabrizi}, \bibinfo{author}{M.~G. Pelizzo}, \bibinfo{author}{A.~J. Corso},
\newblock \bibinfo{title}{Reduction of mechanical losses in ion-beam sputtered tantalum oxide thin films via partial crystallization},
\newblock \bibinfo{journal}{Classical and Quantum Gravity} \bibinfo{volume}{41} (\bibinfo{year}{2024}) \bibinfo{pages}{105009}. \DOIprefix\doi{10.1088/1361-6382/ad3c8a}.
%Type = Article
\bibitem[{He et~al.(2009)He, Wu, Li, Gao, Gan, and Zhao}]{HE2009}
\bibinfo{author}{X.~He}, \bibinfo{author}{J.~Wu}, \bibinfo{author}{X.~Li}, \bibinfo{author}{X.~Gao}, \bibinfo{author}{X.~Gan}, \bibinfo{author}{L.~Zhao},
\newblock \bibinfo{title}{Effects of the post-annealing ambience on the microstructure and optical properties of tantalum oxide films prepared by pulsed laser deposition},
\newblock \bibinfo{journal}{Journal of Alloys and Compounds} \bibinfo{volume}{478} (\bibinfo{year}{2009}) \bibinfo{pages}{453--457}. \URLprefix \url{https://www.sciencedirect.com/science/article/pii/S0925838808020616}. \DOIprefix\doi{https://doi.org/10.1016/j.jallcom.2008.11.077}.
%Type = Article
\bibitem[{Masse et~al.(2006)Masse, Szymanowski, Zabeida, Amassian, Klemberg-Sapieha, and Martinu}]{MASSE2006}
\bibinfo{author}{J.-P. Masse}, \bibinfo{author}{H.~Szymanowski}, \bibinfo{author}{O.~Zabeida}, \bibinfo{author}{A.~Amassian}, \bibinfo{author}{J.~Klemberg-Sapieha}, \bibinfo{author}{L.~Martinu},
\newblock \bibinfo{title}{Stability and effect of annealing on the optical properties of plasma-deposited ta2o5 and nb2o5 films},
\newblock \bibinfo{journal}{Thin Solid Films} \bibinfo{volume}{515} (\bibinfo{year}{2006}) \bibinfo{pages}{1674--1682}. \URLprefix \url{https://www.sciencedirect.com/science/article/pii/S0040609006006882}. \DOIprefix\doi{https://doi.org/10.1016/j.tsf.2006.05.047}.
%Type = Article
\bibitem[{Ren et~al.(2021)Ren, Yang, Feng, Miao, Xia, and Wang}]{Ren2021}
\bibinfo{author}{W.~Ren}, \bibinfo{author}{G.-D. Yang}, \bibinfo{author}{A.-L. Feng}, \bibinfo{author}{R.-X. Miao}, \bibinfo{author}{J.-B. Xia}, \bibinfo{author}{Y.-G. Wang},
\newblock \bibinfo{title}{Annealing effects on the optical and electrochemical properties of tantalum pentoxide films},
\newblock \bibinfo{journal}{Journal of Advanced Ceramics} \bibinfo{volume}{10} (\bibinfo{year}{2021}) \bibinfo{pages}{704--713}. \URLprefix \url{https://doi.org/10.1007/s40145-021-0465-2}. \DOIprefix\doi{10.1007/s40145-021-0465-2}.
%Type = Article
\bibitem[{Santin et~al.(2005)Santin, Ivanchenko, Evans, Nieminen, and Daly}]{SIE05}
\bibinfo{author}{G.~Santin}, \bibinfo{author}{V.~Ivanchenko}, \bibinfo{author}{H.~Evans}, \bibinfo{author}{P.~Nieminen}, \bibinfo{author}{E.~Daly},
\newblock \bibinfo{title}{Gras: a general-purpose 3-d modular simulation tool for space environment effects analysis},
\newblock \bibinfo{journal}{IEEE Transactions on Nuclear Science} \bibinfo{volume}{52} (\bibinfo{year}{2005}) \bibinfo{pages}{2294--2299}. \DOIprefix\doi{10.1109/TNS.2005.860749}.
%Type = Article
\bibitem[{J. et~al.(1975)J., C., and D.}]{TPK75}
\bibinfo{author}{T.~M. J.}, \bibinfo{author}{P.~B. C.}, \bibinfo{author}{K.~B. D.},
\newblock \bibinfo{title}{Transient radiation effects in optical materials},
\newblock \bibinfo{journal}{Intelcom Rad Tech Inc. Report No. SAMSO-TR-75-174}  (\bibinfo{year}{1975}).
%Type = Article
\bibitem[{R. and H.(1992)}]{WWR92}
\bibinfo{author}{W.~G. R.}, \bibinfo{author}{W.~R. H.},
\newblock \bibinfo{title}{Effects of ionizing radiation on selected optical materials: An overview},
\newblock \bibinfo{journal}{Report No. UCRL-ID-111453}  (\bibinfo{year}{1992}).
%Type = Article
\bibitem[{{Katz} and {Penfold}(1952)}]{KPA52}
\bibinfo{author}{L.~{Katz}}, \bibinfo{author}{A.~S. {Penfold}},
\newblock \bibinfo{title}{{Range-Energy Relations for Electrons and the Determination of Beta-Ray End-Point Energies by Absorption}},
\newblock \bibinfo{journal}{Reviews of Modern Physics} \bibinfo{volume}{24} (\bibinfo{year}{1952}) \bibinfo{pages}{28--44}. \DOIprefix\doi{10.1103/RevModPhys.24.28}.
%Type = Misc
\bibitem[{{Berger}(1992)}]{NIST92}
\bibinfo{author}{M.~J. {Berger}}, \bibinfo{title}{{ESTAR, PSTAR, and ASTAR: Computer programs for calculating stopping-power and range tables for electrons, protons, and helium ions}}, \bibinfo{year}{1992}.
%Type = Misc
\bibitem[{SPENVIS(2018)}]{spenvis}
\bibinfo{author}{SPENVIS}, \bibinfo{title}{Royal belgian institute for space aeronomy \& european space agency, space environment information system (spenvis)}, \bibinfo{year}{2018}.
%Type = Article
\bibitem[{ISO(2010)}]{iso-15856}
\bibinfo{author}{ISO},
\newblock \bibinfo{title}{Space systems - space environment - simulation guidelines for radiation exposure of non-metallic materials},
\newblock \bibinfo{journal}{ISO 15856}  (\bibinfo{year}{2010}).
%Type = Article
\bibitem[{ECSS(2019)}]{ecss-e-st-20-06C}
\bibinfo{author}{ECSS},
\newblock \bibinfo{title}{Space engineering, spacecraft charging},
\newblock \bibinfo{journal}{ECSS-E-ST-20-06C Rev.1}  (\bibinfo{year}{2019}).
%Type = Article
\bibitem[{ECSS(2008)}]{ecss-q-st-70-06C}
\bibinfo{author}{ECSS},
\newblock \bibinfo{title}{Space product assurance, particle and uv radiation testing for space materials},
\newblock \bibinfo{journal}{ECSS-Q-ST-70-06C}  (\bibinfo{year}{2008}).
%Type = Article
\bibitem[{ECSS(2010{\natexlab{a}})}]{ecss-e-hb-10-12A}
\bibinfo{author}{ECSS},
\newblock \bibinfo{title}{Space engineering, calculation of radiation and its effects and margin policy handbook},
\newblock \bibinfo{journal}{ECSS-E-HB-10-12A}  (\bibinfo{year}{2010}{\natexlab{a}}).
%Type = Article
\bibitem[{ECSS(2010{\natexlab{b}})}]{ecss-e-st-10-12C}
\bibinfo{author}{ECSS},
\newblock \bibinfo{title}{Space engineering, methods for the calculation of radiation received and its effects, and a policy for design margins},
\newblock \bibinfo{journal}{ECSS-E-ST-10-12C}  (\bibinfo{year}{2010}{\natexlab{b}}).
%Type = Article
\bibitem[{{Sorensen} et~al.(2000){Sorensen}, {Rodgers}, {Ryden}, {Latham}, {Wrenn}, {Levy}, and {Panabiere}}]{SRR00}
\bibinfo{author}{J.~{Sorensen}}, \bibinfo{author}{D.~J. {Rodgers}}, \bibinfo{author}{K.~A. {Ryden}}, \bibinfo{author}{P.~M. {Latham}}, \bibinfo{author}{G.~L. {Wrenn}}, \bibinfo{author}{L.~{Levy}}, \bibinfo{author}{G.~{Panabiere}},
\newblock \bibinfo{title}{{ESA's tools for internal charging}},
\newblock \bibinfo{journal}{IEEE Transactions on Nuclear Science} \bibinfo{volume}{47} (\bibinfo{year}{2000}) \bibinfo{pages}{491--497}. \DOIprefix\doi{10.1109/23.856470}.
%Type = Book
\bibitem[{Sze and Ng(2007)}]{SZE07}
\bibinfo{author}{S.~M. Sze}, \bibinfo{author}{K.~K. Ng}, \bibinfo{title}{Physics of Semiconductor Devices}, \bibinfo{edition}{3} ed., \bibinfo{publisher}{John Wiley \& Sons}, \bibinfo{address}{Hoboken, New Jersey}, \bibinfo{year}{2007}. \DOIprefix\doi{10.1002/0470068329}.
%Type = Article
\bibitem[{Robertson(2004)}]{ROB04}
\bibinfo{author}{J.~Robertson},
\newblock \bibinfo{title}{High dielectric constant oxides},
\newblock \bibinfo{journal}{The European Physical Journal Applied Physics} \bibinfo{volume}{28} (\bibinfo{year}{2004}) \bibinfo{pages}{265--291}. \DOIprefix\doi{10.1051/epjap:2004206}.
%Type = Article
\bibitem[{Klein and Gafni(1966)}]{KG66}
\bibinfo{author}{N.~Klein}, \bibinfo{author}{H.~Gafni},
\newblock \bibinfo{title}{The maximum dielectric strength of thin silicon oxide films},
\newblock \bibinfo{journal}{IEEE Transactions on Electron Devices} \bibinfo{volume}{ED-13} (\bibinfo{year}{1966}) \bibinfo{pages}{281--289}. \DOIprefix\doi{10.1109/T-ED.1966.15681}.
%Type = Article
\bibitem[{Moshnikov et~al.(2026)Moshnikov, Muratova, Vrublevsky, Maximov, Ryabko, Gagarina, and Kozodaev}]{MMV26}
\bibinfo{author}{V.~A. Moshnikov}, \bibinfo{author}{E.~N. Muratova}, \bibinfo{author}{I.~A. Vrublevsky}, \bibinfo{author}{A.~I. Maximov}, \bibinfo{author}{A.~A. Ryabko}, \bibinfo{author}{A.~Y. Gagarina}, \bibinfo{author}{D.~A. Kozodaev},
\newblock \bibinfo{title}{Electrical properties and charge transfer mechanisms in nanoscale anodic tio2 films at low applied voltages},
\newblock \bibinfo{journal}{Inorganics} \bibinfo{volume}{14} (\bibinfo{year}{2026}). \DOIprefix\doi{10.3390/inorganics14010029}.
%Type = Article
\bibitem[{Dutta et~al.(2014)Dutta, Leeladhar, Pandey, Thakur, and Pal}]{DLP14}
\bibinfo{author}{S.~Dutta}, \bibinfo{author}{Leeladhar}, \bibinfo{author}{A.~Pandey}, \bibinfo{author}{O.~P. Thakur}, \bibinfo{author}{R.~Pal},
\newblock \bibinfo{title}{Electrical properties of ultrathin titanium dioxide films on silicon},
\newblock \bibinfo{journal}{Journal of Vacuum Science \& Technology A} \bibinfo{volume}{33} (\bibinfo{year}{2014}) \bibinfo{pages}{021507}. \DOIprefix\doi{10.1116/1.4904978}.
%Type = Article
\bibitem[{Chen et~al.(2025)Chen, Niang, Bakhit, Jeon, Driscoll, and Flewitt}]{CNB25}
\bibinfo{author}{X.~Chen}, \bibinfo{author}{K.~M. Niang}, \bibinfo{author}{B.~Bakhit}, \bibinfo{author}{Y.~Jeon}, \bibinfo{author}{J.~Driscoll}, \bibinfo{author}{A.~Flewitt},
\newblock \bibinfo{title}{Atomic layer deposited ta2o5: From process optimization to thin film characterization},
\newblock \bibinfo{journal}{AIP Advances} \bibinfo{volume}{15} (\bibinfo{year}{2025}) \bibinfo{pages}{045008}. \DOIprefix\doi{10.1063/6.0004254}.

\end{thebibliography}

\providecommand{\noopsort}[1]{}\providecommand{\singleletter}[1]{#1}%

\end{document}